\documentclass[nofootinbib,eqsecnum,tightenlines,10pt]{revtex4}

\usepackage{graphicx}
\usepackage{amsmath,amssymb,amsfonts,amsthm,stmaryrd,mathtools,bm}
\usepackage{mathrsfs}
\usepackage{color}
\usepackage{multirow,bigdelim}
\usepackage{hyperref}
\usepackage{dsfont}
\usepackage{booktabs}
\allowdisplaybreaks[0]

\usepackage{pstricks}
\usepackage{tikz}
\usepackage{ulem}

\newcommand{\JP}[1]{{\leavevmode\color{black}{#1}}}
\newcommand{\JPp}[1]{{\leavevmode\color{black}{#1}}}

\def\be#1\ee{\begin{align}#1\end{align}}

\def\ba{\begin{eqnarray}}
\def\ea{\end{eqnarray}}
\def\nn{\nonumber}
\def\q{\quad}

\begin{document}

\title{Effective Spin Foam Models for Lorentzian Quantum Gravity}

\author{Seth K. Asante}
\affiliation{Perimeter Institute, 31 Caroline Street North, Waterloo, ON, N2L 2Y5, CAN}
\author{Bianca Dittrich}
\affiliation{Perimeter Institute, 31 Caroline Street North, Waterloo, ON, N2L 2Y5, CAN}
\author{Jos\'e Padua-Arg\"uelles}
\affiliation{Perimeter Institute, 31 Caroline Street North, Waterloo, ON, N2L 2Y5, CAN}

\begin{abstract}
Making the Lorentzian path integral for quantum gravity well-defined and computable has been a long standing challenge. In this work we adopt the recently proposed effective spin foam models to the Lorentzian case. This defines a path integral over discrete Lorentzian quantum geometric configurations, which include metric and torsion degrees of freedom. The torsion degrees of freedom arise due to an anomaly, which is parametrized by the Barbero--Immirzi parameter. Requiring a semi-classical regime  constrains this parameter, but the precise bound has to be determined by probing the dynamics.
The effective models provide the computationally most efficient  spin foam models yet, which allows us to perform first tests for determining the semi-classical regime. This includes explorations specific to the Lorentzian case, e.g. investigating quantum geometries with null lengths and null areas as well as geometries that describe a change of spatial topology.

\end{abstract}

\maketitle

\section{Introduction}

General relativity describes the gravitational force as a property of space-time geometry. A promising strategy to obtain a theory of quantum gravity is to first construct a notion of quantum geometry, and then to impose a gravitational dynamics for these quantum geometries. 

This is the starting point of loop quantum gravity \cite{LQG,ThiemannBook}, which offers a very precise, but in some aspects also surprising,  notion of quantum geometry. One key property is that the spectra of spatial area operators in loop quantum gravity are discrete \cite{DiscreteGeom1,Conrady}. We will see that this has important implications for the type of dynamics that can be imposed on such quantum geometries.

Spin foam models are path integrals for quantum gravity, defined as sums over these quantum geometries. These quantum geometries are easiest to define by basing them on a given triangulation.  The triangulation can be understood as a regulator, which can be removed by taking the refinement limit \cite{Improved,DittrichReview14}.  To this end one has to be able to understand and partially solve the dynamics of these models \cite{Dittrich12}. 

In contrast to so-called Euclidean quantum gravity approaches, spin foam models do not rely on a Wick rotation to define the path integral. They thus have also the potential to avoid one of the  main obstacles of Euclidean quantum gravity, the conformal factor problem \cite{Confactor}. Spin foam amplitudes can be defined for geometries of Euclidean signature or Lorentzian signature \cite{EPRL-FK}. As the spin foam construction is based on the canonical quantization method of loop quantum gravity, the EPRL-FK models \cite{EPRL-FK} were initially restricted to triangulations of Lorentzian four-dimensional geometries, in which all tetrahedra, and thus all triangles and edges, are space-like. This restriction has been removed by work of Conrady and Hnybida \cite{Conrady}, which also derives a quantization condition for  time-like areas.

We will see in this work that there are many interesting open questions for the Lorentzian path integral for quantum gravity, which apply to many quantum gravity approaches.  But the properties of  the Lorentzian path integral for quantum gravity have been barely explored \cite{Job}. One reason is that (standard) Monte-Carlo simulations cannot be applied, and there are only few \JPp{techniques} available to evaluate  highly oscillatory integrals over many variables. Spin foams \JPp{somewhat alleviate this situation}, as the integrals are actually sums over the discrete area values.  At the same time the amplitudes of the EPRL-FK models are highly complicated and even for the simplest building block, difficult to compute \cite{Dona:2019dkf}. 

It is thus important to develop techniques to be able to tackle these complicated models. Recent work includes the improvement of direct summations algorithms \cite{Gozzini}, the adaptation of new techniques combining Monte-Carlo simulations with Lefschetz thimbles \cite{MCL} to compute the simplex amplitude \cite{Han-MCL}, or the development of tensor network algorithms for gauge systems \cite{TNW}. 

The recently proposed effective spin foam models \cite{EffSF1,EffSF2} tackle the problem from the other direction: keeping the key dynamical principles of spin foams they constitute a considerable simplification of their structure.  This has led to the most efficient spin foam models so far and  to the first computation of expectation values for geometric observables in simple triangulations of Euclidean geometries \cite{EffSF1,EffSF2}. In particular,  \cite{EffSF2} provides an explicit numerical proof that spin foams can implement a discretized version of the equation of motion of general relativity, and that the so-called flatness problem \cite{flatness,ILQGS} can be avoided.

The flatness problem \JPp{states} that spin foam amplitudes in the semi-classical limit -- understood as limit of large quantum numbers -- suppress configurations with curvature. If true, it could mean that spin foams do not admit a suitable large scale limit. The flatness problem has for a long time been seen as a specific problem for the EPRL-FK models. But the analysis of \cite{EffSF1,EffSF2} shows that the issues underlying this  problem are much more general, and rooted in a key property of loop quantum gravity, namely the discreteness of the areas.  This prevents  a sharp imposition of the so called simplicity constraints, that are necessary for a proper implementation of a gravitational dynamics, but constitute \JPp{Diophantine} equations for the discrete areas. More precisely, the constraints are  \JPp{anomalous}. The anomaly is parameterized by the Barbero-Immirzi \JPp{parameter} \cite{BarberoImmirzi,Perez,DittrichRyan1and2}, which also parameterizes the  gap for the spatial area spectra. 

The EPRL-FK models implement the simplicity constraints therefore weakly\footnote{{The  pioneering work \cite{Reis} is probably the first to propose the weak implementation of constraints in the context of spin foams. In contrast to the EPRL-FK models the work \cite{Reis} suggest a Gaussian implementation of all the primary simplicity constraints.}}, that is allow for an uncertainty, that is determined by the non-commutativity of the constraints. The effective spin foam models proceed in the same way.\footnote{The constraint sets are different however, the EPRL-FK model implements the primary simplicity constraints. There are also secondary simplicity constraints, which are usually implemented automatically with the primary ones. Whether this applies also for anomalous constraint algebras is not known. In contrast, the effective spin foam models operates with variables for which the primary simplicity constraints are already implemented \cite{DittrichRyan1and2}. There remains a subset of the secondary constraints, which restricts the areas, and it is this set the effective spin foam models deal with.}
But, as pointed out in \cite{EffSF2}, with a weak implementation of anomalous or more generally second-class, constraints, one cannot expect that the usual semi-classical limit $\hbar \rightarrow 0$ \JPp{works}.\footnote{An exception is the Gupta-Bleuler formalism for electromagnetism. There one implements (non-\JPp{commutative}) constraints which fix a gauge. But in this case the variation of the action along the constraint gradients is vanishing. This prevents the mechanism at play for systems where the constraints are needed  to suppress variations of the actions in certain directions, as is the case in spin foams. } Rather one has also to scale the anomaly to be small. In the case of spin foams, this anomaly is parametrized by the Barbero-Immirzi parameter, and this parameter has therefore to be scaled to zero in the semi-classical limit. See also \cite{Perini,HanSBI} for related arguments. 

There is however a further caveat: The smallness of the Barbero--Immirzi parameter ensures that the weak implementation of the  constraints  is not impeded by the oscillations of the amplitudes resulting from the variation of the action in the constraint gradient directions.  On the other hand one needs the oscillations of the amplitude in the unconstrained directions to ensure that the equations of motions are implemented in the semi-classical limit. These two conditions could have been mutually exclusive. The examples in \cite{EffSF2} show that this is not the case and that, contrary to expectations,  a suitable semi-classical regime allows for finite (but relatively small) Barbero-Immirzi parameter and even for relatively large curvature per building block.

Spin foams are thus able to impose a suitable gravitational dynamics for discrete geometries. Having clarified this point, we will construct in this paper the effective spin foam model for Lorentzian geometries. The set-up for the effective spin foam model allows us to define at once a model that sums over all possible\footnote{Null triangles appear as limit of the space-like and time-like case.} signatures for the edges, triangles and tetrahedra of the triangulation. (One can also  allow for space-like 4-simplices.)

For the definition of the Lorentzian model we proceed in the same way as for the Euclidean model, incorporating the key dynamical principles of spin foams:  we define a path integral that implements the discreteness of the area spectra, and where the (simplicity) constraints between the areas are imposed weakly, but as strongly as allowed by the uncertainty relations resulting from the non-commutativity of the constraints.

The Lorentzian case does offer many new nuances. The first of these is \JPp{that} the spectra for the time-like areas are different from those of the space-like areas \cite{Conrady}: whereas the (asymptotic) spectrum for time-like areas is $a_S\sim\gamma \ell_p^2 j$ with $j\in \tfrac{1}{2} \mathbb{N}$, the one for time-like areas is given by $a_T\sim\ell_p^2 j$.  But the anomaly in the constraint algebra of the constraint is still parametrized by $\gamma$.

This highlights yet another role of the Barbero-Immirzi parameter $\gamma$: As it gives the difference between the space-like and time-like area gaps, it can be understood as a kind of anisotropy\footnote{This does not necessarily imply a Lorentz symmetry violation: first of all\JPp, the areas are defined space-time areas, that is are invariant under local Lorentz-rotation. Secondly, there are examples, namely the 3D BF (or Ponzano-Regge) model for $\text{SO}(2,1)$ where the spectra of time-like and space-like lengths are also different, namely discrete and continuous respectively. The spectra are constructed out of the Casimir of the local Lorentz-group and are therefore by construction, invariant. The model itself is topological and triangulation invariant and therefore diffeomorphism symmetric \cite{DittrichBahrSteinhaus}.} parameter, which in related \JPp a form appears also in Causal Dynamical Triangulations \cite{CDT}. In addition, it \JPp{parametrizes} the anomaly in the simplicity constraint algebra. Thus\JPp, $\gamma$ \JPp{determines} how strongly one can impose the simplicity constraints. Or in other words, how much fluctuations one \JPp{allows} for the torsion degrees of freedom, that are suppressed by the simplicity constraints.

In Lorentzian triangulations we can also have, e.g. null triangles, whose area is defined to be zero. In the usual understanding of associating the classical regime to large quantum numbers, such null configurations seem to be deeply quantum. But we can easily construct configurations  where all other areas are large and therefore \JPp`classical' and we have only one null area. Here we should also be able to obtain a suitable semi-classical limit.

Another  very interesting feature of the Lorentzian case is that the Regge action can become complex \cite{Sorkin2019}. These configurations with a complex action signify an irregular light cone structure, and describe a topology change of the spatial manifold during time evolution. There are two different types of such topology changes, trouser-like and yarmulke-like configurations, which are, due to the imaginary parts in the Regge action, suppressed\footnote{Which one is suppressed or enhanced \JPp{depends} on a choice of root for the definition of the Lorentzian angles appearing in the Regge action \cite{Sorkin2019}. We will follow here the choice of \cite{Sorkin2019}, which is supported by a calculation for the 2D continuum case \cite{LoukoSorkin}.}, or respectively enhanced, in the path integral. It is not clear yet how the EPRL-FK model \JPp{treats} such configurations. But one can expect that including or excluding such configurations would have defining consequences for the large scale physics. In fact a key principle of Causal Dynamical Triangulations is to forbid such configurations \cite{CDT}, and that leads to a very different continuum limit than for  Dynamical Triangulations.

The effective spin foam model offers sufficient flexibility to study the various choices for the path integral, e.g. allowing or not allowing irregular causal structure. We will perform a first test, which indicates that one should rather not include irregular causal structures, if these are enhanced in the path integral. 

\vspace{3mm}

The paper is fairly self-contained and does not rely on background material from \cite{EffSF1,EffSF2}. The organization of the paper is as follows:  We start in section \ref{Sec:qg} with a review of the relevant aspects of quantum geometry, going from quantized triangles to quantized tetrahedra to quantized 4-simplices. This will explain the various area spectra, define the configuration space for the path integral, and also derive the non-commutativity of the constraints. In section \ref{Sec:Regge} we will review various forms of the Regge action. In this section we will in particular discuss peculiarities that appear for the Lorentzian case, i.e. imaginary parts for the curvature angles, and therefore the Regge action.  Appendices \ref{app:triangle}, \ref{app:angles} include useful material for the computation of the Regge action and the path integral in the Lorentzian case: Appendix \ref{app:triangle} explains the generalized triangle inequalities and the definition of volumes  for Lorentzian simplices, and Appendix \ref{app:angles} provides a definition of Lorentzian (wedge) angles, and provides explicit formulas for the computation thereof. Appendix \ref{app:eom} derives the equations of motions for constrained Area Regge calculus, which provides the classical action principle for the effective spin foam models.
Equipped with this background material, we define the effective spin foam model for Lorentzian geometries in section \ref{sec:effsf}. Here we also point out a number of choices \JPp{that} can be taken for the precise definition of the model. Most of these concern the specification of the configuration space one sums over, e.g. should the path integral include a sum over all possible signatures, over irregular causal structures or over orientations.  In section \ref{sec:semi} we discuss a bound on the parameters of the system in order to lead to a suitable  semi-classical regime. 

We provide first numerical tests for the Lorentzian models in section \ref{Sec:Examples}. Here we consider triangulations with an inner triangle, testing in particular  the imposition of the constraints. In this work we explain a number of puzzling features in the behaviour of the expectation values as functions of $\gamma$, that were first identified in \cite{EffSF2}.  We consider first an example in which we only need to sum over configurations where all tetrahedra are space-like. This will allow a comparison with the (standard) EPRL-FK model, if analogous results become available.  We then consider an example where we have time-like and space-like triangles, and in particular a sum over a time-like area. This allows us to see the effect of the different area spectrum for this case. Finally we can, in this example, also choose boundary data, such that the path integral is peaked on a null area. The sum includes bulk areas of time-like signature, coming with a regular causal structure, as well as irregular causal structure. This allows us to test the semi-classical behaviour for a configuration with null triangles, as well as a probe of different choices regarding the  inclusion or exclusion of configurations with irregular causal structure.

We close with a discussion and outlook in section 
\ref{sec:discuss}.

\section{Aspects of quantum geometry}\label{Sec:qg}

To explain the construction of the effective spin foam models, we will review some aspects of quantum geometry in loop quantum gravity. This type of quantum geometry can be \JP{most easily} understood as arising from a quantization of the geometric data associated to a tetrahedron \cite{Barbieri:1997ks,ConradyFreidel}. One way to define the geometry of a tetrahedron is to specify its six length variables. An alternative way, is to specify the normals to its triangles, whose mutual inner products give rise to six independent parameters. The geometric data \JP{will, of course,} not change if all the normals undergo a rotation (or boost), leading to a $\text{SO}(3)$ or $\text{SO}(2,1)$ gauge symmetry for space-like and time-like tetrahedra respectively -- and a related closure or Gau\ss~\JP{constraint} for the normals. 

These normals are quantized as ($\gamma$ times the) generators of the corresponding symmetry group. This explains already many properties of the quantum geometry used in loop quantum gravity, in particular the discreteness for space-like areas, and the non-commutativity between the components of the triangle normals. 

\JP{When} gluing the tetrahedra to a 4-simplex one needs to impose further constraints on these data to obtain a 4-simplex with a well-defined length \JP{assignment} to its edges. These are the simplicity constraints. The non-commutativity of the tetrahedral variables will lead to a  non-commutativity between these constraints \JP{parametrized by $\gamma$}, and renders the constraints to be partially second class. 

In the following subsections we will explain these aspects in more detail, starting with the quantization of the triangles, which we will then combine to tetrahedra, and finally to 4-simplices.

\subsection{Spectrum for space-like and time-like areas}\label{Sec:spectrum}

The spin foam partition function can be defined for a fixed triangulation, whose triangles carry area variables. In addition one has degrees of freedom parametrizing 3D dihedral angles associated to the tetrahedra. For the definition of the effective spin foam models we will integrate out these angle degrees of freedom.  The partition function can then be understood as an integral over the areas. But this integral is actually a sum, as one only allows for area values that appear in the spectra for the corresponding area observables. 

The discreteness of the spectra for spatial areas is a well-known result in loop quantum gravity \cite{DiscreteGeom1}. The case for time-like areas, requiring a canonical formalism for time-like hypersurfaces, has been less thoroughly discussed. An exception is \cite{Conrady}, which defines EPRL-FK like models for triangulations including time-like tetrahedra and triangles. We will review here the relevant parts of  \cite{Conrady} concerning the quantization for space-like and time-like areas. 


A key result of canonical loop quantum gravity is that the spectra of area operators for surfaces in a spatial hypersurface are discrete, and given by 
\be\label{ASpecSS}
a_{SS}(j) = \gamma \ell_{P}^2 \sqrt{j(j+1)} \sim  \gamma \ell_{P}^2 j
\ee
where $j \in {\mathbb N}/2$ is  often referred to as spin (representation or label). $\ell_{P} = \sqrt{ 8\pi \hbar G/c^3}$ is the Planck length (times $\sqrt{8\pi}$) and $\gamma$ is the dimensionless Barbero-Immirzi parameter \cite{BarberoImmirzi}. In the following we will set $8\pi  G/c^3=1$ and thus  $\ell_{P}^2=\hbar$.
Here we will focus on the large $j$ regime and often approximate the spectrum with $a_{SS}(j)\sim \gamma \ell_{P}^2 j$. 

For a triangulated hypersurface  we can understand  this spectrum to arise in the following way: Consider a (non-degenerate) tetrahedron in the triangulation of the spatial hypersurface and embed this tetrahedron isometrically into four-dimensional Minkowski space ${\mathbb M}^4$. {This allows to define the tetrahedral normal, that is a four-vector, which is orthogonal to all the edges  of the tetrahedron, and whose norm can be chosen to be given by the volume of this tetrahedron.}  The geometry of this tetrahedron can be encoded in the normals $\vec{n}$ to its triangles.  { The normal $\vec{n}$ to a triangle $t$ is defined to be a   vector in ${\mathbb M}^4$, which is orthogonal to the triangle $t$ but lies inside the subspace spanned by the edges of the tetrahedron. The length of the triangle normal is defined to be the area of the corresponding triangle.} As the tetrahedron is spatial one can gauge fix the tetrahedral normal to be parallel to $(\pm1,0,0,0)$, and therefore the triangle normals to be of the form $(0,n^1,n^2,n^3)$. That is, we can treat the normals as spatial 3-vectors, and these 3-vectors are quantized as $\hat n^i=\gamma J^i$, where $J^i,i=1,2,3$ are the generators of $\text{SU}(2)$. The area of a triangle, which is given by the norm of $\vec{n}$, then amounts to $\gamma$ times the square root of the $\text{SU}(2)$-Casimir, leading to (\ref{ASpecSS}). One can also adopt instead of the group $\text{SU}(2)$ the group $\text{SO}(3)$, which leads to $j\in {\mathbb N}$. (We will do so in the numerical simulations.) 

If one has a time-like tetrahedron, that is a tetrahedron which can be embedded into three-dimensional Minkowski space ${\mathbb M}^3$,  one cannot apply anymore the gauge fixing above. Instead one chooses the normals to be of the form $(n^0,n^1,n^2,0) \in {\mathbb M}^4$ and they are now quantized as $\hat n^i=\gamma F^i$, where $F^i, i=0,1,2$ are the generators of $\text{SU}(1,1)$ \cite{Conrady}. 

The group $\text{SU}(1,1)$ has a discrete and a continuous representation. The Minkowski inner product $\eta_{ij}n^in^j$ leads to the Casimir of $\text{SU}(1,1)$, which for the discrete representation gives an area spectrum
\be\label{ASPecST}
a_{ST}(j) = \gamma \ell_{P}^2 \sqrt{j(j-1)} \sim  \gamma \ell_{P}^2 j   
\ee
where a priori $j \in {\mathbb N}/2$. The area-square is positive for $j\geq 1$ and one therefore restricts to these values and associates the discrete representation labels to the case of space-like triangles in a time-like tetrahedron.

Now a space-like triangle can be shared by a space-like tetrahedron and a time-like tetrahedron, the spectra for these cases should therefore coincide. This can be achieved by matching $j_{S}$ for the space-like tetrahedra with $j_T=j_S+1$ for the time-like tetrahedra. We will work with the asymptotic spectrum $a_{ST}\sim \gamma \ell_P^2 j$ where this difference disappears. 

For the continuous series of $\text{SU}(1,1)$ representations the Casimir is negative, one therefore associates these representations to time-like triangles.  But for this case the (primary) simplicity constraints come into play and restrict the continuous representation label to a discrete one \cite{Conrady}. These primary simplicity constraints impose a relation between the labels of the $\text{SL}(2,{\mathbb C})$ representations $(\rho \in {\mathbb R},n \in {\mathbb Z}_+ )$ and the representations of the  subgroups $\text{SU}(2)$ resp. $\text{SU}(1,1)$ \cite{EPRL-FK,Conrady}. For space-like tetrahedra and for space-like triangles in time-like tetrahedra these constraints rather restrict the $\text{SL}(2,{\mathbb C})$ representation parameters as a function of the discrete spin parameter $j$. But for time-like triangles the primary simplicity constraints fix the continuous parameter $\rho$ of the $\text{SL}(2,{\mathbb C})$ and the continuous $\text{SU}(1,1)$ parameter\footnote{There are slight ambiguities in this procedure, which do not matter for large areas. We will adopt here the choices made in \cite{Conrady}.} in terms of the discrete parameter $n \in {\mathbb Z}$, as well as $\gamma$. The area spectrum for time-like triangles can then be computed to be \cite{Conrady}
\ba
a_{TT}= \ell_P^2\,\,  n/2
\ea
and is, importantly, independent of the Barbero--Immirzi parameter $\gamma$.  Similar to the restriction $j\in {\mathbb N}$ for space-like triangles, for the numerical simulations we will restrict to $n\in 2{\mathbb N}$ for time-like triangles, so we work with a spectrum $a_{TT}=\ell_P^2n$ where $n\in {\mathbb N}$.

In summary we will work with a spectrum $a_S=\gamma\ell_{P}^2 j$ with $j\in {\mathbb N}$ for space-like triangles and a spectrum $a_T=\ell_{P}^2 j$ with $j\in {\mathbb N}$ for time-like triangles.

~\\
\emph{Remark:}  The Barbero-Immirzi parameter, apart from parametrizing an anomaly (see below) and an extension of the phase space of geometries \cite{DittrichRyan3},  can therefore be also seen as an anisotropy parameter. Such a parameter also appears for Causal Dynamical Triangulations \cite{CDT} and is treated there as a (tunable) coupling. This turns out to be quite important for finding phase transitions, at which one can expect interesting continuum physics. In Causal Dynamical Triangulations one constructs a quantum gravitational path integral as a sum over all possible triangulations with a certain regular causal structure. The length square of all space-like edges are equal to a lattice parameter $a^2$ in these triangulations, whereas the length squares of time-like edges are equal to $-\alpha a^2$, with $\alpha$ being the anisotropy parameter.

~\\
\emph{Remark:}
The discreteness of the spectrum for time-like areas, $a_T=\ell_P^2 j, j\in {\mathbb N}$, is caused by the following fact: areas are conjugated to extrinsic curvature angles, and for a time-like triangle, this curvature angle is compact, more precisely $\text{U}(1)$--valued. The conjugated area variable can therefore be expected to have a discrete and equidistant spectrum.  

In contrast, the discreteness of the space-like areas arises in a much more subtle way and seems to require the introduction of the Ashtekar variables \cite{Ashtekar}.  One way to derive the  Ashtekar variables  \cite{ThiemannBook} is to start with a canonical pair for general relativity, given by the extrinsic curvature $K$ and triad variables.  From the latter  one can define  the triangle normals for a triangulated hypersurface and then apply a canonical transformation that shifts the momentum from being the extrinsic curvature $K$ to the Ashtekar-Barbero connection $\Gamma +\gamma K$, where $\Gamma$ is the spin connection of the spatial hypersurface.  The surprising feature of this transformation is that these new variables are still canonically conjugated to (a rescaling by $\gamma^{-1}$ of) the triad variables.  The addition of $\Gamma$ makes this new variable transform as a $\text{SU}(2)$--connection, which are quantized as $\text{SU}(2)$--holonomies.  One thus compactifies the extrinsic curvature into $\text{SU(2)}$, with $\gamma$ appearing as a parameter in this compactification procedure. The spectra of the areas, which are  conjugated to the extrinsic curvature, appear therefore discrete and $\gamma$--dependent.  This compactification, which introduces curvature in connection space, leads to a non-commutativity in the conjugated variables and can be considered as the source for the $\gamma$-dependent anomaly in the algebra of the simplicity constraints.

\subsection{Quantum geometry associated to a tetrahedron}\label{Sec:Tetrahedron}

 The geometry of a tetrahedron is determined by six parameters, e.g. the length (squared) of its six edges.  A tetrahedron has four triangles, the  set of triangle areas alone is therefore insufficient to fix the geometry. To the set of areas one can add two of the six dihedral angles of the tetrahedron. The geometry can then be determined uniquely from these data, if the dihedral angles are chosen to be from adjacent edges. 

The dihedral angle between two triangles $t,t'$ in a tetrahedron $\tau$ can be computed from the inner product $\vec{n}_t \cdot \vec{n}_{t'}$ between the normal vectors to the triangles.  (We, again, consider the tetrahedron $\tau$ as embedded in four-dimensional Minkowski space ${\mathbb M}^4$ with signature $(-,+,+,+)$. The normals $n_t$ to a triangle $t\in \tau$ are defined to be orthogonal to the normal to the tetrahedron.)  The length square of the triangle normals is defined to be    $\vec{n}_t \cdot \vec{n}_{t}= \text{sign}(\text{Vol}_\tau^s) A_t^s$. We define signed squared  volumina $\text{Vol}_\tau^s$ for tetrahedra, and signed squared areas for triangles $A_t^s$ in appendix \ref{app:triangle}. These signed squared volumina are positive for space-like simplices and negative for time-like simplices.  The $\text{sign}(\text{Vol}_\tau^s)$ appears because for a time-like tetrahedron, the normal to a space-like triangle is time-like, and can only have negative norm square,  and the normal to a time-like triangle is space-like and can have only positive norm-square. 
The (e.g. outward pointing) normals are then \JPp{satisfying} the closure relation 
\ba\label{Closure}
\sum_{t \in \tau}\vec{n}_t =0 \q .
\ea
Taking the inner product of this closure relation (\ref{Closure}) with each of the normal vectors gives four equations between areas and dihedral angles. These determine the remaining four dihedral angles from the four areas and two dihedral angles, see e.g. \cite{ADH} for the explicit solutions.

Quantizing the normals by associating them to (Euclidean or Lorentzian) rotation group generators, the closure constraint (\ref{Closure}) acts as a generator of global rotations for the four normals.  The areas and the two dihedral angles give a complete set of independent invariants, which can be constructed from four vectors satisfying the closure relation (\ref{Closure}).

But this quantization leads also to the non-commutativity between the two dihedral angles (attached to two adjacent edges) of a given tetrahedron.

  Instead of the dihedral angles\footnote{As explained in appendix \ref{app:angles} the definition of the Lorentzian angles requires to distinguish between space-like and time-like tetrahedra and edges, making it quite involved.}, it will be more convenient to keep working with the (Euclidean or Minkowskian) inner products $p^\tau_{tt'}:= \text{sign}(\text{Vol}^s_\tau) n_t\cdot n_{t'} $ between the normals. Here $(t,t')$ is the pair of triangles  in $\tau$, which therefore define an edge $e$ in $\tau$.   $p_{tt'}^\tau$  is a rotation invariant quantity and can be expressed as a function of the signed\footnote{The signed squared lengths is negative for time-like vectors and positive for space-like vectors.} squared lengths $l_e^s$ of the edges of the tetrahedra, see appendix \ref{app:angles}. The  $\text{sign}(\text{Vol}^s_\tau)$ results from the normalization condition $n_t\cdot n_t=\text{sign}(\text{Vol}^s_\tau) A^s_t$ for the triangle normals. With the inclusion of this sign in the definition of $p^\tau_e$, they define the same function of the signed squared edge lengths for a time-like and a space-like tetrahedron. 

The set of $p^\tau_{tt'}$ for different pairs of triangles $t,t'\subset \tau$ and the set of $p^\tau_{tt}= A^s_t$ satisfy four equations resulting from the closure relations (\ref{Closure}). For each $t \subset \tau$ we have
\ba\label{Closure2}
\sum_{t' \in \tau} p_{tt'}^\tau =0 \q .
\ea
Choosing a triple of triangles $(t_1,t_2,t_3)$ and correspondingly fixing the angle parameters $p_1^\tau=p^\tau_{t_1t_2}$ and $p_2^\tau=p^\tau_{t_1 t_3}$, we can express all other quantities $p^\tau_{tt'}$ as a  linear function of $p_1^\tau$ and $p_2^\tau$ as well as the (signed) area squares  $p_{tt}^\tau$.

As discussed above, the normals are quantized as generators of the $\text{SU}(2)$ group for a space-like tetrahedron, $\vec{n}=\gamma \vec{J}$, and as generators of the $\text{SU}(1,1)$ group for a time-like tetrahedron, $\vec{n}=\gamma \vec{F}$ .  We can write the structure constants for both groups as ${c^{ij}}_k=\epsilon^{ijm}\kappa_{mk}$ where $\epsilon^{ijm}$ is the Levi-Civita tensor and $\kappa_{ij}=\delta_{ij}$ is the Euclidean metric for space-like tetrahedra and  $\kappa_{ij}=\eta_{ij}$ is the Minkowskian metric for time-like tetrahedra. The Poisson bracket between $p^\tau_{tt'}=\text{sign}(\tau) n_t \cdot n_{t'}$ and $p^\tau_{tt''}=\text{sign}(\tau) n_t \cdot n_{t''}$  is therefore given by
\ba\label{PB1}
\{   p^\tau_{tt'}, p^\tau_{tt''}      \} \,= \,   \kappa_{ii'} \kappa_{jj'} n^i_{t'} n^j_{t''}. \{ n^{i'}_t, n^{j'}_t\}
\,=\, \gamma \epsilon^{i'j'k'} \kappa_{ii'} \kappa_{jj'}  \kappa_{kk'}   n^i_{t'} n^j_{t''} n^k_{t}\,=\, \pm \frac{9}{2} \gamma\text{Vol}^s_\tau
\ea 
where $\text{Vol}^s_\tau$ is the signed squared volume for the tetrahedron.

If one keeps the areas of a tetrahedron $\tau$ fixed the two remaining degrees of freedom can be parameterized by $p^\tau_{tt'}$ and $p^\tau_{tt''}$, which determine the shape of the tetrahedron. This phase space for the shapes of a tetrahedron with fixed areas is known (in the Euclidean case) as Kapovich-Millson phase space \cite{KM}. 

In case we are considering a tetrahedron inside a spatial hypersurface, this means that spatial geometric quantities do not commute. This non-commutativity is parametrized by the Barbero--Immirzi parameter $\gamma$. In the following we will see that this non-commutativity can be understood as an anomaly, and $\gamma$ therefore as an anomaly parameter.

 \subsection{Fitting tetrahedra into 4-simplices}
 
We have seen that in loop quantum gravity, the geometry of a tetrahedron $\tau$ is encoded into four area parameters $a_t$ and two angle parameters $p_1^\tau=p_{tt'}^\tau$ and $p_2^\tau=p_{tt''}^\tau$. Given these six parameter we can compute the six edge lengths $l_e$ of $\tau$. 
 
We now glue 5 tetrahedra into a 4-simplex $\sigma$. To this end we need to identify the areas of triangles that are shared by pairs of tetrahedra.  In the end one is left with 10 area parameters for the 10 triangles of a 4-simplex. But in addition to these 10 area parameters, we also have 10 (angle) parameters $p_i^\tau$, namely two per tetrahedron. We thus have 20 parameters. But the geometry of a 4-simplex is determined by 10 parameters only, e.g. by the lengths of its 10 edges. 
 
In fact,  loop quantum gravity does describe a more general type of geometry \cite{AreaAngle,DittrichRyan1and2} as compared to e.g. length Regge configurations, which describe piecewise flat geometries  and  where one uses the lengths of the edges as parameters. The additional degrees of freedom can be understood as  a certain type of torsion degrees of freedom \cite{ADH,BFCG2,TorsionConnection}.

The constraints that enforce the reduction of the configuration space to the one of length Regge calculus can however be easily deduced. For this it is sufficient to consider a single 4-simplex $\sigma$.  A 4-simplex has 10 triangles and 10 edges. Its 10 edge lengths determine uniquely the 10 areas. This relation can be, locally in configuration space, inverted, that is the 10 areas  $a_t$ also determine (locally in configuration space) the 10 edge lengths $l_e=L_e^\sigma(a)$.  From these edge lengths we can compute the angle parameters $p_{i}^\tau=P_{i}^\tau(l)$. We can thus determine the $p_{i}^\tau$ (locally in configuration space) from the 10 areas of the 4-simplex by $p_{i}^\tau=P_{i}^{\tau,\sigma}(a):=P_{i}^\tau(L^\sigma(a))$.

That is the five tetrahedra $\tau=1,\ldots 5$ in a given 4-simplex give rise to 10 area parameters and 10 parameters $(p_{1}^\tau,p_{2}^\tau)_{\tau=1}^5$. But these data have to satisfy the constraints
\ba\label{match2}
p^\tau_{i} - P_{i}^{\tau,\sigma}(a)  \,\stackrel{!}{=}\,0\,\, \q i=1,2 \, ,
\ea
where $P_{i}^{\tau,\sigma}(a)$ are the angle parameters in the tetrahedron $\tau$ as determined by the areas of the simplex $\sigma$.   There is also a more local version of these constraints \cite{AreaAngle,DittrichRyan1and2}, which only involves the data of two neighbouring tetrahedra, and can therefore be applied to general triangulations of  three-dimensional hypersurfaces.

One thus has two constraints for each tetrahedron. The non-commutativity of the tetrahedral geometry (\ref{PB1}) make these constraints second class (note that the areas $a_t$ are Casimirs and do commute with the $p^\tau_{e_i}$):
\ba\label{CRC}
\{ p^\tau_{1} - P_{1}^{\tau,\sigma}(a) , p^\tau_{2} - P_{2}^{\tau,\sigma}(a) \} &=&
\pm\gamma  \,\tfrac{9}{2} \,\text{Vol}^s_\tau\q .
\ea
The non-commutativity of the constraints is parameterized by the Barbero--Immirzi parameter $\gamma$, which can therefore be interpreted as an anomaly parameter. 

To summarize, we quantized the geometry of a three-dimensional or four-dimensional triangulation by representing the triangle normals in a tetrahedron as ($SU(2)$ or $SU(1,1)$) angular momentum operators, scaled by the Barbero--Immirzi parameter $\gamma$. These normals, reduced by the closure relation (\ref{Closure}) and the associated global rotation, give rise to the four areas $a_t$ and two angle parameters $p^\tau_{i},i=1,2$ per tetrahedron and thus do provide a parametrization of the tetrahedral geometry. Gluing tetrahedra, e.g. to a 4-simplex, we encounter however an over-parametrization. That is we need to impose the constraints (\ref{CRC}) so that the tetrahedral data lead to well-defined  lengths for the edges of the 4-simplex. There are two constraints per tetrahedron, which do not commute and, more precisely are second class.

The uncertainty relation prevents a sharp imposition of these second class constraints on the Hilbert space resulting from this quantization procedure. 

The effective spin foam models \cite{EffSF1} \JP{implement} these constraints therefore weakly, but as strongly as allowed by the uncertainty relations. This is following a strategy that was first employed for the EPRL-FK models \cite{EPRL-FK}.  For these models one implements the primary simplicity constraints, which \JP{turn out to also} have an anomaly in their algebra \JP{that} is parametrized by the Barbero--Immirzi parameter \cite{Perez}.

The constraints (\ref{match2}), which \JP{involve} the areas and dihedral angles of pairs of tetrahedra, can be seen as part of the secondary simplicity constraints \cite{DittrichRyan1and2}.  We will see that they are essential for obtaining a dynamics that reproduces general relativity.

~\\
\emph{Remark:}  A very different way of proceeding would be to impose the constraints on the classical phase space, determine the Dirac brackets and then quantize the resulting phase space. In fact the Barbero--Immirzi parameter disappears from the Poisson brackets \cite{DittrichRyan3} and one re-discovers the phase space of length-Regge calculus \cite{DittrichHoehn1}.\JP{ However,} this phase space has not be quantized yet, as it has a very complicated topology. In fact, one motivation of using an enlarged phase space, as in loop quantum gravity, is that one starts with a simpler phase space topology, for which one can find a quantization.

\subsection{Gluing 4-simplices }

For a given 4-simplex $\sigma$ the constraints (\ref{match2}) lead to a reduction of variables: starting with area and angle parameters, we are left with area parameters only.  If we consider a gluing of several 4-simplices\JP{, (\ref{match2}) also imposes} constraints between the area variables.  The reason is the following: If we consider a bulk tetrahedron $\tau$, it is shared by two simplices $\sigma$ and $\sigma'$.  \JP{Therefore, the 3D angle parameters $p^\tau_i$ in (\ref{match2}) appear for both simplices and we can then integrate them out to} remain with the two constraints per tetrahedron
\ba\label{match1}
P_{i}^{\tau,\sigma}(a)  - P_{i}^{\tau,\sigma'}(a)  \,\stackrel{!}{=}\,0\,\,\JP, \q i=1,2.
\ea  
These constrain the areas to configurations which lead to a consistent assignment of lengths to the edges of the triangulation, and provide the so-called area constraints for constrained Area Regge calculus \cite{AreaRegge}. In the formulation (\ref{match1})\JP, the constraints involve variables from pairs of neighbouring 4-simplices, whereas the version (\ref{match2}) involves variables of one 4-simplex only.  Indeed, it \JP{was} first realized in \cite{AreaAngle} that using the 3D angles as such, one can convert the non-local constraints of Area Regge calculus into constraints that are localized on the 4-simplices.

 Adopting the discrete area spectra $a_S(j)\sim \gamma j$ and $a_T(n)\sim n$, these constraints will constitute Diophantine conditions for the parameters $j,n \in {\mathbb N}_+$.  There will be only very few solutions (of high symmetry) to these relations \cite{MV}, preventing a suitable quantum dynamics, see the discussion in \cite{EffSF1}. Imposing the constraints weakly\JP, but as strongly as allowed by the commutation relations (\ref{CRC}), does lead to a suitable density of states \cite{EffSF1}.

\section{Lorentzian Regge  action}\label{Sec:Regge}

\subsection{Length Regge calculus} 

Regge calculus \cite{Regge} provides an action for a triangulated, piecewise flat manifold. One usually uses lengths as variables in Regge calculus, as these determine uniquely the geometry of each simplex, and therefore the piecewise flat geometry of the triangulation. But other choices for the variables are also possible \cite{AreaAngle,ReggeOther}. Although each simplex is flat, that is can be embedded in Minkowski space, one can get curvature by gluing the simplices together. The curvature is  concentrated in 4D on the triangles (or bones). For a given triangle $t$ this curvature is given by the so-called deficit angle $\epsilon_t$, which can be defined in a plane orthogonal to the triangle $t$. This deficit angle  measures how much the sum of the dihedral angles $\sum_{\sigma \supset t}\theta^\sigma_t$ at the triangle $t$ does differ from the flat space value. 

Here we follow the convention of Sorkin in \cite{Sorkin2019}, which we review in Appendix \ref{app:angles}.  Importantly, in Sorkin's definition angles in 2D Minkowski space are not just associated to a pair of vectors, but rather to the wedge bounded by the two vectors. That is the angle is different if the wedge extends  in clockwise direction from the first vector to the second vector or in clockwise  direction from the second vector to the first vector.  On the other hand  \cite{Sorkin2019} does define non-oriented angles, i.e. the angle for the wedge that goes from the first vector to the second vector in clockwise direction is the same as the angle for the wedge that goes from the second to the first vector in anti-clockwise direction. 

The dihedral angles at space-like triangles can have an imaginary part ${\rm Im}(\theta^\sigma_t)= -\imath N^\sigma_t \pi/2$, where $N^\sigma_t$ is a positive integer.  To define $N^\sigma_t$ embed the 4-simplex into flat Minkowski space. The dihedral angle can be defined in the plane orthogonal to the triangle $t$ and describes the wedge formed by the two tetrahedra sharing the triangle $t$. For a space-like triangle  $t$ the plane orthogonal to $t$  is the Minkowski plane. $N_t^\sigma$ then counts the number of light cone crossings, that are included in the wedge. E.g. it is $N_t=2$ if the two adjacent tetrahedra to $t$ are space-like and the wedge contains one light-cone. For a more detailed explanation we refer to appendix \ref{app:angles}.  Importantly the convention of \cite{Sorkin2019} makes the angles additive. The flat space value of the sum of the dihedral angles around a space-like triangle is given by $-2\pi \imath$, corresponding to a wedge containing two light cones (that is four light cone crossings). 

For a time-like triangle $t$ the plane orthogonal to the triangle is Euclidean. The dihedral and deficit angles can therefore be defined in the usual way, and the flat space value of the sum of the dihedral angles around a time-like triangle is given by $2\pi$. 

The Lorentzian Regge action is then given by
\ba\label{LRegge}
S_{LR} =  \sum_t  Z_t 2\pi  A_t(l)  - \sum_\sigma \sum_{t \subset \sigma}  A_t(l) \theta^\sigma_t(l)   \q .
\ea
We do define the areas to be positive valued for both space-like and time-like triangles. 
Here $Z_t=1$ for time-like triangles in the bulk and $Z_t=-\imath$ for space-like triangles in the bulk. For time-like triangles or space-like triangles  in the boundary we have $Z_t=1/2$ or $Z_t=-\imath/2$ respectively.\footnote{
One can also introduce a more refined distinctions for the values $Z_t$ for the boundary triangles. This would take into account whether the wedge angle attached to a space-like boundary triangle is `expected' to be thick or thin, and for the space-like triangles, the `expected' number of light cone crossings the wedge includes. We say `expected' because this number is determined by the bulk lengths and could therefore fluctuate. Given a lengths assignment to the boundary and bulk edges, which is causally regular, i.e. where all the bulk deficit angles are real, we can adjust he values $Z_t$ for the space-like boundary triangles, so that the boundary terms in the action are also real.
The 'expected' values can be considered part of the boundary data, but their choice do not affect the equations of motions. The choice of the $Z_t$ does matter if one considers the gluing of two  or more triangulations. With our choice we assume that one glues along a given boundary only two triangulations.}

For the variation of the action we keep the boundary lengths fixed, the equation of motion do therefore not depend on the precise choice of the boundary types $Z_t$.  The equation of motion are given by (see appendix \ref{app:eom})
\ba\label{LREOM}
 \sum_{t \supset e } \frac{\partial A_t(l)}{\partial l_e} \epsilon_t(l) \,=\, 0 \q .
\ea

~\\
\emph{Remark (Oriented angles):}  
Alexandrov \cite{Alexandrov2001} defines oriented angles between ordered pairs of vectors. We can use that to defined oriented angles for oriented wedges, by defining the angles for clockwise oriented wedges to be minus the ones for anti-clockwise oriented wedges.  

There is still an important difference between (oriented) angles for wedges and (oriented) angles for pairs of angles. For wedges we can differentiate between a wedge with vanishing extend, that is $\theta=0$ and a "full" wedge with $\theta=-2\pi\imath$, or even a wedge that includes a number $N$ of windings in the Minkowski plane, that is we can have angles $\theta=-2N\pi \imath$. Working with a pair of vectors one has to identify $-2\pi \imath=0$, that is one neglects the winding number.

\subsection{Imaginary parts in the action} \label{Sec:ImPart}

As discussed in \cite{Sorkin2019}\JP, the action can have imaginary parts. These are non-vanishing if either $(a)$ there are space-like bulk triangles to which there are less or more than two light cones attached or $(b)$ the actually geometry of the wedge attached to a space-like boundary triangle includes less or more than the expected number of light cone crossings. In a path integral with amplitude $\exp( \frac{\imath}{\hbar} S_{LR})$  the cases with more light cones lead to an exponential suppression, whereas the cases with less light cones lead to an enhancement. 

Such configurations \JP{leading to complex deficit angles can be interpreted as describing} a geometry where the spatial topology changes in time. \JP{Trouser type topology changes of the spatial hypersurface} lead to singular points which have  more than two light cones attached, whereas a topology change \JP{of} yarmulke type \JP{leads}  to points with less than two light cones attached \cite{Sorkin2019}. 
That is\JP, trouser type singularities are suppressed whereas yarmulke like singularities are enhanced in the path integral. Sorkin does point out that this conclusion \JP{depends} on a choice of root for $(-1)$ in the construction for the dihedral angles, but that the choice in \cite{Sorkin2019} does conform with a continuum calculation for a topology changing process in  two space-time dimensions \cite{LoukoSorkin}. 

The appearance of imaginary parts in the action has important implications for the physics defined by the various versions of the path integral. Firstly\JP{, we see} that Regge calculus can easily accommodate spatial topology change\JP{, but} it also delivers a mechanism through which trouser type singularities are suppressed.  This type of singularity can be associated with the splitting \JP{off} of baby universes. In Euclidean Dynamical Triangulations (thought of as a Wick-rotated path integral of Lorentzian geometries) this splitting off of baby universes \JP{dominates} in the weak gravity phase. This leads\JP, in the continuum limit\JP, to fractal\JP-like geometries with a Hausdorff dimension of around two. 

Causal Dynamical Triangulations  \cite{CDT} do explicitly forbid such splitting off processes \JP{by} introducing a slicing structure. This leads to a phase diagram \JP{which} does include a phase that is suitable to describe smooth manifolds in the continuum limit \cite{CDT4DResult}. A more general version of Causal Dynamical Triangulations \cite{LollS} \JP{replaces} the slicing structure with the condition of a regular causal structure, which includes that all space-like (bulk) triangles have exactly  two light cones attached. That is, all deficit angles have to be real.\footnote{If one has a space-like boundary, one can similarly demand the reality of the boundary angles.} Implementing such a condition for the Regge path integral would prevent the appearance of imaginary parts. 

The dominance of baby universe splittings in Euclidean Dynamical Triangulations is related to the conformal factor problem of Euclidean quantum gravity, which is that  the conformal mode is associated with a kinematic term with the `wrong sign' in the Euclideanized action. In Euclidean Regge gravity this conformal mode problem appears as the problem of {\it{spikes} }\cite{AmbjornSpikes}, which are vertices where all adjacent edges have very  large lengths. The Euclidean action for these configurations is large and negative, leading to \JP{their} domination in the Euclidean Regge path integral. 

Such a problem might not appear for the Lorentzian Regge path integral: Firstly\JP, we deal here with a real quantum mechanical path integral, \JP{i.e.} have complex amplitudes $\exp( \frac{\imath}{\hbar} S)$. Secondly\JP, demanding either a regular light cone structure or relying on the suppression of trouser like singularities might also suppress such spike configurations. 

\JP{Given} these considerations it would be important to know \JP{whether} the semi-classical limit of the EPRL-FK models \cite{SFLimit,SFtimelike} does include such imaginary parts in the action, associated to an irregular light cone structure. Unfortunately this is not clear yet, as most works focus on the amplitude for one simplex, for which the boundary types can be defined such that the imaginary parts cancel by construction. 

One important feature, in which spin foams do differ from Regge calculus, is that there are two types of space-like tetrahedra in spin foams --- one type has future directed normals, the other type past directed normals.\footnote{We thank W. Kaminski for pointing out this fact to us.} This differentiation results from using normals to encode the (quantum) geometries, and the fact that one cannot transform a future directed vector into a past directed one, using only proper Lorentz rotations. This is however not sufficient to prevent all causal irregularities.

The question \JP{of} whether such imaginary parts can appear is connected to \JP{whether one should allow} non-trivial winding numbers in the (Lorentzian) deficit angles or not.  The dihedral angles are encoded in $\text{SO}(3,1)$ group elements, \JP{so} one might therefore think that \JP{such} winding numbers are not allowed. Gluing many simplices together, one obtains angles for wedges composite of many simplices. Here one can resort to a definition of the angles, which does not rely on the multiplication of group elements, but computes the dihedral angles for each simplex and then adds these together. This \JP{then allows} for winding numbers. Such winding numbers have been found to be relevant for the semi-classical asymptotics of the 3D loop quantum gravity partition function for a triangulation with many building blocks,  describing the (Euclidean) solid torus \cite{3DHol}.

\subsection{Area Regge calculus}

As a 4-simplex $\sigma$ \JP{carries} 10 triangles and 10 edges we can, locally in configuration space\footnote{There are global ambiguities, which can be resolved with the help of the 3D angle parameters.}, invert the 10 areas $A_t(l)$ for the 10 lengths, and in this way obtain 10 functions $L_e^\sigma(a)$, where $a$ \JP{denotes} the 10 area parameters of the simplex $\sigma$. The geometric quantities appearing in the Length Regge action (\ref{LRegge}) are the areas $A_t(l)$  and the dihedral angles $\theta^\sigma_t(l)$, which are defined separately for each simplex. We can therefore define $ \theta^\sigma_t(a)=\theta^\sigma_t( \{L^\sigma_e(a)\})$.
This leads to the Area Regge action \cite{AreaRegge, ADH}, in which the areas are the fundamental variables
\ba\label{SAR}
S_{AR}(a)= \sum_t  Z_t 2\pi  a_t  - \sum_\sigma \sum_{t \subset \sigma} a_t \theta^\sigma_t(a)   \q .
\ea
In a typical triangulation one has however  more triangles than edges. \JP{By varying} the Area Regge action with respect to the areas one can therefore expect to obtain stronger conditions than the Lengths Regge equations of motion. In fact, the area variations lead to the equations of motion $\epsilon_t(a)=0$, which impose flatness \cite{AreaRegge}. This does not mean that the Area action is topological -- there are propagating degrees of freedom, which can be identified with a certain type of torsion \cite{ADH}.

To obtain the same equations of motion as in Length Regge calculus (which do provide a discretization of the Einstein equations), we need to constraint the area variables such that they describe a consistent lengths assignment. We defined such constraints in (\ref{match1}).  Adding these constraints with Lagrange multipliers to the Area Regge action, one now obtains equations of motion which reproduce the ones of Length Regge calculus (see  appendix \ref{app:eom}).

\section{The effective spin foam model}\label{sec:effsf}

The effective spin foam model has been introduced in \cite{EffSF1} as a path integral which (a) implements the area spectrum of loop quantum gravity and (b) does also incorporate the non-commutativity of the boundary geometry, as expressed in the commutation relations (\ref{CRC}).  

To implement (a) it is most convenient to work with the Area Regge action (\ref{SAR}), as here the areas are fundamental variables. The discrete spectrum can be easily implemented by summing only over area values consistent with this spectrum. Here,\footnote{We can easily adopt to other versions of the spectrum, e.g. $\gamma \ell_P^2 \sqrt{j(j+1)}$ for the space-like triangles. But this choice does not matter for the semi-classical regime, for which we need large representation labels $j$. } for space-like triangles we sum over  $a^S_t=\gamma \ell_P^2 j_t$   and for time-like triangles    over    $a^T_t=\ell_p^2 n_t$\JP, where here we will take $j_t,n_t \in {\mathbb N}_+$. We can also admit the $a_t=0$ value describing null or degenerate triangles.  

But we have seen that we need to constrain the areas in order to get  (a discrete version of) the dynamics of general relativity. We cannot implement the constraints (\ref{match1}) exactly -- there would be almost no solutions. The form of the alternative constraints (\ref{match2}), which employ the 3D  angle parameters as auxiliary variables, and the non-commutativity of these constraints does imply that we can implement the constraints only weakly, \JP{i.e.} with a minimal uncertainty  as determined by the commutator (\ref{CRC}).

The effective spin foam models can be derived from a path integral over areas and the angle parameters. The  angle parameters will only appear in the constraint implementation \JP{and} can be integrated out independently of the areas. This will give an effective amplitude, which only depends on the areas. 

We will first discuss this process of implementing the constraints via an integration over the angle parameters.

\subsection{Constraint implementation}\label{Sec:CI}

The Hilbert space of loop quantum gravity does include the degrees of freedom corresponding to the angle parameters.  For a given tetrahedron  there are two angle parameters $(p^\tau_{1},p^\tau_{2})$, which are conjugated to each other. Fixing the areas of the tetrahedron, the angle parameters can be taken as coordinates of a two-dimensional phase space, which describes the shape of the tetrahedron. Upon quantization\JP, the two angle parameters, which are conjugated to each other, are encoded into one quantum number. This quantum number is the intertwiner, \JP{i.e.} a choice of \JP{an} invariant tensor in the tensor product of four rotation group representation spaces.  For a space-like tetrahedron\JP, this rotation group is $\text{SU}(2)$. For a time-like tetrahedron the rotation group is $\text{SU(1,1)}$.  (A general rigorous Hilbert space construction for time-like boundaries has not been achieved yet, but see the recent work \cite{Livine2021}.)

For $\text{SU}(2)$ there are various constructions of coherent states for the intertwiner Hilbert space \cite{Coherent, BonzomLivine,FreidelHnybida} and we will assume that analogous states can be also constructed for $\text{SU}(1,1)$. Such coherent states are peaked on a phase space point in shape space, which can be parametrized by the two angle parameters $(p^\tau_{1},p^\tau_{2})$. We will denote such states by ${\cal K}_\tau( \cdot, P^\tau_i)$, where $P^\tau_i, i=1,2$ are labels \JP{describing} on which phase space point the coherent state is peaked. The first entry is for the argument of the coherent state. These can be the two angle parameters $p_i^\tau$ in a Bargmann-Fock like representation  (with associated measure $d \mu_{\cal K}(p^\tau_i)$, which comes with the construction of the coherent states). But the states can \JP{also be} expressed in a intertwiner basis.

We will associate a coherent state ${\cal K}_\tau( \cdot, P^{\sigma,\tau}_i(a))$ to each (positively oriented) tetrahedron $\tau$ in a simplex $\sigma$. This coherent state is peaked on the angle parameters as computed from the areas of the simplex $\sigma$. To negatively oriented tetrahedra we associate the complex conjugated coherent state.  Gluing two simplices $\sigma$ and $\sigma'$ along $\tau$ amounts to integrating two coherent states over the shared bulk variables -- that is the angular parameters. This gives the inner product
\ba\label{IPCS}
G^{\sigma,\sigma'}_\tau \!(a):=\langle {\cal K}_\tau ( \cdot, P^{\sigma,\tau}_i(a)) \,| {\cal K}_\tau ( \cdot, P^{\sigma',\tau}_i(a))\rangle  \q .
\ea
This results into a function of the areas of two simplices $\sigma$ and $\sigma'$, which is peaked on configurations for which the constraints (\ref{match1})
\ba
\frak{C}_i=P^{\sigma,\tau}_{i}(a)-P^{\sigma',\tau}_{i}(a) \q ,\,  i=1,2
\ea
are satisfied. We would like to approximate this function with a Gaussian function
\ba
\exp\left( -\frac{ \frak{C}_1^2+ \frak{C}_2^2 }{4\Sigma^2}\right)
\ea
 of these constraints, whose spread $\Sigma^2$ is determined by the commutation relation (\ref{CRC}) of the constraints (\ref{match2}).  

\JP{However, this approximation introduces}  a  dependence on the choice of our pair of angular parameters $p_1^\tau=p_{tt'}^\tau$ and $p_2^\tau=p_{tt''}$, which corresponds to a choice of non-opposite edges $(e_1,e_2)$ that are shared by the triangles $(t,t')$ and $(t,t'')$ respectively, in the tetrahedron $\tau$. Choosing another pair of non-opposite edges $(e_3,e_4)$, we can express $p_3^\tau$ and $p_4^\tau$ as linear functions of $p_1^\tau$ and $p_2^\tau$ and the areas. This induces a linear transformation between the constraints, e.g. depending on the choice of the pair of edges we can have\footnote{
One can deduce these transformations from the rules that $\frak{C}_e+\frak{C}_{e'}+\frak{C}_{e''}=0$ if $(e,e',e'')$ are the edges of a triangle and $\frak{C}_e=\frak{C}_{\bar{e}}$ if $(e,\bar{e})$ are a pair of opposite edges in the tetrahedron. These rules follow from the closure relations (\ref{Closure2}).
} 
$\frak{C}_3=\frak{C}_1$ and $\frak{C}_4=-\frak{C}_1-\frak{C}_2$ or $\frak{C}_3=\frak{C}_2$ and $\frak{C}_4=-\frak{C}_1-\frak{C}_2$ or $\frak{C}_3=\frak{C}_1$ and $\frak{C}_4=\frak{C}_2$. Note that the absolute value of the Jacobian determinant for all transformations is equal to 1. Indeed all these choices would come with the same spread, as the commutator between the constraints only changes by a sign. To avoid this dependence on the choice of pair of edges, we average over all 12 choices of non-opposite pairs in a given tetrahedron. This amounts to
\ba\label{equ16}
\tfrac{1}{12} \!\!\!\sum_{(v:e,e')\subset \tau} (\frak{C}^2_e+\frak{C}^2_{e'}) \,\,=\,\, \tfrac{1}{3} \sum_ {e\subset \tau} \frak{C}^2_e \,\,\,=\,\, \,
\tfrac{4}{3} \left(\frak{C}^2_{e_1} +\frak{C}^2_{e_2} + \frak{C}_{e_1}\frak{C}_{e_2}\right)
\ea
where in the first sum we include  all pairs of non-opposite edges (i.e. pairs of edges sharing a vertex) in $\tau$ and in the second sum we include all six edges of $\tau$. For the last equation in (\ref{equ16}) we fix some choice of non-opposite edges $(e_1,e_2)$. That is, the expression $(\frak{C}^2_{e_1} +\frak{C}^2_{e_2} + \frak{C}_{e_1}\frak{C}_{e_2})$ is independent of the choice of pair of non-opposite edges $(e_1,e_2)$.

We will therefore approximate the inner product (\ref{IPCS}) with the following Gaussian
\ba\label{Gfct1}
G^{\sigma,\sigma'}_\tau\!(a) \sim  \delta(\text{sign}(\tau\subset\sigma) ,\text{sign}(\tau\subset\sigma')) \,\, \exp\left( -\frac{ \frak{C}_{e_1}^2+ \frak{C}_{e_2}^2 +\frak{C}_{e_1} \frak{C}_{e_2} }{3\Sigma^2}\right)
\ea
where $(e_1,e_2)$ is a choice of non--opposite edges in the tetrahedron $\tau$. We included a factor  $\delta(\text{sign}(\tau\subset\sigma) ,\text{sign}(\tau\subset\sigma'))$: it vanishes, if the signature of the geometry of the tetrahedron $\tau$ as induced by $\sigma$ differs from the one induced by $\sigma'$, e.g. if $\tau\subset \sigma$ is of Euclidean signature but $\tau\subset \sigma'$ is of Lorentzian signature.  It is equal to one, if the two signatures agree. The reason for including this factor, is that in the case of differing signatures the coherent states are originating from different groups, namely $\text{SU}(2)$ for Euclidean signature, and $\text{SU(1,1)}$ for Lorentzian signature.  Finally, $\Sigma^2$ is determined by the commutation relations (\ref{CRC}) and hence proportional to the squared volume of the tetrahedron
\ba\label{SigmaF}
\Sigma^2 = \,  \ell_P^2 \,\gamma \, \tfrac{9}{4} \,\,(|\text{Vol}^s_\tau(\sigma)| + |\text{Vol}^s_\tau(\sigma')|)
\ea
which we average over the two geometries induced by $\sigma$ and $\sigma'$ respectively. 

We have so far discussed how the $G$-factors  (\ref{Gfct1}) do result from an integration over the angular parameter in a {\it bulk} tetrahedron. The same kind of factors can arise for the {\it boundary} tetrahedron:
For a simplex $\sigma$, glued via a tetrahedron $\tau$  to the boundary, one obtains the inner product of the coherent state associated to the pair $(\tau,\sigma)$ with the boundary state. Here we will assume that this boundary state is an eigenstate for all the area operators of the boundary triangles and is given as a product of  coherent state  $ \prod_\tau {\cal K}_\tau(\cdot, P^\tau_i)$ in the tetrahedral intertwiner degrees of freedom. (Here we assume that all tetrahedra in the boundary are positively oriented.) That is the boundary data include the areas of all triangles in the boundary as well as a pair of angular parameters $(P_1^\tau, P_2^\tau)$ for each boundary tetrahedron.  In this way we obtain a factor
\ba 
G^{\sigma}_\tau(a; P^\tau)=\langle {\cal K}_\tau ( \cdot, P^{\sigma,\tau}_i(a)) \,| {\cal K}_\tau ( \cdot, P^{\tau}_i)\rangle \q 
\ea
for each boundary tetrahedron.

Choosing these coherent state labels $P^\tau_i,i=1,2$ freely might lead to a set of boundary data, for which there is no consistent bulk configuration that satisfies the constraints (\ref{match1}). The product of $G$-factors would then suppress such configurations. To avoid such a suppression the boundary data have to satisfy constraints, also known as gluing or shape matching constraints \cite{AreaAngle,DittrichRyan1and2}. These ensure that the geometric data of the two tetrahedra sharing a given triangle $t$ do induce the same geometry (or shape) for this triangle.\footnote{These constraints, if applied to all pairs of tetrahedra in all simplices of the 4D triangulation, are equivalent to the constraints (\ref{match1}).} {We will approximate the boundary G-factors in the same way as the bulk G-factors in  \eqref{Gfct1}, that is with Gaussians in the constraints $\frak C_i = P^\tau_i-P^{\tau,\sigma}_i$, and  with a deviation $\Sigma^2 = \,  \ell_P^2 \,\gamma \, \tfrac{9}{4} \,\,(|\text{Vol}^s_\tau(\sigma)| + |\text{Vol}^s_\tau(\text{bdry})|)$, where $\text{Vol}^s_\tau(\text{bdry})$ is the volume of the boundary tetrahedron as determined by the boundary data.}

~\\
\emph{Remark:} In \cite{EffSF1,EffSF2} {the authors} used the 3D dihedral angles  to define the constraints ${\frak C'}_e= \Phi^{\sigma,\tau}(a)- \Phi^{\sigma',\tau}(a)_e$, where $\Phi^{\sigma,\tau}(a)$ is the 3D dihedral angle at the edge $e$ in $\tau$ as computed from the areas in the 4-simplex $\sigma$. One can also use  these constraints for the Gaussian approximation of the inner product between the coherent states in (\ref{equ16}).
Although only two out of the six constraints ${\frak C'}_e$ are independent, the Gaussian approximation to the inner product of the coherent states can introduce a dependence on this choice. This can be avoided by averaging over all 12 pairs of non-opposite edges in a tetrahedron. One then obtains an alternative approximation to the $G$-functions given by
\ba
G^{\sigma,\sigma'}_\tau\!(a) \sim  \delta(\text{sign}(\tau\subset\sigma) ,\text{sign}(\tau\subset\sigma')) \,\, \exp\left( - \frac{1}{12} \sum_{(e,e')\, \text{adj in}\, \tau }\frac{ |\frak{C}'_{e}|^2+ |\frak{C}'_{e'}|^2  }{4\Sigma^2(e,e')}\right)
\ea
where the variance is now determined by the commutation relations between the ${\frak C'}_e$ and given by
\ba
\Sigma^2(e,e')=\ell_p^2 \gamma \left( \frac{1}{L^\sigma_{e} L^{\sigma}_{e'}}+ \frac{1}{L^{\sigma'}_{e} L^{\sigma'}_{e'}} \right)\, .
\ea
Here we used absolute values $|\frak{C}_e|^2$ as the Lorentzian angles can become complex. 

The two different approximations to the $G$-functions can be understood to result from a variable transformation from the $\frak{C}_e$ to $\frak{C}'_e$, which is implemented in an expansion to second order in the constraints. That is the approximations will almost coincide in the semi-classical regime, where the Gaussians have a small spread.

\subsection{The path integral}\label{Sec:PI}

We can now define the path integral. As we mentioned before, in this path integral we integrate over the (bulk) areas, or more precisely sum over the spectral values for these areas. {We have as boundary data the boundary areas and angular parameters $P^\tau$ for each boundary tetrahedron.} The amplitude is \JP{built} from two kind of factors, firstly the oscillating factor given by the exponential of the Area Regge action (\ref{SAR}) and secondly the $G$-factors (\ref{Gfct1}), which implement the constraints (\ref{match1}) between the areas weakly. 

That is, we define
\ba\label{EffModel}
Z&=&\sum_{\text{Sig}}\sum_{a_t}\mu(a) \exp\left( \frac{\imath}{\hbar} S_{AR}(a)\right) \prod_\sigma \Theta^{\text{Sig}}_\sigma(a) 
\prod_{\tau \in \text{bulk}} G^{\sigma,\sigma'}_t\! (a)  \prod_{\tau \in \text{bdry}} G^{\sigma}_\tau (a;P^\tau) \q .
\ea
Here  
\begin{itemize}
\item We sum over $a_t=\gamma j_t$ with $j_t\in {\mathbb N}_+$ for space-like triangles and $a_t=n_t$ with $n_t\in {\mathbb N}_+$ for time-like triangles. We can also include the value $a_t=0$ for a degenerate or null triangle.

\item {We can sum over signatures $\text{Sig}$ of the various types of sub-simplices, or restrict to a certain type, see the discussion further below.}

\item  $\Theta^{\text{Sig}}_\sigma(a)$ implements the generalized triangle inequalities, see Appendix \ref{app:triangle}. It is equal to 1, if they are satisfied for a 4-simplex $\sigma$ with areas $\{a_t\}_{t\subset \sigma}$  , and vanishing otherwise. The super-index $\text{Sig}$ stands for signature, as the triangle inequalities are different for the different signatures of the tetrahedra, triangles and edges included in $\sigma$.

\item $\mu(a)$ is a measure term, that can be fixed by e.g. demanding invariance under coarse graining or implementing an approximate version of diffeomorphism invariance \cite{PImeasure}. Below we will compute expectation values for which the measure term does (approximately) cancel out in the semi-classical regime. We will therefore set this measure term to $\mu(a)=1$. 
\end{itemize}

As we have  already seen in section \ref{Sec:ImPart} there are several further specifications and alternative definitions for the path integral. These include
\begin{itemize}
\item {\it Restricting signatures or sum over signatures:} We can either sum over all possible signatures of edges, triangles and tetrahedra or restrict to a prescribed signature for all these elements. Note that the  original EPRL-FK models \cite{EPRL-FK} only allowed for space-like tetrahedra and therefore also only for space-like triangles and edges.  If one \JP{allows} for different signatures, one might have configurations in which the signatures for an edge differ in the geometries induced from the different 4-simplices it is contained in. (This cannot happen for tetrahedra, as we did define the $G$-function to be vanishing in this case. It \JP{also cannot} happen for triangles, as we \JP{specify} the triangle areas and with \JP{these} their signatures.) 
\item {\it Allowing or restricting irregular light cone structure:}  As we reviewed in section \ref{Sec:ImPart} the Regge calculus set-up, and thus spin foams, allow for irregular light cone \JP{structures}. That is\JP, (bulk) triangles where less or more than two light cones meet.  Such configurations can be interpreted to describe  a change for the topology of the spatial hypersurface in time. We can decide to exclude such configurations, as is done in Causal Dynamical Triangulations \cite{CDT}. In case we allow such configurations  the Regge action will include imaginary parts. Such imaginary parts \JP{in the path integral lead} to either a suppressing or enhancing factor for such configurations. The enhanced configurations could turn out to be dominating, it is therefore important to understand \JP{when} such configurations can appear.
\item {\it Including \JP{contributions} from Euclidean 4-simplices:}  Another possibility is to allow the signature of the 4-simplices to be Euclidean. The signature of a given simplex can be decided from the sign of its \JPp{Cayley}-Menger determinant (see Appendix \ref{app:triangle}), which is a polynomial in the (signed) squared edge lengths and also defines the signed squared volume. The generalized triangle inequalities for a Lorentzian 4-simplex demand in particular that the signed squared 4-volume is negative. For a Euclidean 4-simplex the signed squared 4-volume is positive and such simplices are a priori not included in the sum (\ref{EffModel}). We can however decide to include such Euclidean simplices, but use an exponentially suppressed amplitude $\exp(-\frac{1}{\hbar} |S_{AR}|)$. The reason to do so \JP{is} that analogous contributions do appear in the asymptotic analysis \cite{BarrettFoxon} of the  Ponzano-Regge model \cite{PR}, which can be understood as spin foam model for three-dimensional Euclidean geometries. In this case the amplitude for a tetrahedron with edge lengths \JP{violating} the Euclidean triangle inequalities, but \JP{that} define a tetrahedron with Lorentzian geometry is not vanishing, and given by an exponentially decaying factor, determined from the Lorentzian Regge action.  
    The situation is unfortunately more involved for the asymptotic analysis of four-dimensional  Lorentzian spin foam models \cite{SFLimit,SFtimelike}: there Euclidean configurations can appear with amplitudes that are not exponentially suppressed\JP{, but} these contributions result from so-called vector geometries, which is a  sector of solutions to the simplicity constraints resulting from a certain set of degenerate configurations, which one rather might want to exclude.

    But there is another argument to include such Euclidean configuration even within Regge calculus:  This argument relies on the demand that the amplitudes associated to a simplex should be ideally invariant under refinement \cite{Improved,DittrichReview14}. That is, if we refine a 4-simplex, e.g. by subdividing it into five 4-simplices, and compute the partition function for this new configuration, we would like to obtain back the amplitude for the original 4-simplex. As an example we can choose all edges of the original 4-simplex to have the same positive length square. Such a configuration would not be allowed by the Lorentzian triangle inequalities, and the corresponding amplitude would be zero \JP{if} we do not allow for Euclidean simplices. But we can subdivide this  4-simplex into five  4-simplices \JP{with} geometric data that are \JP{allowed} by the Lorentzian triangle inequalities.  The amplitude for such a configuration would therefore be a priori non-vanishing.  We \JP{expect} that there  does not exist a classical solution for the lengths of the bulk edges using the Lorentzian Regge action\JP{, but} starting with time-like bulk edges, one can  `Wick-rotate' these to Euclidean signature and with it the Lorentzian Regge action to the Euclidean Regge action.  For the Euclidean Regge action one will find extrema. That is\JP, the Lorentzian path integral for the subdivided simplex with Euclidean boundary data has saddle points in the complex plane. These might lead to exponentially suppressed contributions, which are controlled by the Euclidean Regge action.
    
The area spectrum for Euclidean 4-simplices is also $\gamma$--independent (and appears in this from in the asymptotics \cite{SFLimit}), that is given (asymptotically) by $a_{t}^{\text{Eucl}} \sim n_t$. The reason is that, although canonical loop quantum gravity does suggest a $\gamma$--dependent spectrum also in this case, the Euclidean EPRL-FK model does actually impose constraints on the values of $\gamma$, which effectively lead to $a_{t}^{\text{Eucl}} \sim n_t$. Another way to argue for such a spectrum is that the areas are conjugated to the angles. These angles are compact for time-like triangles in Lorentzian 4-simplices, and for space-like triangles in Euclidean 4-simplices. The $\gamma$--dependent spectrum  for space-like triangles in Lorentzian 4-simplices  can be understood as a consequence of the canonical transformation underlying the construction of the Ashtekar-Barbero connection.

\item {\it Sum over orientations:} Another variation consists of including a sum over orientations for the 4-simplices.  The sum over positive and negative orientation for a given simplex would then lead to an oscillatory factor given by the cosine of the Regge action instead of the exponential of $\imath$ times the Regge action. This is the usual choice for spin foams, but \cite{Engle} argues for a version of EPRL-FK model in which one restricts to one orientation. The effective model does allow for a more straightforward implementation of the different versions, and facilitates in this way a test for the various proposals.

\end{itemize}

We expect that these choices have  important physical implications, but we will leave the explorations of most of these options for future work. Here we will only test some aspects of summing over different signatures in section \ref{Sec:Examples}. The effective spin foam model has the advantage that these differences in the definition of the model can be much more easily tested and explored than for the EPRL-FK models. 

The effective spin foam model in the form of (\ref{EffModel}) does require only a summation over the areas. This summation range can be furthermore constrained using the $G$--factors. This is one reason why this model is much more numerically efficient than previous spin foam models.

The EPRL-FK models \cite{EPRL-FK} are defined as \JP a path integral over both the area as well as the 3D angle degrees of freedom. Integrating out the angles one \JP{also gets} an effective amplitude \JP{which} does only depend on the areas and on the 3D angles in the boundary. It is not clear yet \JP{whether} these amplitudes would, in the limit of large quantum numbers, coincide with the amplitude of the effective model. In particular, it is not fully clear yet \JP{to} which degree the EPRL-FK \JP{models implement} the area constraints (\ref{match2}).  But as the large \JP{$j$--limits} of the EPRL-FK models \JP{also lead} to the Regge action \cite{SFLimit}, we expect that the difference in the actions can be only involving terms that are proportional to the area constraints.

\section{Semiclassical regime}\label{sec:semi}

The semiclassical limit is usually associated with the limit of large quantum numbers, or equivalently, the limit $\hbar\rightarrow 0$.  In this \JP{limit} the amplitudes in the path integral oscillate very strongly except around stationary points of the action. Such stationary points will therefore dominate.

There is however a caveat in this argument if one \JP{has} a system where one has weakly implemented constraints. In the case of the Area Regge action, as well as the Plebanski action underlying spin foams, the unconstrained action is not stationary in all the directions at configurations describing gravitational solutions.  Indeed\JP, the constraints \JP{disallow} variations in certain directions\JP{. Therefore, the} (unconstrained) action \JP{needs} to be \JP{stationary only} in the remaining directions. 

In the path integral with weakly implemented constraints, e.g. with Gaussians, the constrained directions can still contribute over the widths of these Gaussians. If the amplitudes \JP{oscillate} very strongly along these constrained directions it can still lead to destructive interference that would prevent the classical solutions from dominating. Indeed\JP, the number of oscillations grows \JP{as} $1/\hbar$ whereas the widths of the Gaussians, determined by the uncertainty relations, \JP{grow as} $\sqrt{\hbar}$. That is, we can expect that the oscillations \JP{`overtake'} the Gaussians in the $\hbar\rightarrow 0$ limit. This is the origin of the so-called  flatness problem for spin foams \cite{flatness,ILQGS}: in this limit one obtains the condition that the (unconstrained) action is stationary in all directions, which gives both for the Area Regge and Plebanski action, flat solutions. 

In the case of (effective) spin foams the non-commutativity of the constraints can be seen as an anomaly, which is parameterized by the Barbero-Immirzi parameter $\gamma$. It is therefore reasonable to demand that $\gamma$ should be small. This can indeed ensure a semi-classical regime: Demanding that the number of oscillations over the widths of the Gaussian factors is less \JP{than} or of order one we obtain the condition \cite{EffSF1,EffSF2}
\ba\label{CondSL}
\frac{\sqrt{\gamma a_t}}{\ell_P}  \epsilon_t  \leq {\cal O}(1) \q ,
\ea
where $a_t$ is the area (either given by $\gamma j_t$ or $n_t$) and $\epsilon_t$ the deficit angle. The argument in \cite{EffSF1} was derived for triangulations with space-like tetrahedra only, but \JP{goes} through in the same way for triangulations with time-like triangles. 
This condition did first arise from a  semi-classical analysis of the EPRL/FK models using micro-local analysis techniques \cite{HanSBI}. But it can be \JP{explained} quite simply from the scaling properties for the frequency of the oscillating factor, which is determined by the variation of the action,  and for the widths  of the Gaussian factors, which are determined by the commutator of the simplicity constraints \cite{EffSF1}.

In section \ref{Sec:Examples} we will compute expectation values as functions of $\gamma$ for a number of examples.  For each \JP{one} we will identify a semi-classical regime, \JP{i.e.} a range of $\gamma$'s where the expectation values do reproduce the classical values. These examples will give some support to the condition (\ref{CondSL}).  But we will also see behaviour which deviates from what the bound (\ref{CondSL}) suggests.

In this work we will consider examples where we have only one bulk triangle. For Euclidean signature, more involved examples have been already considered \cite{EffSF2}, and it has been shown that there is a semi-classical regime which implements the (Length) Regge equations of motion. There, for small curvature angles, the semiclassical regime is not strictly controlled by the bound in (\ref{CondSL}). The bound (\ref{CondSL}) \JP{implies} that with fixed $\gamma$ and $\epsilon_t$ the semi-classical regime appears only for a bounded range of areas. In contrast\JP{, \cite{EffSF2} suggests} that for sufficiently small curvatures the matching of the expectation values to the classical value does actually improve with growing scale $a_t$. This \JP{suggests} that for examples requiring an integration over more variables\JP, the semi-classical properties of the weakly constrained path integral  can actually improve.

\section{Examples}\label{Sec:Examples}

\subsection{Expectation values}

As a first test of the Lorentzian effective spin foam models, we will consider triangulations \JPp{having only} one bulk triangle. {We will discuss these examples in detail in the next section.}
To test whether the models admit a semi-classical regime, we will compute expectation values of some geometric observables, in particular of the area and the deficit angle of the bulk triangle. The expectation values of a given observable ${\cal O}$ is defined as
\ba\label{Expvalue}
\langle {\cal O} \rangle&=&\frac{ \sum_{a_t} {\cal O}(a_t) {\cal A}(a_t)} {\sum_{a_t}{\cal A}(a_t)}
\ea
where  the amplitudes ${\cal A}(a_t)$ are defined in (\ref{EffModel}), with the understanding that $Z=\sum_{a_t} {\cal A}(a_t)$.

  \begin{figure}[ht!]
\begin{picture}(500,175)
\put(70,7){ \includegraphics[scale=0.44]{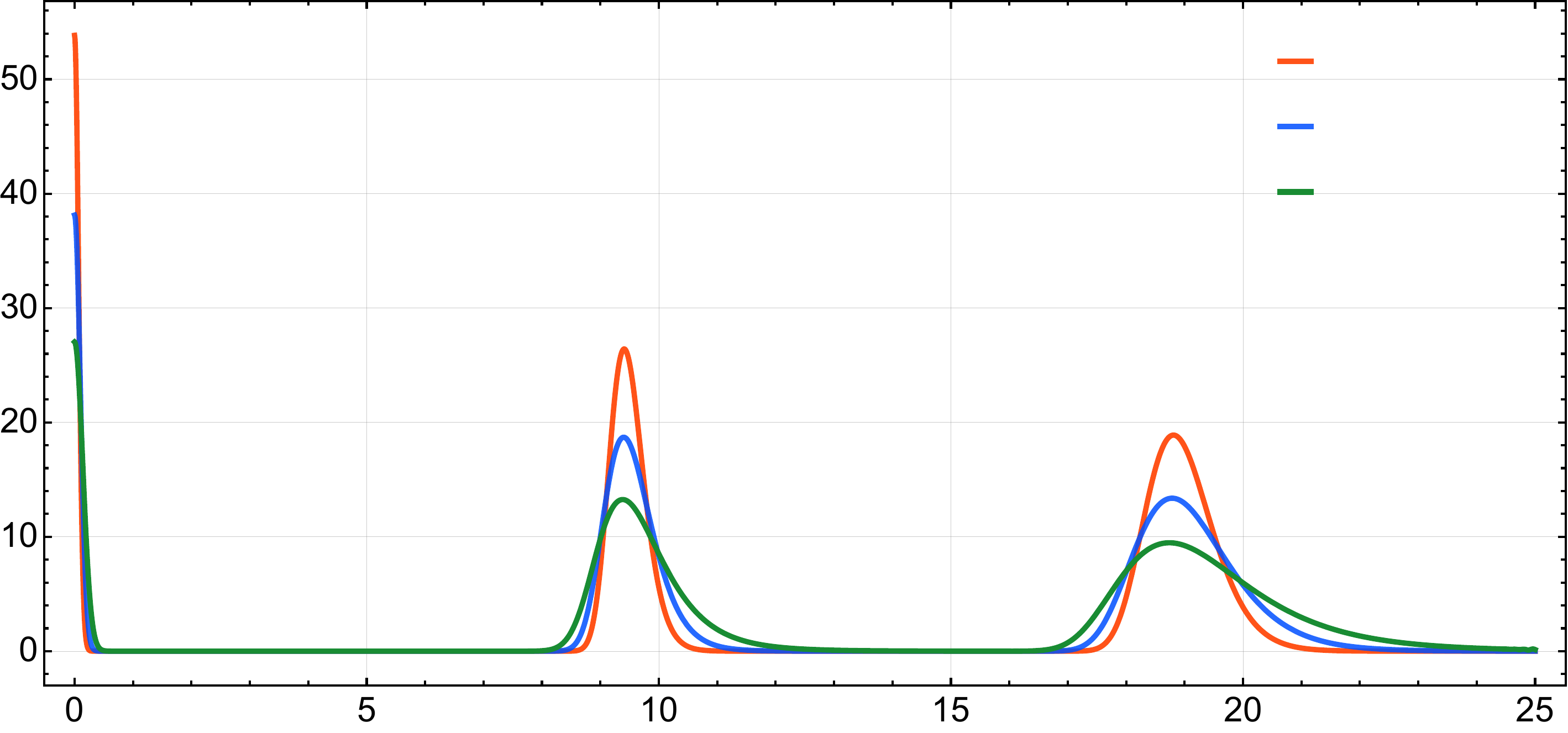} }

\put(280,0){$\gamma$}
\put(50,168){${\rm Abs}({Z})$}

\put(375, 160){ \scriptsize   $\lambda = 20 $ }
\put(375, 145){ \scriptsize   $\lambda = 10 $ }
\put(375, 130){ \scriptsize   $\lambda = 5 $ }

\end{picture}
\caption{ This plot shows the absolute value of the partition function for a configuration with one bulk triangle, discussed in section \ref{Sec:spacelike}.  The $G$-factors are peaked on an area value, leading to a curvature $\epsilon^c \simeq 0.667$. We show ${\rm Abs}({Z})$ for three different sets of boundary data, which just differ by a global scaling factor $\lambda$. Although ${\rm Abs}({Z})$ appears to be zero over large ranges of $\gamma$, it is actually just extremely small, see Figure \ref{Fig:Instab} for a zoom into $\gamma \in (0.5,1)$, where ${\rm Abs}({Z})$ values range from approximately $10^{-11}$ to $10^{-9}$. \label{Fig1}}
\end{figure}
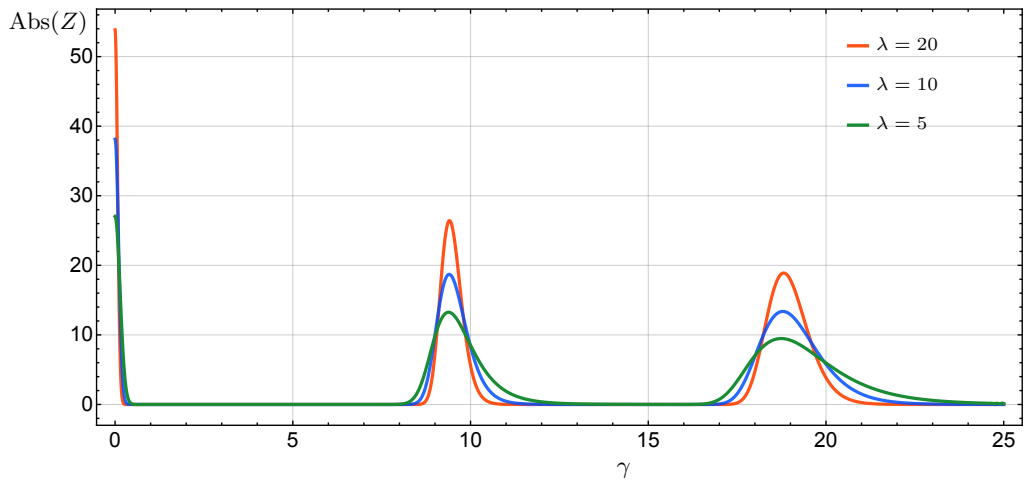

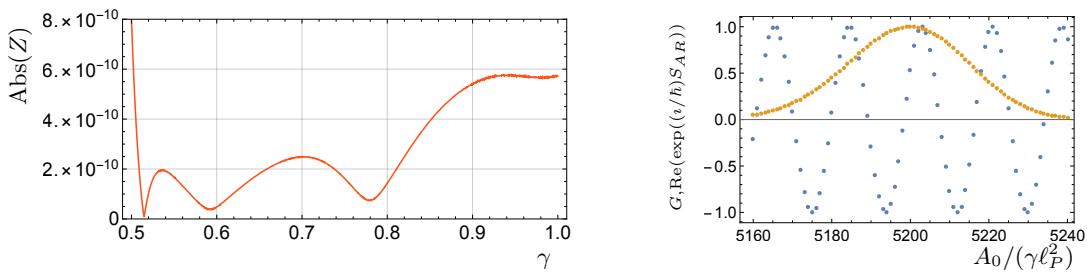
\begin{figure}[ht!]
\begin{picture}(500,100)
\put(30,7){ \includegraphics[scale=0.4]{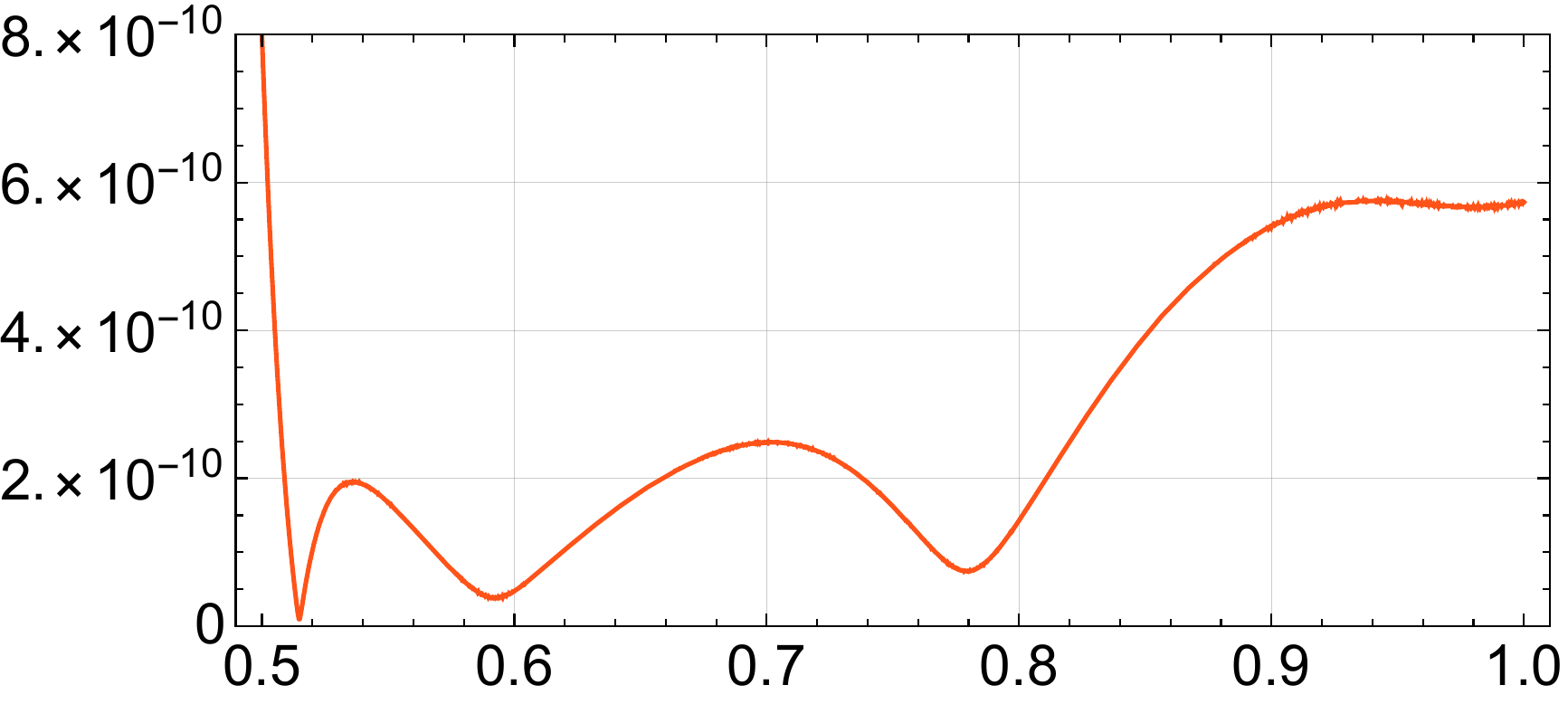} }
\put(280,7){ \includegraphics[scale=0.4]{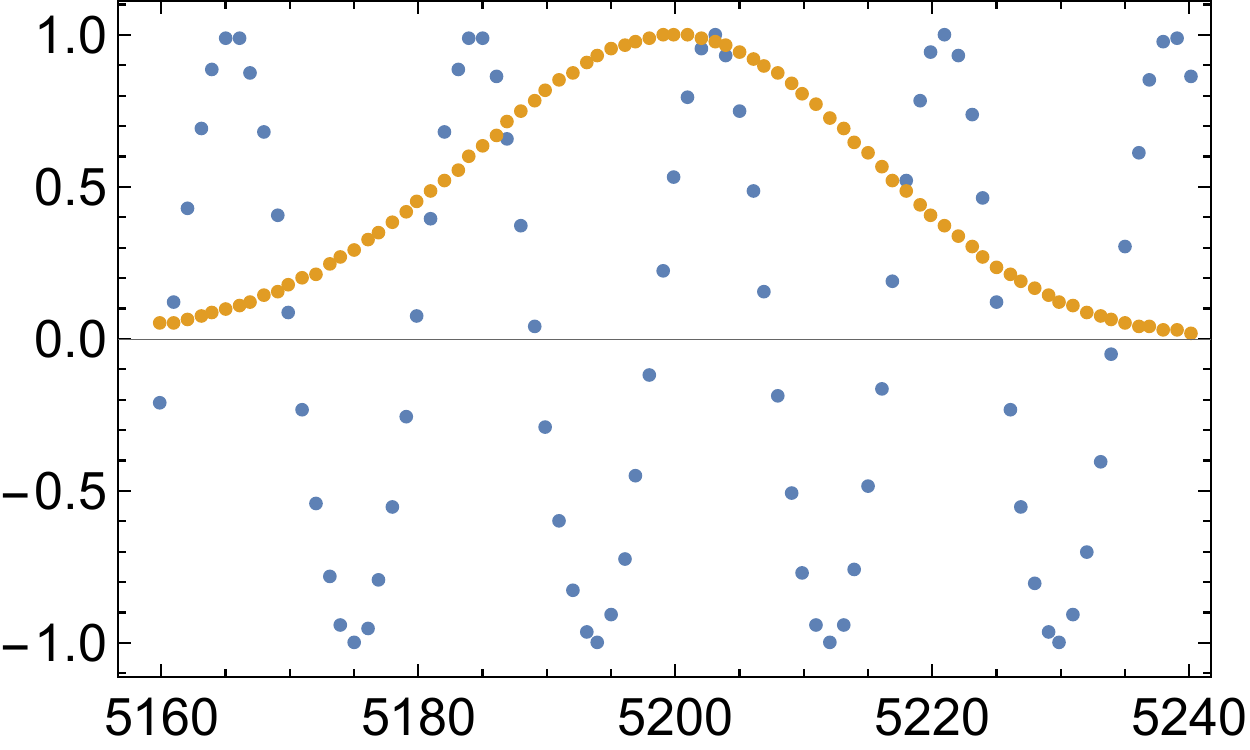} }

\put(220,0){$\gamma$}
\put(385,0){$A_0/(\gamma \ell_P^2)$}

\put(20,60){\rotatebox{90}{$ {\rm Abs}({ Z})$}}
\put(270,20){\rotatebox{90}{${\scriptstyle G, \text{Re}(\exp((\imath/\hbar )S_{AR} ))} $}}

\end{picture}
\caption{ The left panel shows  ${\rm Abs}({Z})$ for $\lambda=20$ for the same configuration as in Figure \ref{Fig1}, but for a smaller range of $\gamma$'s. This reveals a  fine structure in ${\rm Abs}({Z})$, and a particular a series of minima, the first of which occurs around  $\gamma \approx 0.5148$. The right panel shows the product of the $G$-factors and $\exp(\tfrac{\imath}{\hbar} S_{AR})$ for this value of $\gamma=0.5148$.  For this configuration the oscillations average the product of the $G$-factors to almost zero.  \label{Fig:Instab}}
\end{figure}

As was noted in \cite{EffSF2}\JPp, the behaviour of the absolute value of the partition function as a function of $\gamma$ gives already an indication  of how the expectation values behave.  Note that the absolute value of the partition function will typically drop sharply and monotonously from $\gamma=0$  up to some configuration dependent threshold value. Figure \ref{Fig1} shows an example where the absolute values drop by a factor of around $10^{-10}$.  This drop is more pronounced for configurations with larger curvature (compare with Figure \ref{Fig3} which shows an example with lower curvature) and can be interpreted as a `suppression' of configurations with curvature, indicative of the flatness problem for spin foams \cite{flatness}. This drop alone is actually not so problematic for the expectation values, as these are defined via the quotient  in (\ref{Expvalue}). We indeed find that the semi-classical regime includes $\gamma$--values for which ${\rm Abs}(Z)$ is relatively small.  But the interplay between the $G$-factors, the oscillations of the amplitude, and the discreteness of the sum, can lead to quite pronounced minima in ${\rm Abs}(Z)$. Figure \ref{Fig:Instab} shows one such example, where a `sharp' minimum occurs around $\gamma\approx 0.515$.

Here we have that the oscillations of the amplitude, which occur over the widths of the Gaussian factor,  average this Gaussian factor to almost zero, see the right panel of Figure \ref{Fig:Instab}. This can be considered to be an anti-resonance effect. Such an effect will typically appear for the examples with larger curvature. The  smallest $\gamma$--value leading to such an effect will give an upper bound for the semi-classical regime. Indeed we will see that such minima lead to sudden jumps for the expectation values, see Figure \ref{Fig2}.

Figure \ref{Fig1} shows that for quite large $\gamma$--values we have also very pronounced maxima. These maxima can be understood as a resonance effect, resulting from the interplay of the discreteness of the area values and the phase of the oscillations in the amplitude. We will explain this effect in the next section.

\subsection{A triangulation with space-like triangles}\label{Sec:spacelike}

In this section we will consider a triangulation with four 4-simplices glued around a triangle. The chosen boundary data will  lead to 4-simplices \JPp{that} are Lorentzian, but the tetrahedra, and therefore triangles and edges, are all space-like. This will allow to compare the results obtained here with the standard EPRL-FK-models, once similar numerical simulations for these models are available.

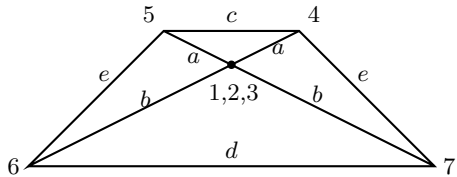
\begin{figure}[ht!]
\begin{tikzpicture}[scale=0.6]
\draw[thick] (0,0) --(9,0) -- (6,3)--(3,3)--(0,0)--(6,3) (3,3)--(9,0);
\draw[thick,fill] (4.5,2.25) circle (2pt);

\node[above right] at (6,3) {4}; 
\node[above left] at (3,3) {5}; 
\node[left] at (0,0) {6}; 
\node[right] at (9,0) {7}; 
\node[below right] at (3.8,2) {1,2,3}; 

\node[left] at (2,2) {$e$}; 
\node[right] at (7.1,2) {$e$}; 
\node[above] at (4.5,3) {$c$}; 
\node[above] at (4.5,0) {$d$}; 
\node[left] at (2.9,1.5) {$b$}; 
\node[right] at (6.1,1.6) {$b$}; 
\node[right] at (5.2,2.6) {$a$}; 
\node[left] at (4.,2.4) {$a$}; 

\end{tikzpicture}
\caption{This shows a projection of a complex of four simplices sharing a bulk triangle, to the (Minkowski) plane orthogonal to this triangle. The bulk triangle therefore appears as vertex here labelled by the vertices $(1,2,3)$ of the original triangle. To illustrate the symmetry reduction (\ref{EdgeL1}), we labelled the remaining edges with their  lengths parameters in the non-projected triangulation. All edges are assumed to be spatial.}
\label{SLComplex}
\end{figure}

Figure \ref{SLComplex} shows a dimensionally reduced representation of the triangulation. The inner triangle is shared by four 4-simplices, which we can name as down, up, left and right. We assume that certain edge lengths are equal to each other, this forces also two of the 4-simplices (the left and the right one) to agree.

We describe the simplices by their vertex set, the vertices are enumerated from 1 to 7. The 4-simplices are then given by
\ba\text{up}=(12345), \q \text{left}=(12356), \q \text{down}=(12367), \q \text{right}=(12347) \,.
\ea
These simplices all share the triangle $(123)$. But the 4-simplices are distinguished by the remaining pair of vertices, which are given by $(45),(56),(67)$ and $(47)$ respectively.

We make some symmetry assumptions, which will constrain the boundary data.  These result from setting the following edge lengths to be equal, see also Figure \ref{SLComplex}:
\ba\label{EdgeL1}
x&=& l_{12}=l_{13}=l_{23} \,\, ,\q \nn\\
a&=& l_{14}=l_{15}=l_{24}=l_{25}=l_{34}=l_{35} \,\, ,  \nn\\
b &=& l_{16}=l_{17}=l_{26}=l_{27}=l_{36}=l_{37} \,\, , \nn\\
c&=&l_{45}\,\,,\nn\\
d&=&l_{67}\,\,,\nn\\
e&=&l_{47}=l_{56} \,\,.
\ea
Thus the simplex $\text{up}=(12345)$ has three length parameters $x,a$ and $c$ and also three area parameters $A_0=A(x,x,x),A_1=A(a,a,x)$ and $A_3=A(a,a,c)$. (Here $A(x,y,z)$ is the area for a triangle with edge lengths $x,y,z$.) Similarly we have for the simplex $\text{down}=(12345)$ three length parameters $x,b$ and $d$ and three area parameters $A_0=A(x,x,x),A_2=A(b,b,x)$ and $A_4=A(b,b,d)$. The $\text{left}=
(12356)$ and $\text{right}=(12347)$ simplices have the same geometry, which is described by four lengths parameters $x,a,b,e$ or, alternatively, by four area parameters $A_0=A(x,x,x),A_1=A(a,a,x), A_2=A(b,b,x)$ and $A_5=A(a,b,e)$.

{We have discussed in sections \ref{Sec:CI} and \ref{Sec:PI} that the boundary data for the effective spin foam path integral are given by the boundary areas and the angle parameters $P_i^\tau$ for each boundary tetrahedron $\tau$. With our symmetry assumption on the boundary data, the five parameters $A_1,\ldots, A_5$ do give all the boundary areas. These boundary areas, however, do not fix all the boundary length parameters. For this we need to also consider  the angle parameters $P_i^\tau$ (where $\tau$ are boundary tetrahedra), which add --- within our symmetry assumption --- one additional parameter to the five area parameters. We thus have over-all six parameters describing the boundary data -- and these six parameters do encode the six boundary edge lengths $x,a,\ldots,e$.  These six boundary edge lengths do also fix the ``classical" value $A^c_0=A(x,x,x)$ for the bulk area. The effective spin foam path integral is a sum over the bulk area $A_0$. But its amplitude does include Gaussian factors that are peaked on the value $A^c_0$ for the bulk area $A_0$, with a spread determined by (\ref{SigmaF}).  It also contains an oscillating factor,  determined by the Area Regge action (\ref{SAR}).} 
Whether the expectation value of $A_0$ does approximate well the classical value $A^c_0$ will depend on how much the   amplitude oscillates over the spread of the Gaussian factors, see Figure \ref{Fig:Instab}.


To test this we will consider a certain configuration of boundary areas given by
\ba\label{Bareas}
(A_1,A_2,A_3,A_4,A_5)=  \lambda \,\gamma\, \ell_P^2 \,(99,187,68,190,49)          \q ,
\ea
where $\lambda$  allows us to investigate different scales. Choosing different  values for $A^c_0$ \JPp{will} pick out different deficit angles and therefore curvatures at the bulk triangle. Note that with changing $\gamma$, we change also the scale for the values (\ref{Bareas}) of the boundary areas. What is kept fixed are the spin representation labels $j_i$, which determine the area values by $A_i=\lambda \,\gamma\, \ell_P^2 j_i$.

The range of the allowed values for $A_0$ depends on a choice of root for the area-length system. Different roots are distinguished by their angular parameters, including those for the boundary tetrahedra. By specifying the angular parameters of the boundary tetrahedra, or equivalently all boundary lengths,  and using sufficiently large scales, we can peak the $G$--function so that only one root contributes. In our case is  root is given by:
\ba 
&&x^2=  \frac{4 A_0}{\sqrt{3} }\, ,\q\q
a^2=\frac{A_0^2+3A_1^2}{\sqrt{3} A_0} \, ,\q\q
b^2=\frac{A_0^2+3A_2^2}{\sqrt{3} A_0} \, ,\q \q \nn\\
&&c^2=2\frac{A_0^2+3A_1^2}{\sqrt{3} A_0} - 2\sqrt{  \frac{(A_0^2+3A_1^2)^2}{3 A^2_0}  -4 A_3^2   } \, ,\q\q
d^2=2\frac{A_0^2+3A_2^2}{\sqrt{3} A_0} +2\sqrt{  \frac{(A_0^2+3A_2^2)^2}{3 A^2_0}  -4 A_4^2   }\,,\nn\\
&&e^2=\frac{1}{\sqrt{3}A_0} \left( 
3 A_1^2+3 A_2^2 +2 A_0^2+2 A_0 \sqrt{3 A_1^2 + 3 A_2^2 - 12 A_5^2 +A_0^2 +9 A_1^2 A_2^2 A_0^{-2}  }
\right)
\ea
For the boundary areas given in (\ref{Bareas}) these lengths functions are well defined (and positive) for positive $A_0$, except for a region where the argument in the square root appearing in the $d^2$--solution becomes negative:
\ba
  \frac{(A_0^2+3A_2^2)^2}{3 A^2_0}  <4 A_4^2    \q ,
\ea
that is for $ 270.84\leq \frac{A_0}{\lambda\gamma \ell_P^2}\leq 387.34$.  We will choose $A^c_0<270\lambda\gamma \ell_P^2$, and because of the Gaussian factors the region with $A_0>387 \lambda \gamma \ell_P^2$ becomes irrelevant. Thus we will impose $A_0=270 \lambda  \gamma \ell_P^2$ as an upper bound for our summation range. The generalized triangle inequalities impose furthermore a lower bound $A_0\geq  239  \lambda\gamma\ell_P^2$.  In addition to these bounds the product of the $G$-functions will have extremely small  for values of $A_0$ that are sufficiently distant from $A_0^c$. \JPp{Taking into account only values} where the product of the  $G$-functions is larger than e.g. $10^{-10}$, we can restrict the summation range even further.

In the following we will discuss the results of the numerical evaluation of the partition functions for two different regimes of curvature (as induced by the boundary data): one with larger curvature (where $A_0^c=260 \lambda \gamma \ell_P^2$ and  corresponding deficit angle  $\epsilon^c=0.67$), and one with smaller curvature (where
 $A_0^c= 245 \lambda \gamma \ell_P^2$ and  corresponding deficit angle $\epsilon^c=0.24$).

~\\
\emph{Larger curvature regime:}
Here we will first consider the absolute value of the partition function in dependence of $\gamma$. Note that with $\gamma$ we vary also all area (spectral values), including the boundary areas. Fig. \ref{Fig1} shows the absolute value over a very large range of the Barbero--Immirzi parameter. This is to showcase a characteristic feature of the larger curvature regime\JPp: peaks in the absolute value of the partition function at certain values of $\gamma$, which occur here at $\gamma \approx 9.4$ and $\gamma \approx 18.8$. These peaks occur for $\gamma$--values satisfying the condition $\gamma \times \epsilon^c=2\pi \times N$ with $N\in {\mathbb N}$. Such a condition (with a $4\pi$ factor instead of a $2\pi$ factor, if one also allows for half integer $j$ in the spectrum $a_t=\gamma \ell_P^2 j$) has been also derived in the asymptotic analysis of the EPRL/FK spin foam models \cite{flatness}.  

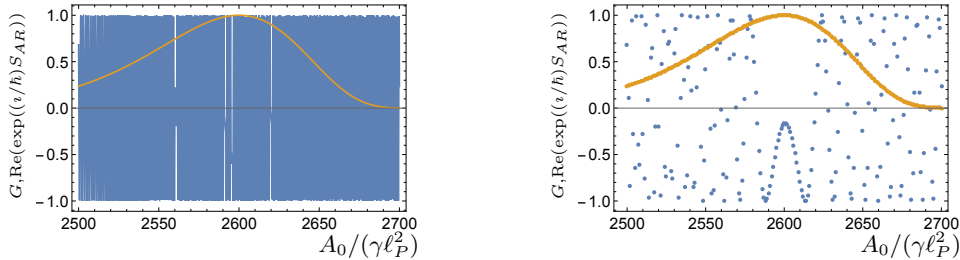
\begin{figure}[ht!]
\begin{picture}(500,75)
\put(50,7){ \includegraphics[scale=0.4]{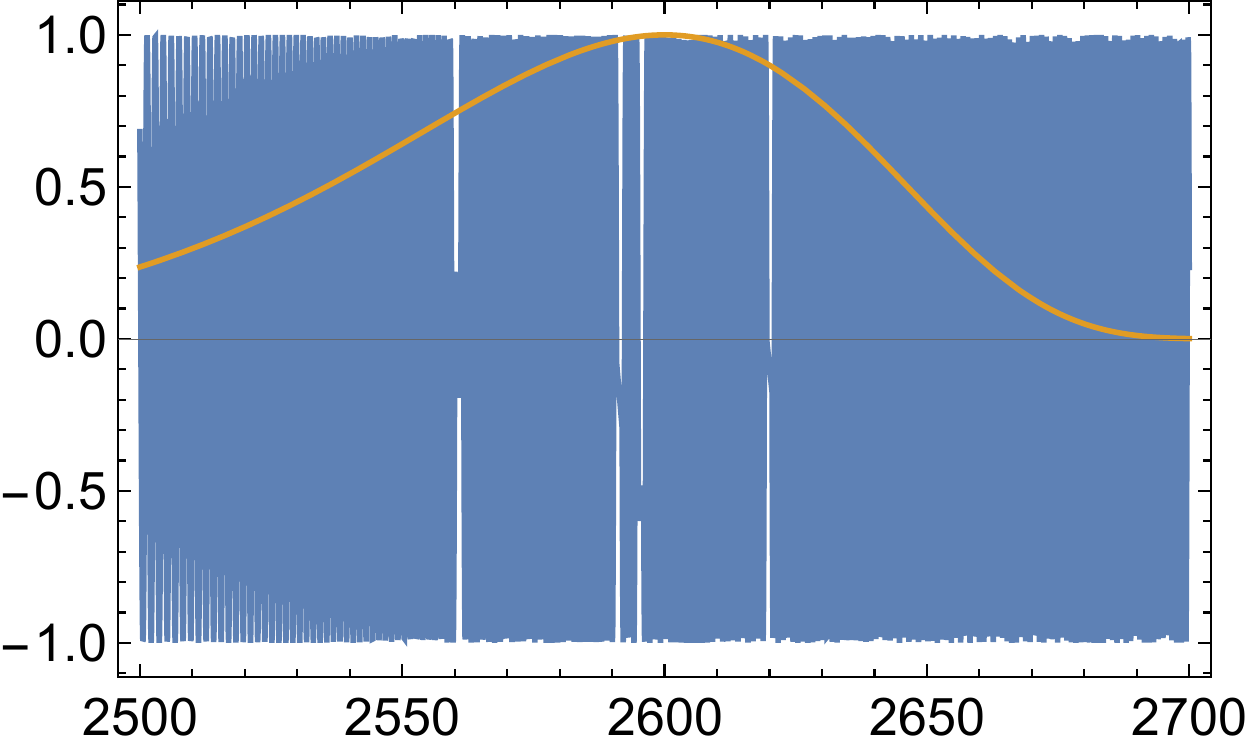} }
\put(255,7){ \includegraphics[scale=0.4]{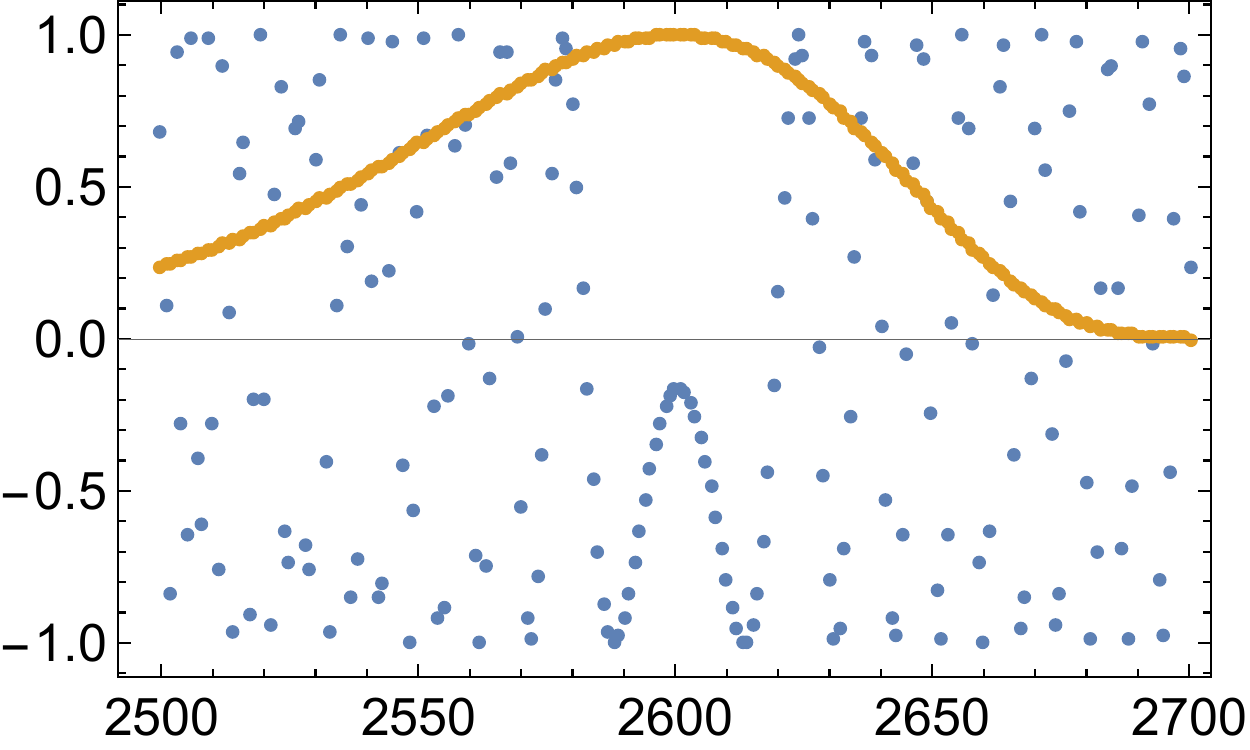} }

\put(160,0){$A_0/(\gamma \ell_P^2)$}
\put(365,0){$A_0/(\gamma \ell_P^2)$}

\put(43,20){\rotatebox{90}{$ {\scriptstyle G, {\rm Re}(\exp((\imath/\hbar )S_{AR} ))}$}}
\put(248,20){\rotatebox{90}{$ {\scriptstyle G, {\rm Re}(\exp((\imath/\hbar )S_{AR} ))} $}}

\end{picture}
\caption{ This figure shows the product of the $G$-factors and the oscillating factor $G, \exp((\imath/\hbar )S_{AR} )$ for $\gamma=9.4$ and $\lambda=10$. The left panel shows these quantities as continuous functions of $A_0$, \JPp{illustrating} the highly oscillatory behaviour of the amplitude. The right panel shows these quantities only for the discrete spectral values of $A_0$. The interplay between the phase of the oscillations and the discreteness leads to an appearance of a `pseudo stationary' point near $A_0^c$, where the $G$-factors are peaked.  \label{Fig:Pseudo}}
\end{figure}

The peaks appear due to an interplay of the discreteness of the sum, i.e. the area spectra $A_0=\gamma \ell_P^2 j$\JPp, and the frequency $\nu$ of the oscillation in the amplitude as a function of $j$, which is  given by $ \nu=(2\pi)^{-1} \gamma \epsilon$. The deficit angle $\epsilon$ is a function of $j$, but the $G$-functions peak on the value $\epsilon^c$. That is the partition functions sums over an oscillating factor with frequency near $(2\pi)^{-1} \gamma \epsilon^c$. If $\gamma\epsilon^c=2\pi \times N$ we will have that the frequency is approximately an integer $\nu \approx N$ and we \JPp{sum} $\exp( 2\pi \imath \nu j )\approx 1$ over $j\in \mathbb{N}$. This  explains the peaks in the absolute value of the partition function. This effect can be understood as the appearance of `pseudo stationary points' due to the discreteness of the variables one integrates over, see Figure \ref{Fig:Pseudo}.

~\\

We should note however \JPp{that} these special values of $\gamma$ are quite large (for reasonable small deficit angles) and well outside a reliable semi-classical regime. That is the area expectation values do not reproduce well the classical values. Indeed\JPp, before this coherence effect we \JPp{have} the opposite effect. E.g. for $\gamma\epsilon^c=2\pi \times \tfrac{1}{2}$ we would sum over values that are approximately (for $j$'s near the Gaussian peak) given by $(-1)^j$, leading to a destructive interference.

\begin{figure}[ht!]
\begin{picture}(500,150)
\put(1,7){ \includegraphics[scale=0.295]{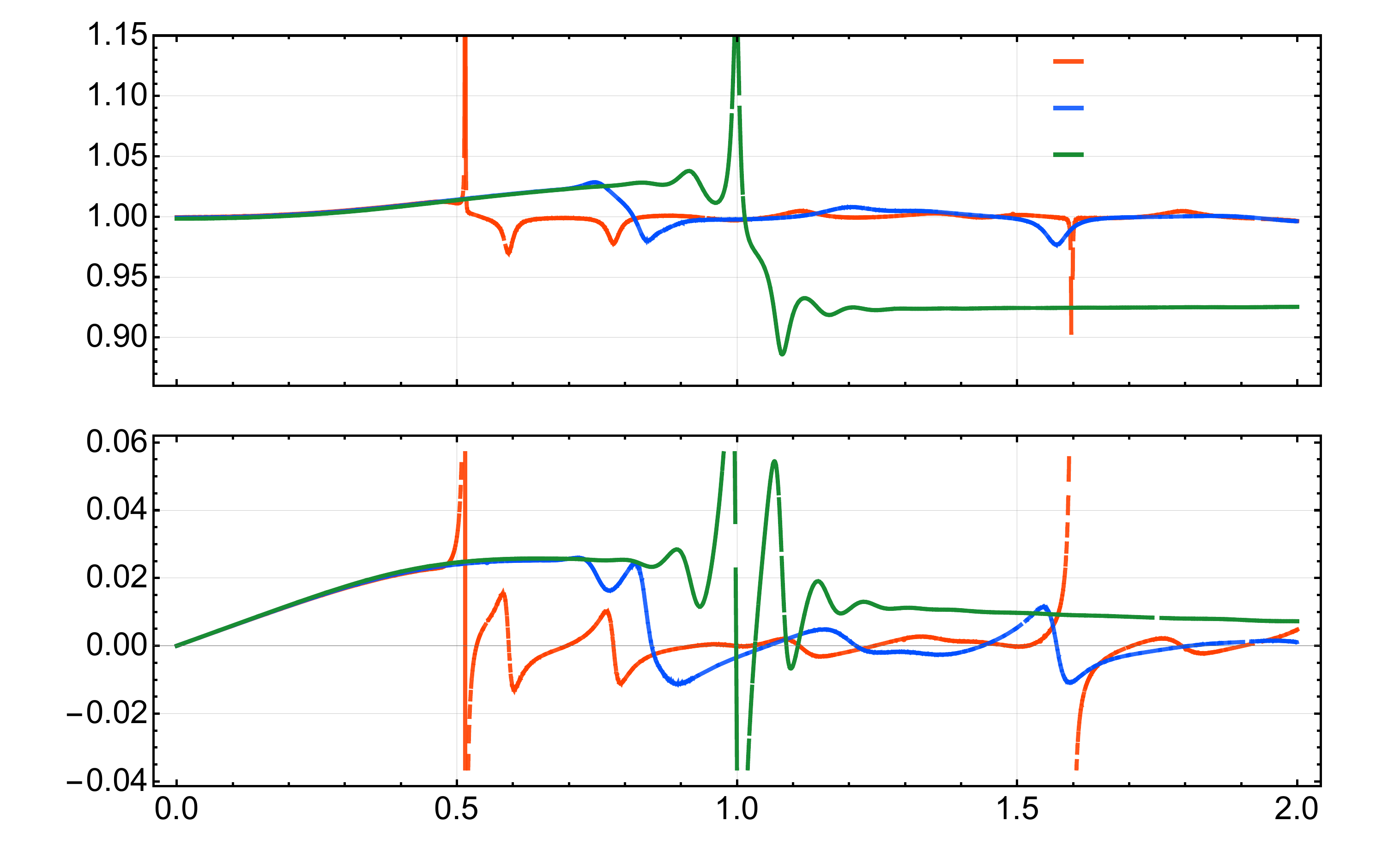} }
\put(255,7){ \includegraphics[scale=0.295]{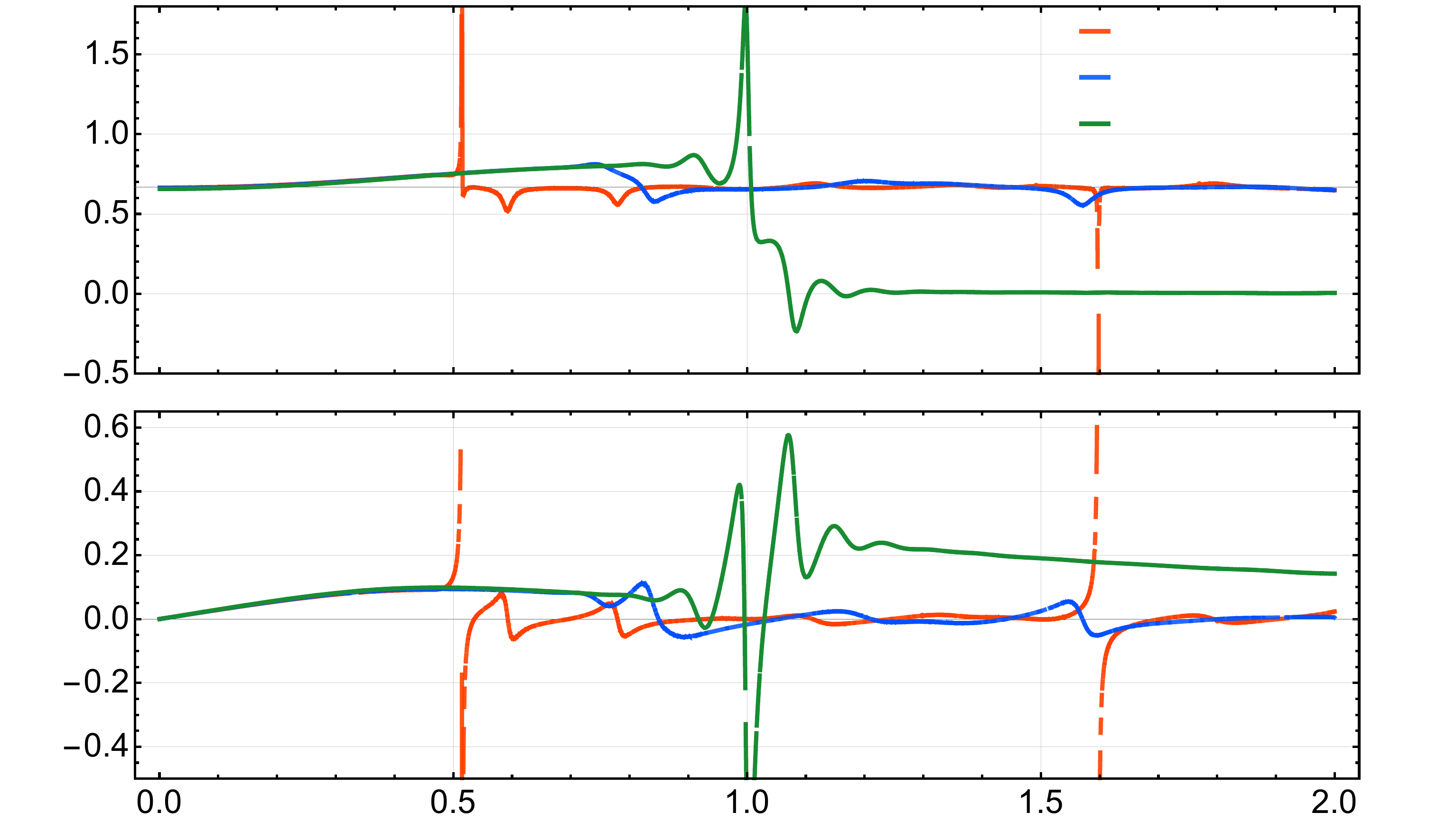} }

\put(150,0){$\gamma$}
\put(430,0){$\gamma$}

\put(6,110){\rotatebox{90}{$ {\rm Re} \langle A_0\rangle/A_0^c $}}
\put(6,37){\rotatebox{90}{$ {\rm Im} \langle A_0\rangle/A_0^c $}}
\put(260,125){\rotatebox{90}{$ {\rm Re} \langle \epsilon\rangle $}}
\put(260,55){\rotatebox{90}{$ {\rm Im} \langle \epsilon\rangle $}}

\put(204, 150){ \scriptsize   $\lambda = 20 $ }
\put(204, 142){ \scriptsize   $\lambda = 10 $ }
\put(204, 133){ \scriptsize   $\lambda = 5 $ }

\put(459, 150){ \scriptsize   $\lambda = 20 $ }
\put(459, 142){ \scriptsize   $\lambda = 10 $ }
\put(459, 133){ \scriptsize   $\lambda = 5 $ }

\end{picture}
\caption{ The left panel \JPp{shows} plots for the expectation value for the bulk area $A_0$ (normalized by $A_0^c$) and the right panel shows plots for the expectation value of the  bulk deficit angle $\epsilon$ for boundary data that induce a classical curvature $\epsilon^c\approx 0.67$.  \label{Fig2}}
\end{figure}
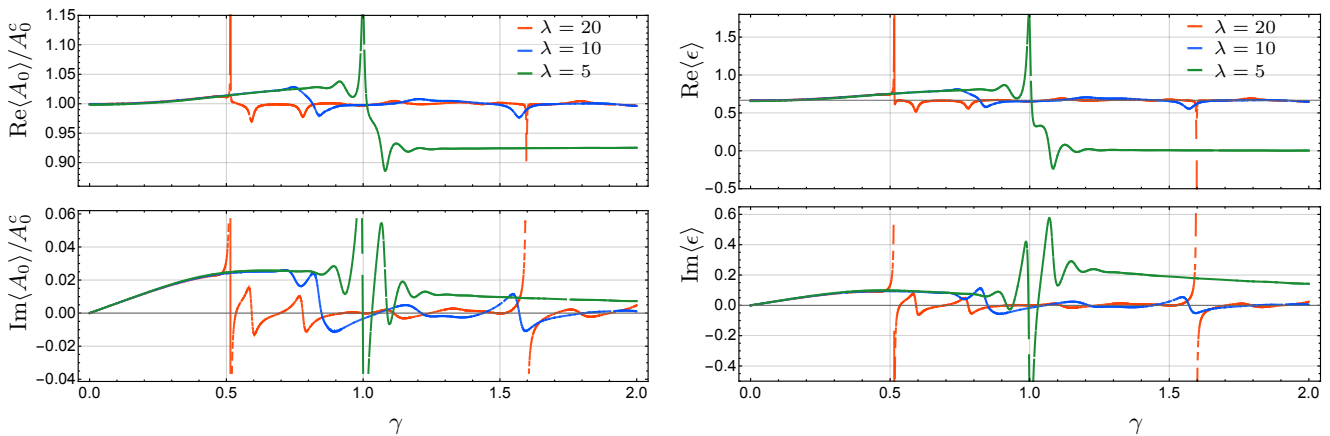

A different kind of destructive interference effect, where the discreteness of the sum plays less of a role, can occur if $\gamma$ is relatively small but sufficiently large to allow for a few oscillations over the widths of the Gaussian factors. This can lead to incidences where the amplitude's oscillations almost perfectly average out the $G$--functions, and we obtain a sharp minimum in the absolute value of the partition function, see Figure \ref{Fig:Instab}. \JPp{And this in turn can conduct} to sudden jumps in the expectation values, see Figure \ref{Fig2}. 

~\\

The expectation values for the bulk area and the deficit angle are shown in Figure \ref{Fig2} \JPp{for} three different scales $\lambda=5,10,20$. The real parts of the expectation values show a very stable behaviour up to a $\lambda$--dependent threshold value $\gamma_T(\lambda) \approx 1, 0.7,0.5$ for $\lambda=5,10,20$ respectively. These values confirm the behaviour predicted by the bound (\ref{CondSL}), which suggest that $\gamma_T(\lambda) \times \sqrt{\lambda}=\text{const}$. 

Note that the expectation values also have a non-vanishing imaginary part. For small $\gamma$ these imaginary parts are relatively small, and their size can be taken as an indicator for how semi-classical the regime is. The appearance of these imaginary parts can be understood if we summarize the oscillating factor and the $G$-functions into an amplitude of the form $\exp(\frac{\imath}{\hbar} S_{\rm eff})$, where $S_{\rm eff}$ has now a real part (given by the Area Regge action) and an imaginary part (resulting from the $G$-functions). Thus, the actual saddle points for this complex actions occur for complexified values of the bulk area \cite{EffSF2}.  As a condition for a semi-classical regime we require small imaginary parts for the expectation values of real observables.

~\\
\emph{Remark:}  The peak structure in the absolute value of the partition function has been first noted in \cite{EffSF2}. This work investigated the effective spin foam model for Euclidean signature, for a configuration with a bulk edge.  \JPp{The} configuration had five bulk triangles. A symmetry reduction was employed, which allowed for three different deficit angles to occur at the five bulk triangles. This example shows also peaks but the $\gamma$-values for these peaks  are {\it not} explained by the condition $\gamma \times \epsilon_i^c=2\pi \times N$, which would have to hold for three different values $\epsilon_i^c, i=1,2,3$. But even in the case where $\epsilon^c_1=0$ and $\epsilon^c_2=\epsilon^c_3$ the condition does not explain the $\gamma$-values for the peaks. A possible reason might be that for this configuration one sums over more variables and the interference effects between the frequencies in the various directions and the discreteness of the summation variables are  more involved. But also in this more complicated triangulation the $\gamma$--values for these peaks are well outside a reliable semi-classical regime.

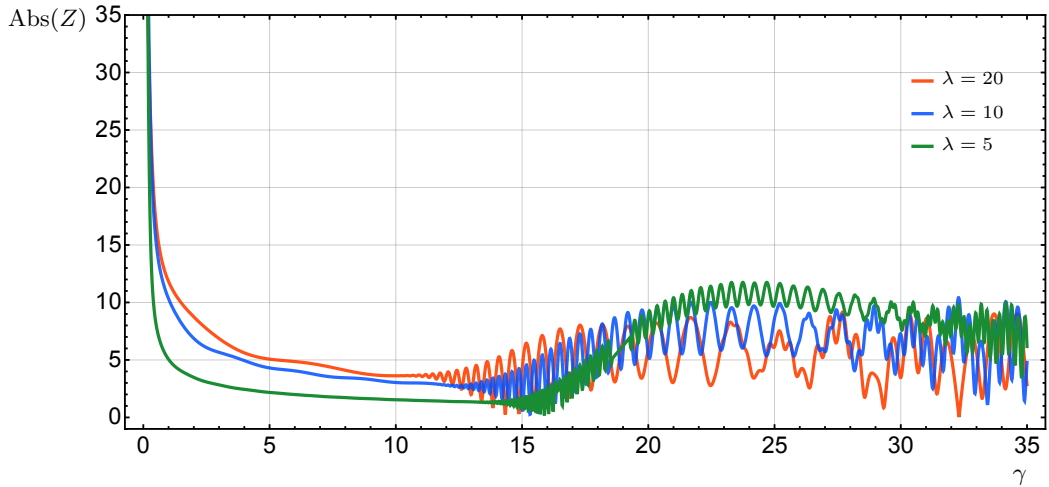
\begin{figure}[ht!]
\begin{picture}(500,190)
\put(70,7){ \includegraphics[scale=0.44]{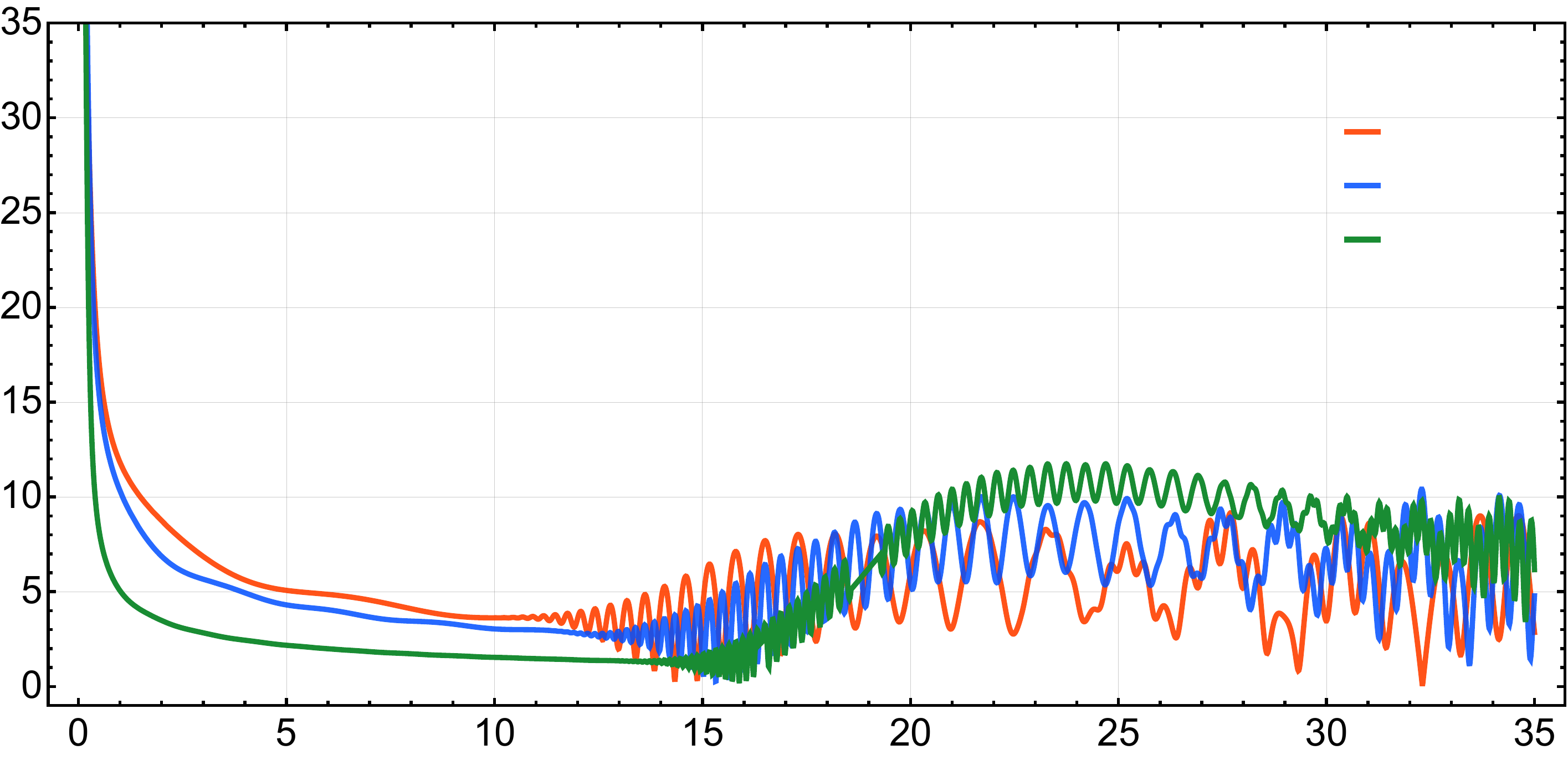} }

\put(420,0){$\gamma$}
\put(40,172){${\rm Abs}({Z})$}

\put(390, 150){ \scriptsize   $\lambda = 20 $ }
\put(390, 137){ \scriptsize   $\lambda = 10 $ }
\put(390, 125){ \scriptsize   $\lambda = 5 $ }

\end{picture}
\caption{ 
Absolute value of the partition function for a configurations peaked on a curvature value $\epsilon^c\approx 0.24$.
\label{Fig3}
} 
\end{figure}

~\\
\emph{Smaller curvature regime:} Next we consider boundary values which lead to a `classical' value  $\epsilon^c\simeq 0.24$ for the deficit angle (as compared to $\epsilon^c\simeq 0.67$ in the previous example). Figure \ref{Fig3} shows  the absolute value of the partition function.  Here we do not see such a pronounced peak structure as for larger curvature. The condition $\gamma \times \epsilon^c=2\pi \times N$ would suggest that the first peak occurs at  $\gamma\simeq 26.2$, and there appears to be a weak peak at this $\gamma$--value, but it is super-imposed with relatively strong oscillations,  which start to occur for $\gamma$-values between 10 and 15.  There will be similar oscillations in the expectation values, and therefore a regime which is not semi-classical. 

\begin{figure}[ht!]
\begin{picture}(500,150)
\put(1,7){ \includegraphics[scale=0.275]{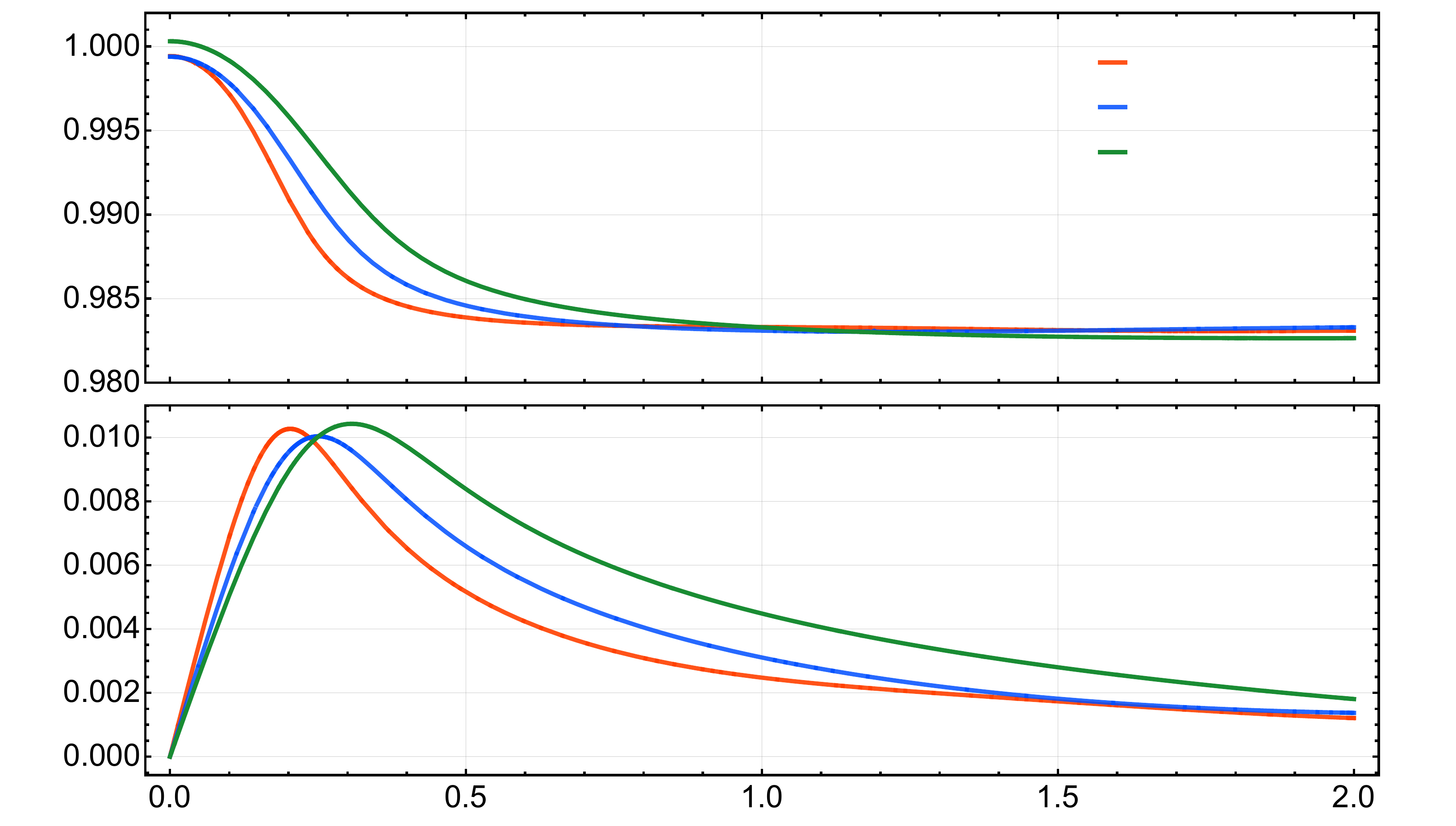} }
\put(255,7){ \includegraphics[scale=0.27]{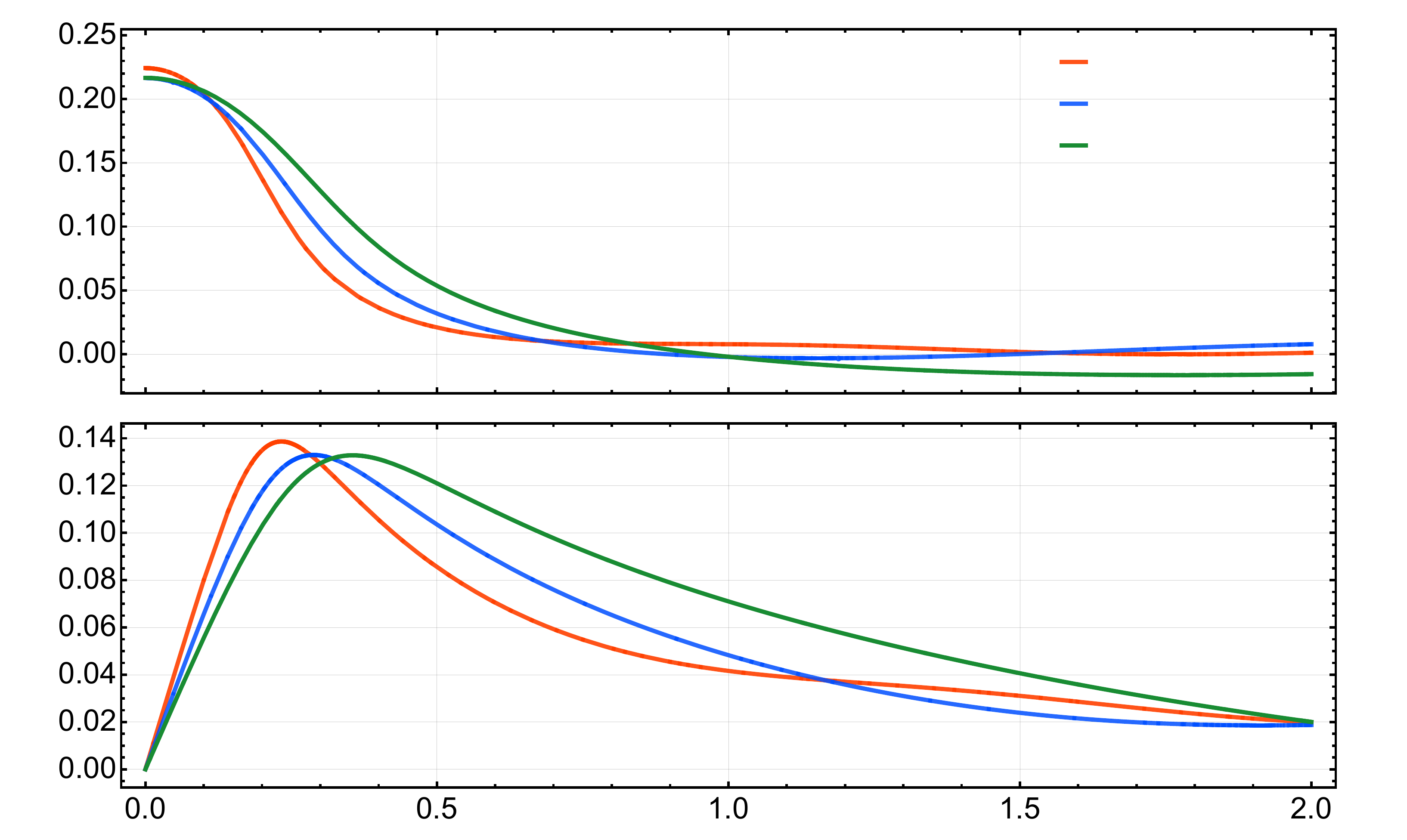} }

\put(150,0){$\gamma$}
\put(430,0){$\gamma$}

\put(3,108){\rotatebox{90}{$ {\rm Re} \langle A_0\rangle/A^c_0 $}}
\put(3,37){\rotatebox{90}{$ {\rm Im} \langle A_0\rangle /A^c_0$}}
\put(258,125){\rotatebox{90}{$ {\rm Re} \langle \epsilon\rangle $}}
\put(258,55){\rotatebox{90}{$ {\rm Im} \langle \epsilon\rangle $}}

\put(204, 143){ \scriptsize   $\lambda = 20 $ }
\put(204, 134){ \scriptsize   $\lambda = 10 $ }
\put(204, 126){ \scriptsize   $\lambda = 5 $ }

\put(456, 147){ \scriptsize   $\lambda = 20 $ }
\put(456, 138){ \scriptsize   $\lambda = 10 $ }
\put(456, 130){ \scriptsize   $\lambda = 5 $ }

\end{picture}
\caption{ The left panel \JPp{shows} plots for the expectation value for the bulk area $A_0$ (normalized by $A_0^c$) and the right panel shows plots for the expectation value of the  bulk deficit angle $\epsilon$ for boundary data that induce a classical curvature $\epsilon^c\approx 0.24$.
\label{Fig4}
} 
\end{figure}
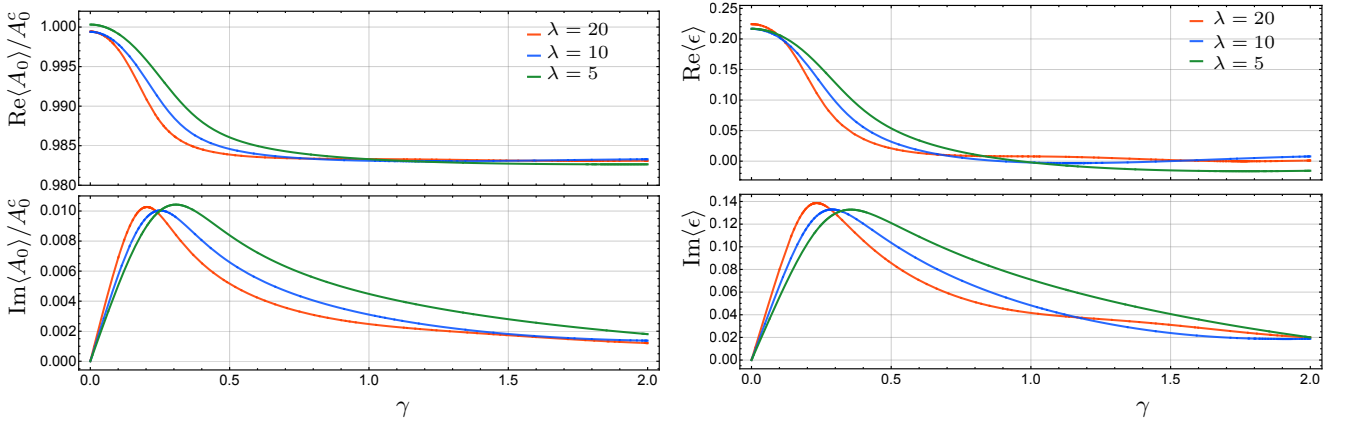

Figure \ref{Fig4} shows the expectation values for the bulk area and the deficit angle. These show a smoother behaviour over the shown range of $\gamma$, as compared for the larger curvature regime.  But for small $\gamma$ the values are not approximately constant (as in the larger curvature regime) in $\gamma$, they rather smoothly approach the flat value for the area  $A_0^f=0.984 \times A_0^c$ and deficit angle $\epsilon=0$ respectively. (Such a smooth behaviour can be reproduced with toy examples of Gaussian form, see \cite{EffSF2}.) 

One reason is that, for this low curvature value, the flat area value is included in the spread of the Gaussian factors, and this value is more and more enhanced with growing $\gamma$. The reason is that with increasing $\gamma$ one also increases the frequency of the oscillations in the amplitude away from the stationary point, given by the flat area value. The behaviour of the expectation values is quite similar for the different scales -- for larger scales (that is larger frequencies) we do have a faster approach to the flat value. 

Since the behaviour of the expectation values is rather smooth, it is difficult to specify a precise semi-classical regime for  $\gamma$. If we accept that the expectation value of the curvature can be up to 50 percent off, the maximal $\gamma$ values range from $\gamma \approx 0.2$ to $\gamma \approx 0.4$ for $\lambda$ decreasing from $\lambda=20$ to $\lambda=5$.

~\\

~\\
\emph{Remark:}  Here we can also compare with the results for the Euclidean effective spin foam model for a configuration with an inner edge \cite{EffSF2}.  There the expectation values were rather constant over quite a large regime of $\gamma$. At certain threshold values for $\gamma$\JPp, oscillations in the absolute value of the partition function and the expectation values do appear, but these \JPp{have} smaller magnitude for larger scales. This suggest that for more complicated triangulations we can expect an improvement in the semi-classical behaviour.

~\\
 \emph{Remark:} 
In this section we computed expectation values as functions of the Barbero--Immirzi parameter $\gamma$. Having only space-like areas in our triangulation with spectral values $A_t=\gamma j_t \ell_p^2$, it was convenient to keep {\it not} the boundary areas fixed, but instead, the boundary spin labels.  That is, for the plots of expectation values versus $\gamma$, different $\gamma$ values lead to different boundary areas.  This way of comparing the expectation values for different $\gamma$'s allowed us to cover a large $\gamma$-range and to highlight some features, e.g. the structure of maxima in $\text{Abs}(Z)$.

We can alternatively fix the boundary areas. This will be more natural if we do have (also) time-like areas, as their spectral values do not scale in $\gamma$. We will therefore use this procedure in the next subsection.

There is a caveat however: keeping a  spatial boundary area $A_b=\gamma  \frac{J_b}{\gamma} \ell_P^2$ fixed \JPp{means} to fix $J_B$. But according to the spectral condition\JPp, $N_b=J_b/\gamma$ has to be an integer\JPp. Thus only discrete values for $\gamma$ are allowed, these become however dense in the limit of vanishing $\gamma$. If we \JPp{have} several spatial boundary areas, we have the condition that $J_b/\gamma, J_{b'}/\gamma, \ldots$ are integers, which could restrict $\gamma$ very much. This is however only a problem of comparing results for different $\gamma$'s, and insisting on holding the boundary areas fixed: allowing a slight variation of the boundary areas up to a maximum of one half Planck area resolves this issue.

\begin{figure}[ht!]
\begin{picture}(500,150)
\put(110,7){ \includegraphics[scale=0.45]{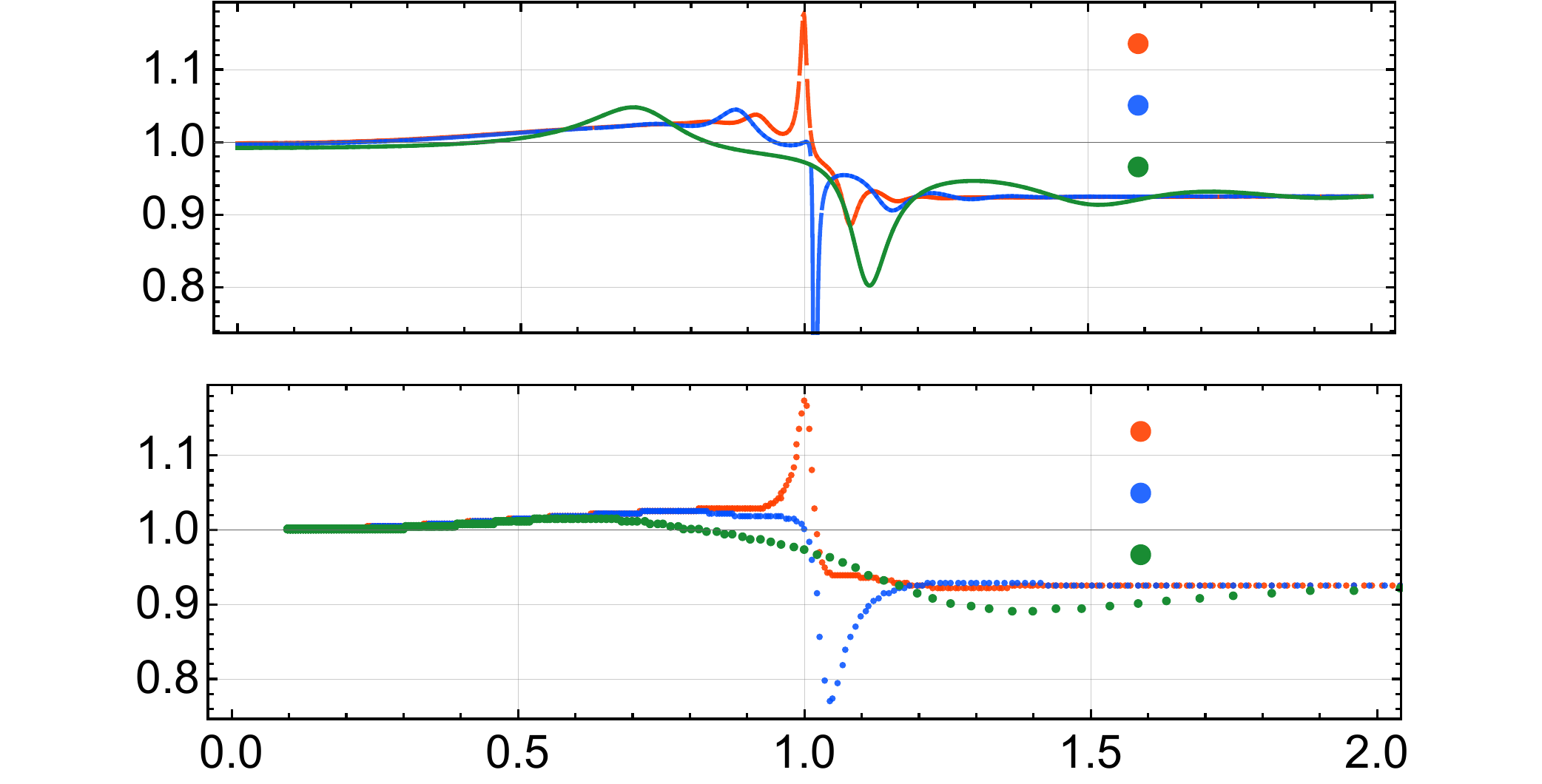} }

\put(340,0){$\gamma$}

\put(125,91){\rotatebox{90}{$ {\rm Re} \langle A_0\rangle/A^c_0 $}}
\put(125,22){\rotatebox{90}{$ {\rm Re} \langle A_0\rangle /A^c_0$}}

\put(315, 135){ \scriptsize   $\Lambda = 5\times\gamma $ }
\put(315, 124){ \scriptsize   $\Lambda = 3\times\gamma $ }
\put(315, 113){ \scriptsize   $\Lambda = 1\times\gamma $ }

\put(315, 67){ \scriptsize   $\Lambda = 5$ }
\put(315, 56){ \scriptsize   $\Lambda = 3$ }
\put(315, 45){ \scriptsize   $\Lambda = 1 $ }

\end{picture}
\caption{  This figure shows the real part of the area expectation values, normalized by the classical value, as two slightly different functions of $\gamma$.  In the upper panel we keep the boundary spin label fixed, that is the boundary areas change linearly in $\gamma$. In the lower  panel we keep the boundary areas fixed. The boundary areas are given in (\ref{Bareas}) with $\lambda=\Lambda/\gamma$. The configuration is peaked on a curvature $\epsilon^c\approx 0.67$.
\label{OldNew}
} 
\end{figure}
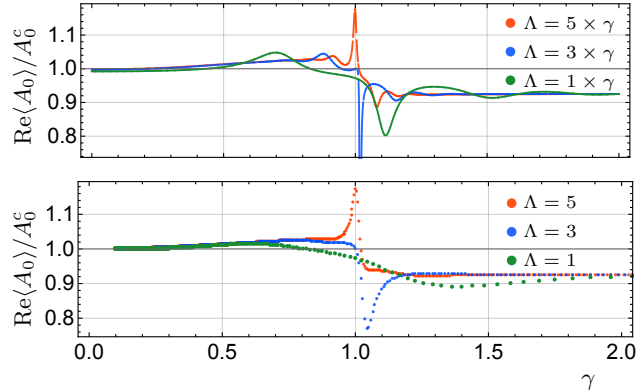

The reader should \JPp{thus be} aware that the expectation values in dependence of $\gamma$ are two different functions in this section and the next \JP{section}. To allow for a comparison between the two different functions \JPp{we} have computed for the larger curvature example of this section both ways of presenting the expectation values. 
As one can see in Figure \ref{OldNew}, for this example  the two functions are  quite similar \JPp{and} show instabilities at very similar $\gamma$ values. But the version \JPp{where} we keep the boundary areas fixed \JPp{tends} to show a more stable behaviour.

~\\

\subsection{A triangulation with space-like and time-like triangles}

In this section we will consider a triangulation with three 4-simplices and a bulk triangle. We will first choose boundary data such that the path integral is over data inducing a time-like geometry for the bulk triangle. We will then also explore boundary data \JPp{such} that the $G$--functions are peaked on a vanishing value for the signed squared area of the inner triangle. 

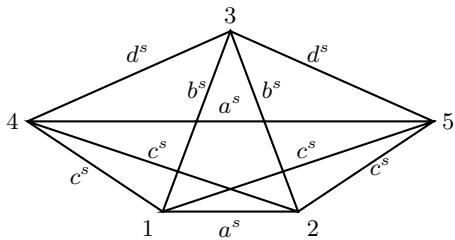
\begin{figure}[ht!]
\begin{tikzpicture}[scale=0.6]
\draw[thick] (0,0) --(4.5,2) -- (9,0)--(3,-2)--(6,-2)--(4.5,2)--(3,-2)--(0,0)--(9,0)--(6,-2)--(0,0);

\node[below left] at (3,-2) {1}; 
\node[below right] at (6,-2) {2}; 
\node[above] at (4.5,2) {3}; 
\node[left] at (0,0) {4}; 
\node[right] at (9,0) {5}; 

\node[below] at (4.5,-2) {$a^s$}; 
\node[above] at (4.5,0) {$a^s$}; 
\node[left] at (2.9,1.5) {$d^s$}; 
\node[right] at (6.0,1.5) {$d^s$}; 
\node[right] at (5.,0.7) {$b^s$}; 
\node[left] at (4.2,0.7) {$b^s$}; 
\node[left] at (1.6,-1.2) {$c^s$}; 
\node[right] at (7.4,-1) {$c^s$}; 
\node[above] at (2.9,-1.) {$c^s$}; 
\node[above] at (6.2,-1) {$c^s$}; 

\end{tikzpicture}
\caption{This shows the squared lengths  parameters for the simplex $(1,2,3,4,5)$. Three such simplices, with the same geometry, are glued around the triangle $(1,2,3)$.  For the first set of examples we consider in this section,  the classical values for $b^s$ and $d^s$ are time-like and $a^s$ and $c^s$ are space-like. We will then consider a configuration, where $d^s$ is almost null. }
\label{41Complex}
\end{figure}

The three 4-simplices of the triangulation share the bulk triangle $(123)$ and are given by 
\ba
\{\sigma_1,\sigma_2,\sigma_3\}=\{(12345),(12346),(12356)\} \q .
\ea
{Our symmetry reduction of the boundary data \eqref{EdgeL1}} will force all three simplices to have the same geometry, and reduce the length and area parameters of a given 4-simplex from  ten to four. For the simplex $(12345)$, which we depict in Figure \ref{41Complex}, we have for instance:
\ba
a^s&=&l_{12}^s=l_{45}^s \, , \q\nn\\
b^s&=&l_{13}^s=l_{23}^s \, ,\nn\\
c^s&=& l_{14}^s=l_{15}^s=l_{24}^s=l_{25}^s \, ,\nn\\
d^s&=& l_{34}^s=l_{35}^s \q .
\ea
Here we work with the squared edge lengths and as a reminder denote the corresponding parameters with a super index $s$. The squared edge lengths can be positive or negative for space-like and time-like edges respectively. We will peak on a geometry where $a^s,c^s>0$ and $b^s,d^s<0$.  We can thus imagine that the vertices $\{1,2,4,5,6\}$ are all at a given time $T$ and the vertex $3$ is at a later time $T'$. 

The squared area parameters are given by 
\ba
A^s=A^s(a^s,b^s,b^s)\, ,\q B^s=A^s(a^s,c^s,c^s)\, , \q C_1^s=A^s(b^s,c^s,c^s)\,, \q C_2^s=A^s(a^s,d^s,d^s) \, .
\ea
In our conventions the squared area is given by  Heron's formula
\ba
A^s(x^s,y^s,z^s) = \frac{1}{16}(2 x^sy^s + 2 x^s z^s + 2 y^s z^s - x^s x^s-y^sy^s -z^s z^s)
\ea
for both time-like triangles and space-like triangles. Note that the (signed)  squared area is negative for time-like triangles and positive for space-like triangles.\footnote{A triangle which has only space-like edges can be either space-like (if the area square is positive) or time-like (if the area square is negative), so allowing for both space-like and time-like signature there is no triangle inequality. A triangle which has at least one time-like edge is time-like. In this case one does have a priori a triangle inequality from demanding that the area square is negative. But it turns out that if one has at least one time-like edge and at least one space-like edge, the area square is automatically negative. The only case for a triangle in 3D or 4D Minkowski space with a triangle inequality is then the case with three time-like edges.}

We will choose boundary data for which $C_1^s=C_2^s=:C^s$. {To compute the Area Regge action as well as the Gaussian factors} we will  work with the following root for the area--length system:
\ba\label{L2}
&&b^s=\frac{1}{4a^s}\left( 16 A^s+ (a^s)^2  \right) \, ,\q  c^s=\frac{1}{4a^s}\left( 16 B^s+ (a^s)^2  \right) \, , \q
 d^s=\frac{1}{4a^s}\left( 16 C^s+ (a^s)^2  \right) 
\ea
and
\ba \label{Solas}
a^s&=&\frac{4}{\sqrt{3}} \sqrt{-A^s-B^s+ 7C^s +2 \sqrt{(A^s)^2 + (B^s)^2 + 13 (C^s)^2   - A^s B^s - 5 A^s C^s - 5 B^s C^s} }
   \q .
\ea
Both the inner and outer square root in (\ref{Solas}) are positive for $B^s>0$ and $C^s<0$. {For our choice of boundary data below, the other roots are not contributing to the path integral. This is either due to the generalized triangle inequalities, or  due to a suppression by the Gaussian factors.}

{
With the symmetry assumptions discussed above, the boundary squared areas are given by $B^s$ and $C^s$. The (symmetry reduced) angular parameters $P^\tau_i$ provide one additional boundary datum, which allows to determine all length (squares) in the boundary. These boundary lengths do determine the ``classical" value for the bulk area $A^s_c$. The effective spin foam path integral is a sum over the bulk area $A^s$, but it includes Gaussian factors that are peaked on $A^s=A^s_c$.}

Using the choice (\ref{L2},\ref{Solas}) of roots we can then study what the generalized triangle inequalities imply for the range of $A^s$, keeping the boundary areas $B^s>0$ and $C^s<0$ fixed.  One finds that, depending on the values of $B^s$ and $C^s$, there is a lower bound with negative value for $A^s$, but no upper bound. In particular $A^s$ can also take positive values. For negative $A^s$, that is a time-like inner triangle, the deficit angle is Euclidean and ranges from $2\pi$ for $A^s=0$ to $-\pi$ for $A^s$ taken the maximal negative allowed value. For positive $A^s$ we have a space-like inner triangle and the deficit angle is Lorentzian. As we discussed in section \ref{Sec:ImPart}, such Lorentzian angles can have imaginary parts. If this is the case the number of light cones at the inner triangle differs from the flat case. Indeed for the case at hand, there are zero light cones, and the deficit angle has an imaginary part $\text{Im}(\epsilon)=-2\pi$.  We can choose to include these  causally irregular configurations into the path integral, or not allow them. In the first case, the imaginary part of the deficit angle will lead to an imaginary part in the action $\text{Im}(S_{AR})=-2\pi \sqrt{A^s}$. This leads to an enhancing factor $\exp( 2\pi \sqrt{A^s})$ for the amplitudes.  The $G$-factors will still suppress configurations which are sufficiently far away from the \JPp`classical' value $A^s_c$, which is induced by the boundary data. But the enhancing factor $\exp( 2\pi \sqrt{A^s})$ will affect considerably the expectation values if $A^s_c$ is negative and small, or even positive. 

\subsubsection{ Configurations peaked on time-like inner triangle  }

We will first discuss examples \JPp{where} $A^s_c<0$ is sufficiently large, so that the $G$-functions completely suppress configurations with \JPp{positive} $A^s$. We therefore do not need to include the sum over such positive values into the path integral. 

 We choose boundary values $B^s=(\gamma   \frac{50}{\gamma} \Lambda \ell_p^2)^2$ and $C^s=-(55\Lambda \ell_p^2)^2$. That is---different from the discussion for triangulation with only space-like triangles in section \ref{Sec:spacelike}---we keep here the values of all the boundary triangles constant with varying $\gamma$. According to the area spectrum (\ref{ASpecSS}) for spatial triangles, we need that $N_B= \Lambda \frac{50}{\gamma}$ is an integer. That is only discrete values $\gamma= 50 \Lambda/N_B $ are allowed, but these values are becoming dense for very small $\gamma$.  
 
 We will choose three different values for $A^s_c$ namely  $A^s_c=-(90)^2\Lambda^2 \ell_P^2$, $A^s_c=-(85)^2 \Lambda^2\ell_P^2$ and $A^s_c=-(70)^2 \Lambda^2\ell_P^2$, corresponding to $\epsilon_c\approx 0.26$, $\epsilon_c\approx 0.62$ and $\epsilon_c\approx 1.71$ respectively.  We use $\Lambda=1,2$ and $4$. The $G$-functions then allow only for a small effective summation range around $A^s_c$, which in particular excludes positive $A^s$ (despite the exponentially enhancing factor). 
 
 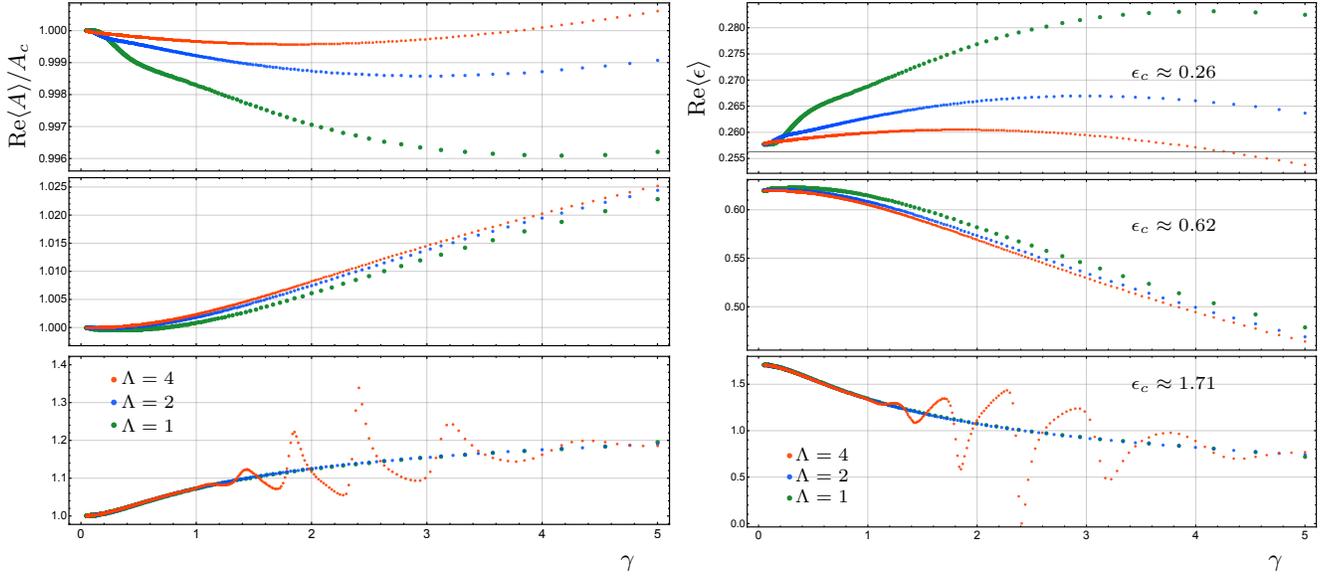
\begin{figure}[ht!]
\begin{picture}(500,215)
\put(1,7){ \includegraphics[scale=0.238]{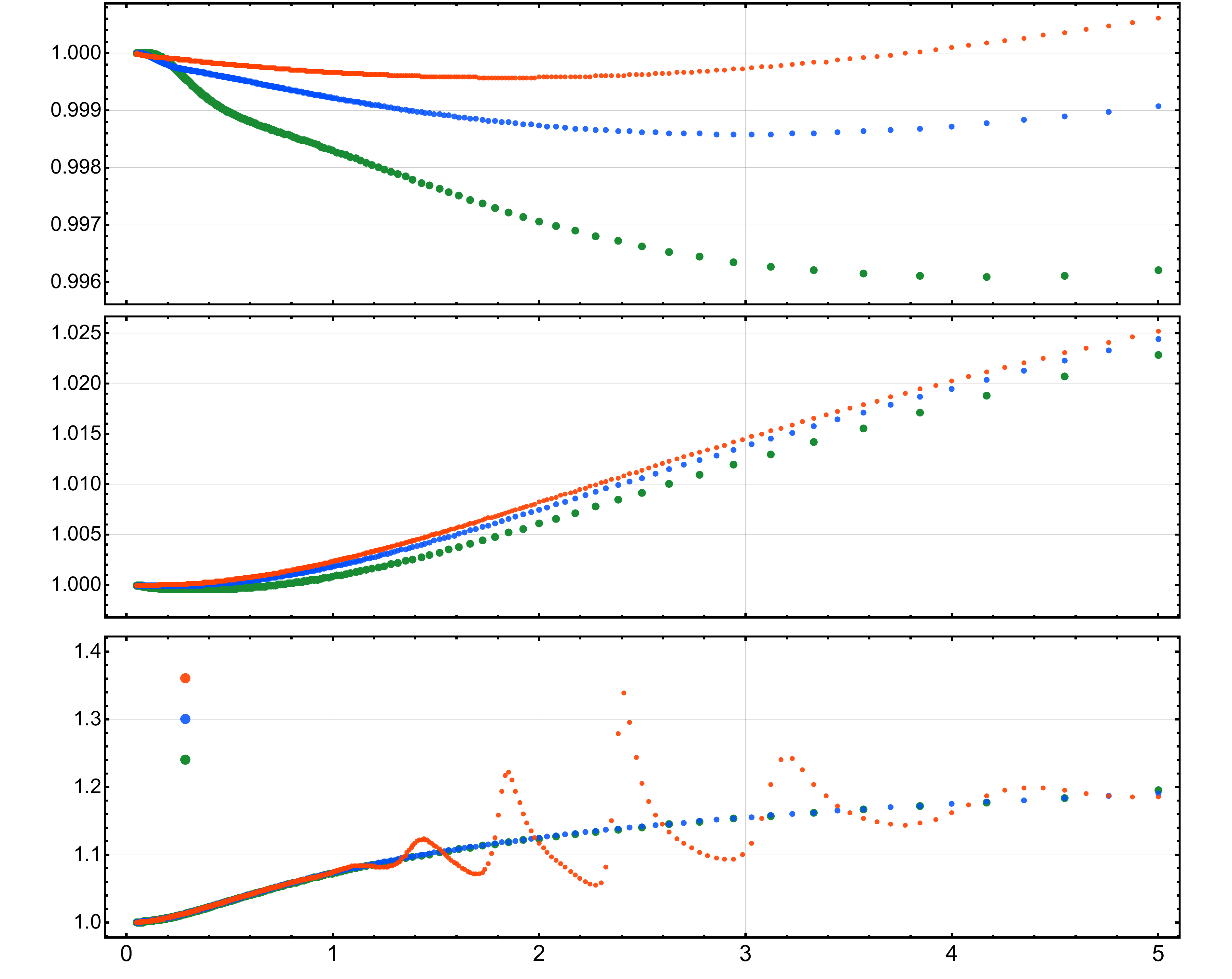} }
\put(259,7){ \includegraphics[scale=0.222]{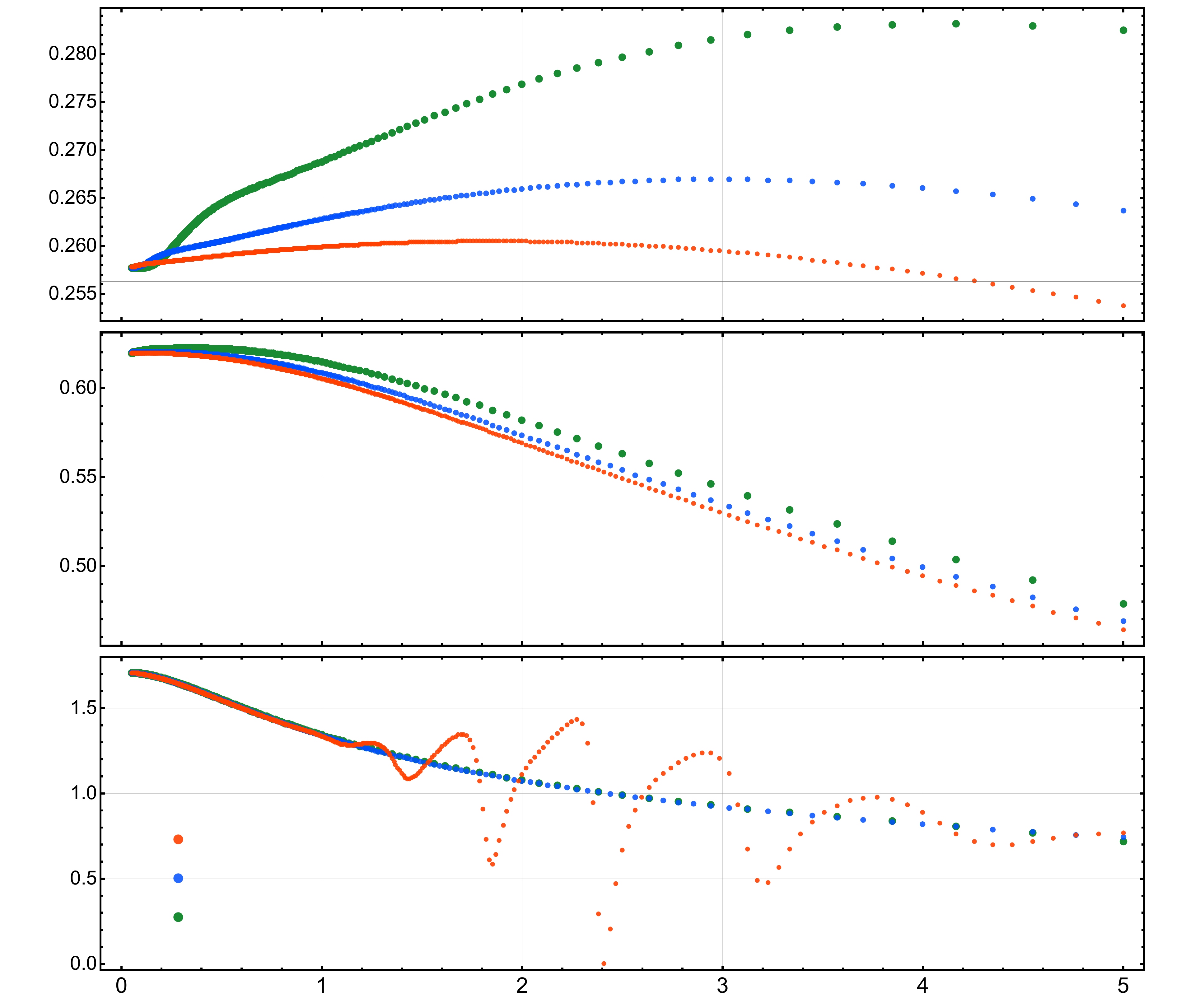} }

\put(235,0){$\gamma$}
\put(480,0){$\gamma$}

\put(3,158){\rotatebox{90}{$ {\rm Re} \langle A\rangle/A_c $}}
\put(260,170){\rotatebox{90}{$ {\rm Re} \langle \epsilon\rangle $}}

\put(43, 69){ \scriptsize   $\Lambda = 4 $ }
\put(43, 60){ \scriptsize   $\Lambda = 2 $ }
\put(43, 51){ \scriptsize   $\Lambda = 1 $ }

\put(298, 40){ \scriptsize   $\Lambda = 4 $ }
\put(298, 32){ \scriptsize   $\Lambda = 2 $ }
\put(298, 24){ \scriptsize   $\Lambda = 1 $ }


\put(425, 185){ \scriptsize   $\epsilon_c \approx 0.26 $ }
\put(425, 127){ \scriptsize   $\epsilon_c \approx 0.62 $ }
\put(425, 67){ \scriptsize   $\epsilon_c \approx 1.71 $ }

\end{picture}
\caption{ The left panel \JPp{shows} plots for the expectation values of the bulk areas $A$ (normalized by $A^c$) and the right panel shows plots for the expectation values of the  bulk deficit angles $\epsilon$. The plots from top to bottom of each panel are for boundary data that induce classical curvatures $\epsilon_c\approx 0.26,\epsilon_c\approx 0.62$ and $\epsilon_c\approx 1.71$ respectively.
\label{Fig7}
} 
\end{figure}

We show in Figure \ref{Fig7}  the expectation values for the inner area and the deficit angle for the various examples. In almost all cases the expectation values show a very stable behaviour as a function\footnote{Note that here we do not change the values for the boundary areas with $\gamma$, whereas the boundary values did change with $\gamma$ in the previous subsection.} of $\gamma$, and we can speak of a semi-classical regime up to $\gamma=1$, which can in some cases even be extended.  There is one example, namely the one with largest curvature $ \epsilon_c\approx1.71$ and largest scale $\Lambda=4$, where we see an onset of oscillations, starting with $\gamma=1$. These are due to a series of minima \JPp{that} occur in the absolute value of the partition function. These indicate values of $\gamma$ where the oscillations in the amplitude over the spread of the Gaussian factors lead to an almost vanishing partition function.  
 
The dependence of the expectation values (which for the areas are divided by the classical value $A_c$) on the scale differs for the various curvatures, but is overall rather weak.  For the smallest curvature case $\epsilon_c\approx 0.26$ the area expectation values first decrease slightly with growing $\gamma$, but then increase again. \JPp{However,} the maximal deviation over the range of $\gamma\in (0,5)$ for $\Lambda=1$ is around $0.4$ percent of the classical value. These deviations are even smaller for the larger scales $\Lambda$.  For the medium curvature case $\epsilon_c\approx 0.62$\JPp, the (normalized) area expectation values are very near to each other,  with the deviations minimally larger for smaller scales and for $\gamma \leq 0.6$ and slightly smaller for $\gamma >1$. Here the deviations reach around $2.5$ percent for $\gamma=5$. For the largest curvature case $\epsilon_c\approx 1.71$ the \JPp{area} and deficit angle expectation values are almost on top of each other, up to a value of $\gamma \approx 1$.  The deviation for e.g. the areas from the classical value is considerably larger than for the smaller curvature cases: it reaches around $10$ percent for $\gamma \approx 1$. For larger values of $\gamma$ the expectation values for $\Lambda=1$ and $\Lambda=2$  are still almost the same for the range of $\gamma <5$.  For $\Lambda=4$ we see however an onset of oscillations, indicating a breakdown of the semi-classical regime.

If we have time-like areas, the spectral values of these areas does not change with $\gamma$. Here we should rather understand the action and the $G$-factors as function of the areas (and not the representation labels). The action \JPp{then does not} depend on $\gamma$, whereas the spread of the $G$-factors now increases with $\gamma$.  Thus the plots of the expectation values in this subsection show just the effect of the spread of the $G$-factors increasing with $\gamma$. This is one of the reasons why the plots depict a more stable behaviour of the expectation values as functions of $\gamma$. 

 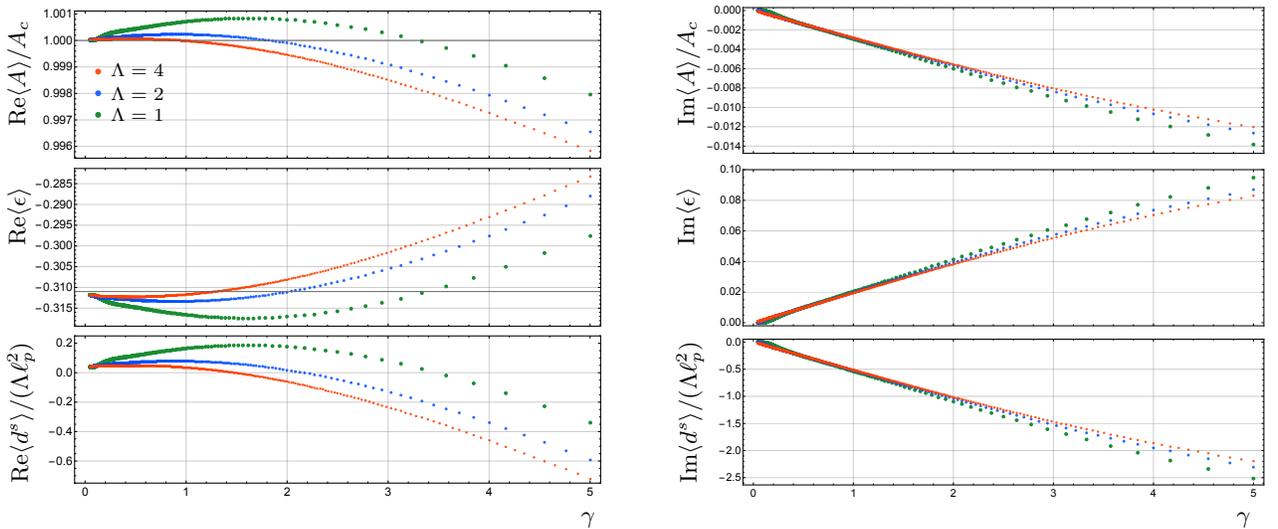
\begin{figure}[ht!]
\begin{picture}(500,215)
\put(1,7){ \includegraphics[scale=0.247]{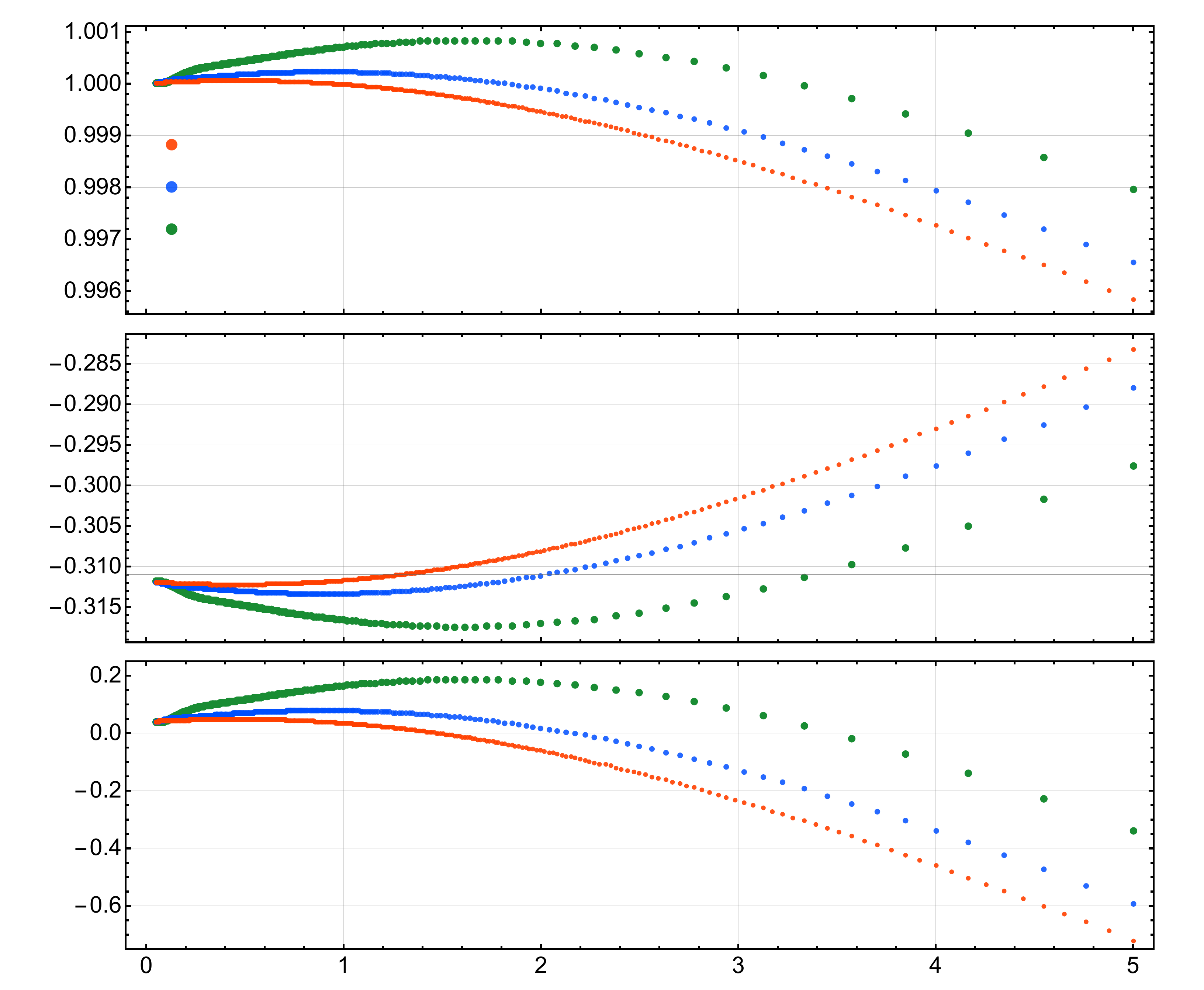} }
\put(255,7){ \includegraphics[scale=0.227]{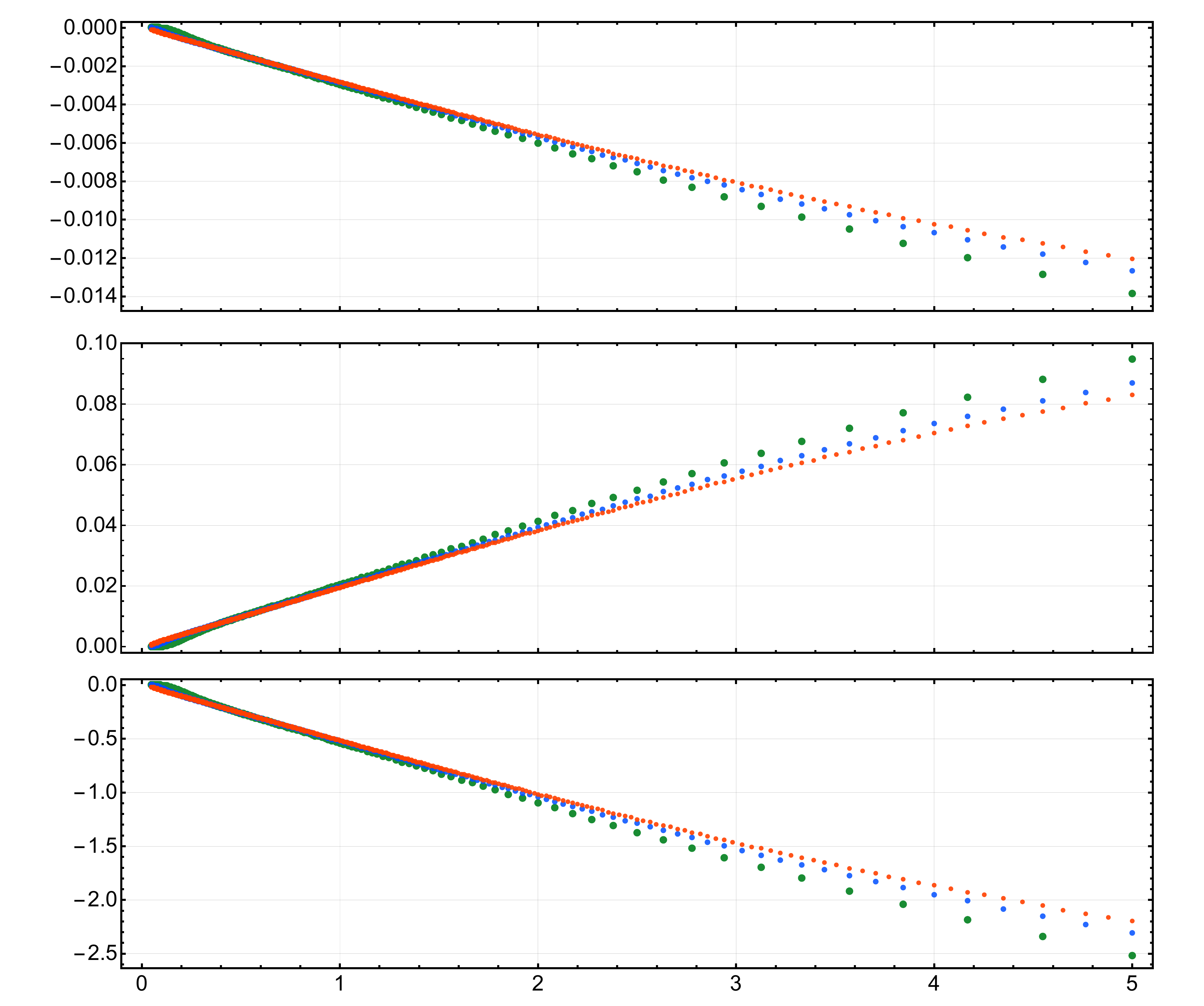} }

\put(220,0){$\gamma$}
\put(468,0){$\gamma$}

\put(3,150){\rotatebox{90}{$ {\rm Re} \langle A\rangle/A_c $}}
\put(3,105){\rotatebox{90}{$ {\rm Re} \langle \epsilon\rangle $}}
\put(3,15){\rotatebox{90}{$ {\rm Re} \langle d^s\rangle/(\Lambda \ell_p^2) $}}

\put(256,150){\rotatebox{90}{$ {\rm Im} \langle A\rangle/A_c $}}
\put(256,105){\rotatebox{90}{$ {\rm Im} \langle \epsilon\rangle $}}
\put(256,15){\rotatebox{90}{$ {\rm Im} \langle d^s\rangle/(\Lambda \ell_p^2)$}}

\put(39, 169){ \scriptsize   $\Lambda = 4 $ }
\put(39, 160){ \scriptsize   $\Lambda = 2 $ }
\put(39, 152){ \scriptsize   $\Lambda = 1 $ }


\end{picture}
\caption{ This figure shows the real and imaginary parts of the expectation values for the bulk area $A$, the deficit angle $\epsilon$ and the signed squared lengths $d^s$ .
\label{Fig8}
} 
\end{figure}

Next we discuss one last example, where we choose $A^s_c=-(98)^2 \Lambda^2 \ell_P^2 $  such that the length square $d^s_c\approx 0.04 \ell_P^2 \Lambda$ is almost null. Figure \ref{Fig8} shows the expectation values for the bulk area (normalized by $A_c$), the deficit angle and the signed squared lengths $d^s$ (normalized by $\Lambda$).  The spread of the $G$-factors includes  positive and negative values of $d^s$, so in the path integral we sum over both signatures for $d^s$.   The deficit angle is $\epsilon_c\approx -0.31$. The expectation values for the  bulk area show again a quite stable behaviour in $\gamma$, with the deviation being smaller than $0.1$ percent for $\gamma \in (0,2)$.  In Figure \ref{Fig8} we show the expectation values for $d^s_c$ divided by $\Lambda\ell^2_P$.  The expectation values reproduce for  small $\gamma$ quite well the classical value. But the relative deviations can become quite large as $d^s_c$ is very small.    For larger $\gamma$ the expectation value goes to negative values. The reason for this behaviour is that the stationary point for the oscillatory factor is given by $\epsilon=0$, which occurs for $d^s\approx -8.2 \Lambda\ell_p^2$. 

This example shows that we can include into the path integral configurations where the signature of  edges changes from time-like to space-like. If one considers the expectation value of the (squared) edge lengths, one might conclude that the configuration is rather not in the semi-classical regime, as the deviations from the classical value, compared to the value of this classical value are quite large. This is a general issue for configurations \JPp{whose} classical solutions include simplices being null or almost null. But we see that in this example the classical area value is very well approximated by the area expectation value, and we should possibly compare the deviations rather with other non-null quantities, e.g. the boundary values. (We did this already in Figure \ref{Fig8} \JPp{by} showing $\langle d^s\rangle/(\Lambda \ell_2^2)$  instead of $\langle d^s\rangle/\ell_2^2$)  Next we consider a configuration peaked on a null area, where the question of semi-classicality becomes even more involved.

 \subsubsection{ Configurations peaked on a null  triangle and with irregular light cone structure  }

Let us  consider an example where the $G$-functions are peaked on $A^s_c =0$. We choose boundary values  $B^s=(\gamma   \frac{10}{\gamma} \Lambda \ell_p^2)^2$ and $C^s=-(11\Lambda \ell_p^2)^2$ keeping the boundary triangle areas constant with varying $\gamma$. 
 As one can see from (\ref{L2}), if $A^s=0$ we have $a^s=4b^s$. The triangle can  be null\footnote{Consider for instance the triangle with vertices $v_1=(0,0,b,0)$, $v_2=(0,0,-b,0)$  and $v_3=(y,y,0,0)$.  It has, independent from the parameter $y$,  squared edge lengths $(b^2,b^2,4b^2)$. For $y=0$ we obtain a degenerate triangle.} or degenerate. For our boundary data  $a^s,b^s$ and $c^s$ are positive,   whereas $d^s$ is negative.  
 
As the $G$-functions are peaked on a null area, both time-like and space-like configurations are included in the spread of the $G$-functions. As we have discussed, the configurations with space-like $A^s>0$ have a irregular light cone structure --- the deficit angle has an imaginary part $\text{Im}(\epsilon)=-2\pi$ leading to an imaginary part $\text{Im}(S_{AR})=-2\pi \sqrt{A^s}$ for the Regge action.  For space-like areas  the absolute value of the amplitudes is therefore given by $|G(A,\gamma)\times \exp(2\pi A)|$ where $G(A,\gamma)$ summarizes all the $G$-factors.  

We have the choice to either forbid such causally irregular configurations, or to include them in the path integral. Let us first discuss the latter option. The results in this case can be understood from the effects of the enhancing factor $ \exp(2\pi A)$. 

\begin{figure}[ht!]
\begin{picture}(500,85)
\put(30,7){ \includegraphics[scale=0.48]{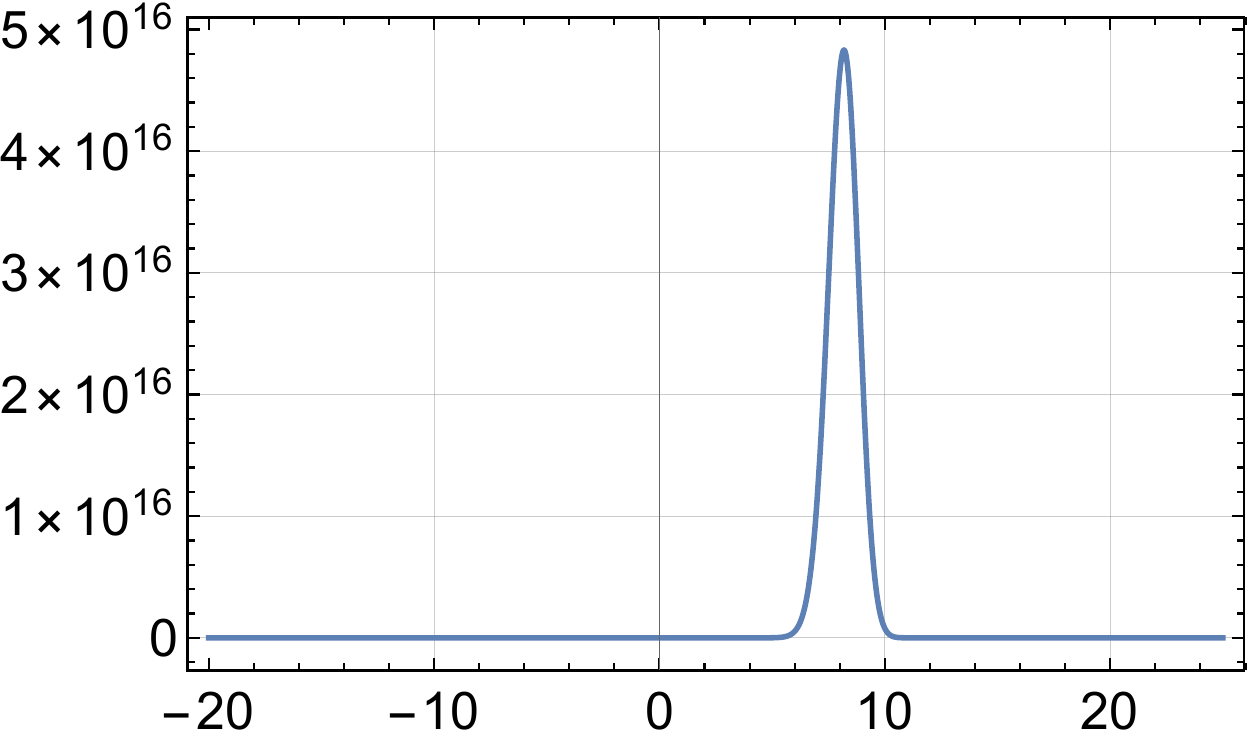} }
\put(275,7){ \includegraphics[scale=0.5]{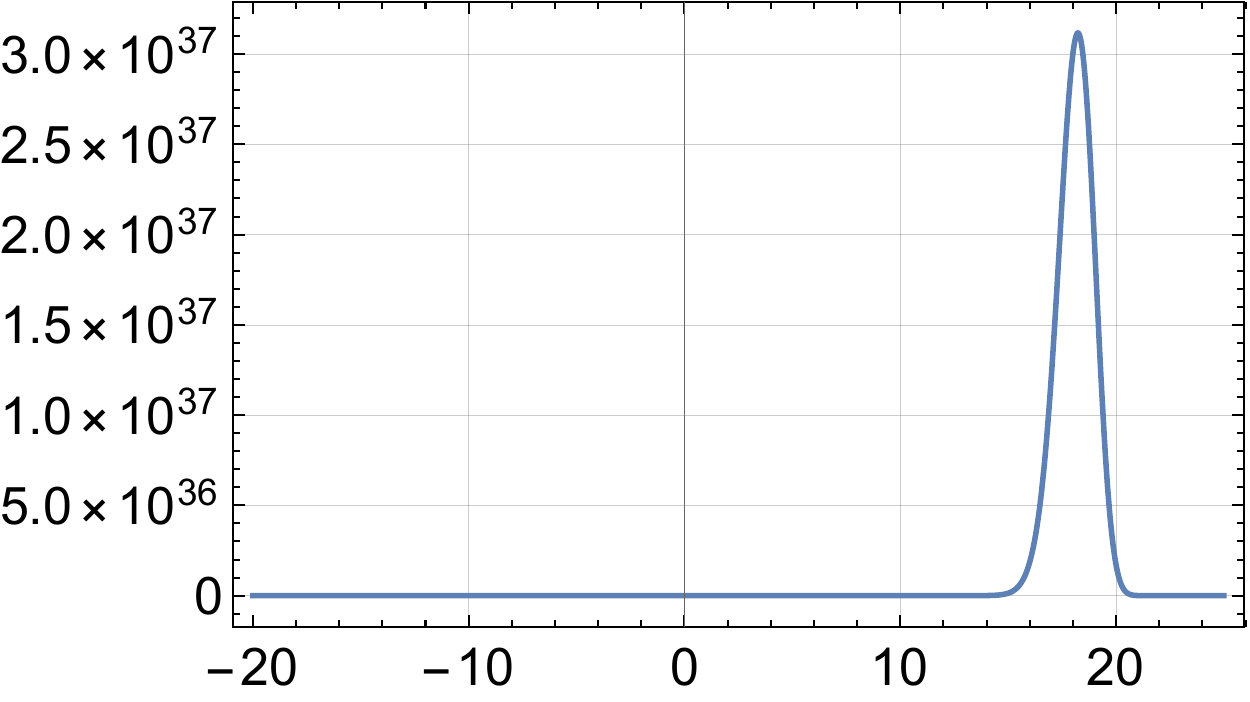} }

\put(180,0){$A/ \ell_P^2$}
\put(430,0){$A/\ell_P^2$}

\put(21,27){\rotatebox{90}{$ |G \exp((\imath/\hbar )S_{AR} )|$}}
\put(260,27){\rotatebox{90}{$  |G \exp((\imath/\hbar )S_{AR} )|$}}

\end{picture}
\caption{ Plots for the absolute value of the amplitude $|G \exp((\imath/\hbar )S_{AR} )|$ for a scale $\Lambda=1$.  In the left and right panels, we set $\gamma$ to be $\gamma=0.1$ and $\gamma=1.0$ respectively. This illustrates the shift of the peaks due to the enhancement from the $\exp((\imath/\hbar )S_{AR} )$ factor. \label{Fig:GAbsS}}
\end{figure}
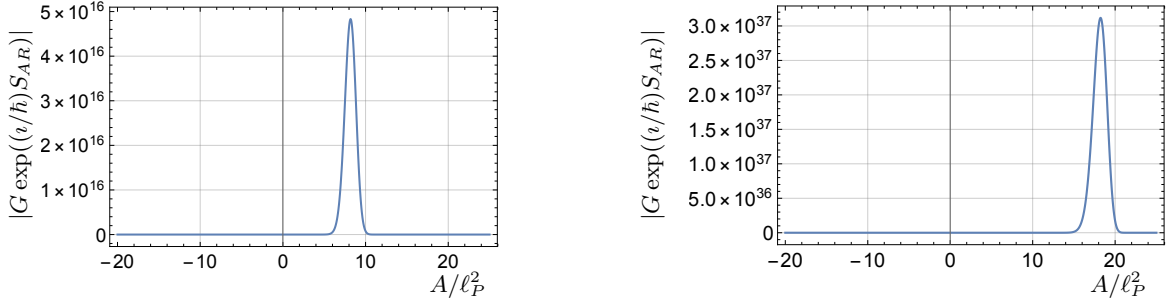

The generalized triangle inequalities allow for an arbitrary large space-like area, and so there is even the danger of encountering a divergence. In the examples at hand the $G$-factors do eventually
over-take the $ \exp(2\pi A)$ factor. But this leads $(a)$ to a shift of the peak of $G(A,\gamma)\times \exp(2\pi A)$ to a $\gamma$-dependent value $A_p(\gamma)$, and $(b)$ to a huge $\gamma$-dependent maximal value  $G(A_p,\gamma)\times \exp(2\pi A_p)$ for the absolute value of the \JPp{amplitude}, which  completely dwarfs the amplitude for the configurations with negative area, see Fig. \ref{Fig:GAbsS}. Thus the sum over time-like areas \JPp{basically does not} contribute.

~\\

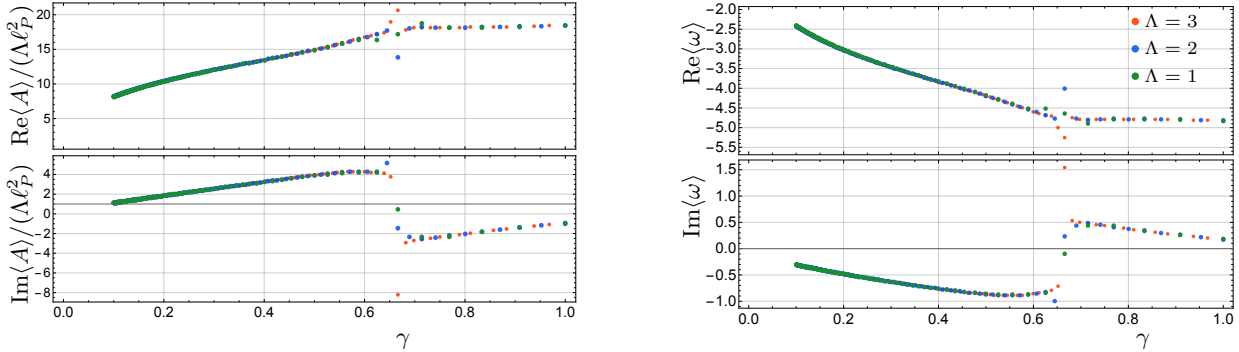
\begin{figure}[ht!]
\begin{picture}(500,130)
\put(1,7){ \includegraphics[scale=0.273]{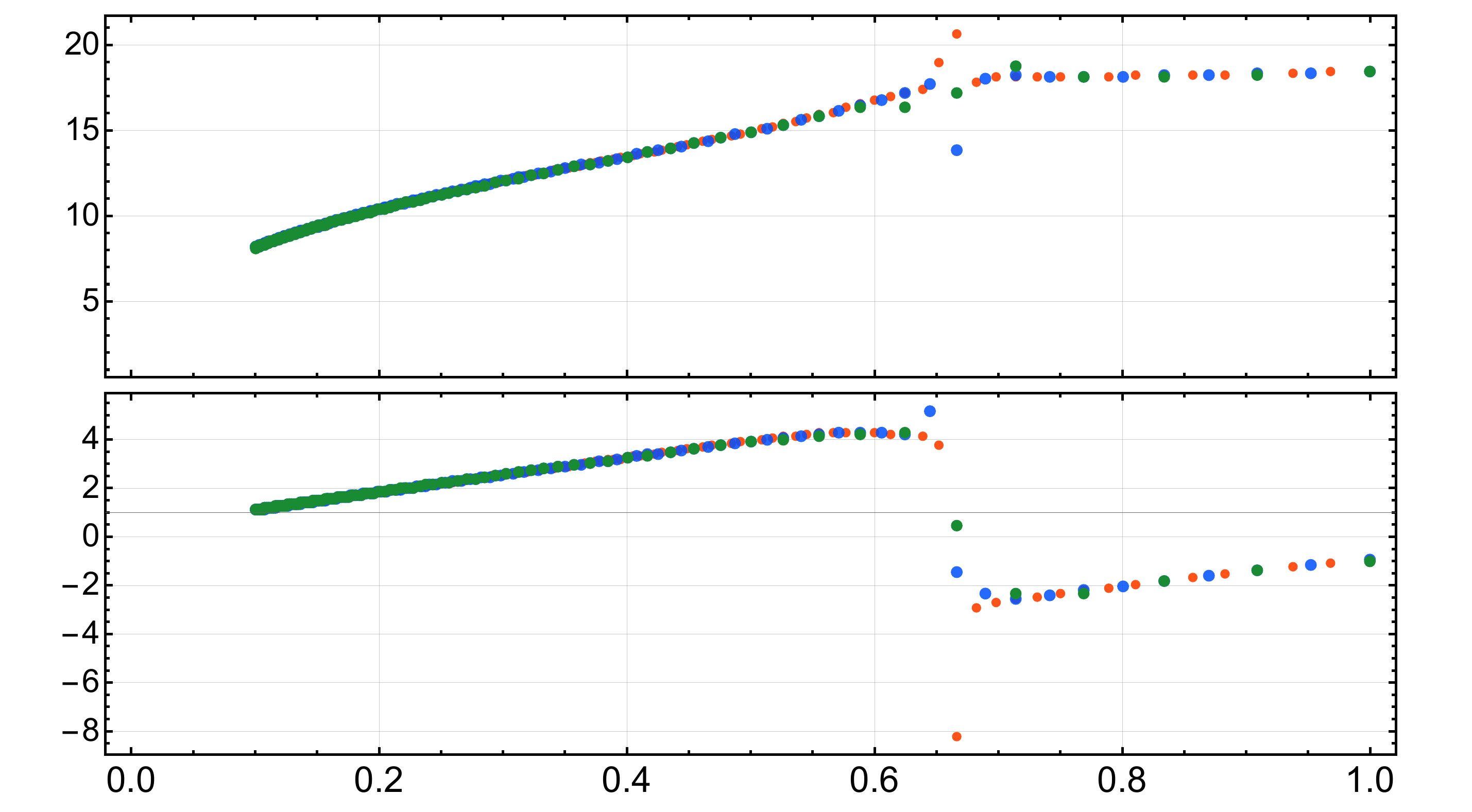} }
\put(255,7){ \includegraphics[scale=0.313]{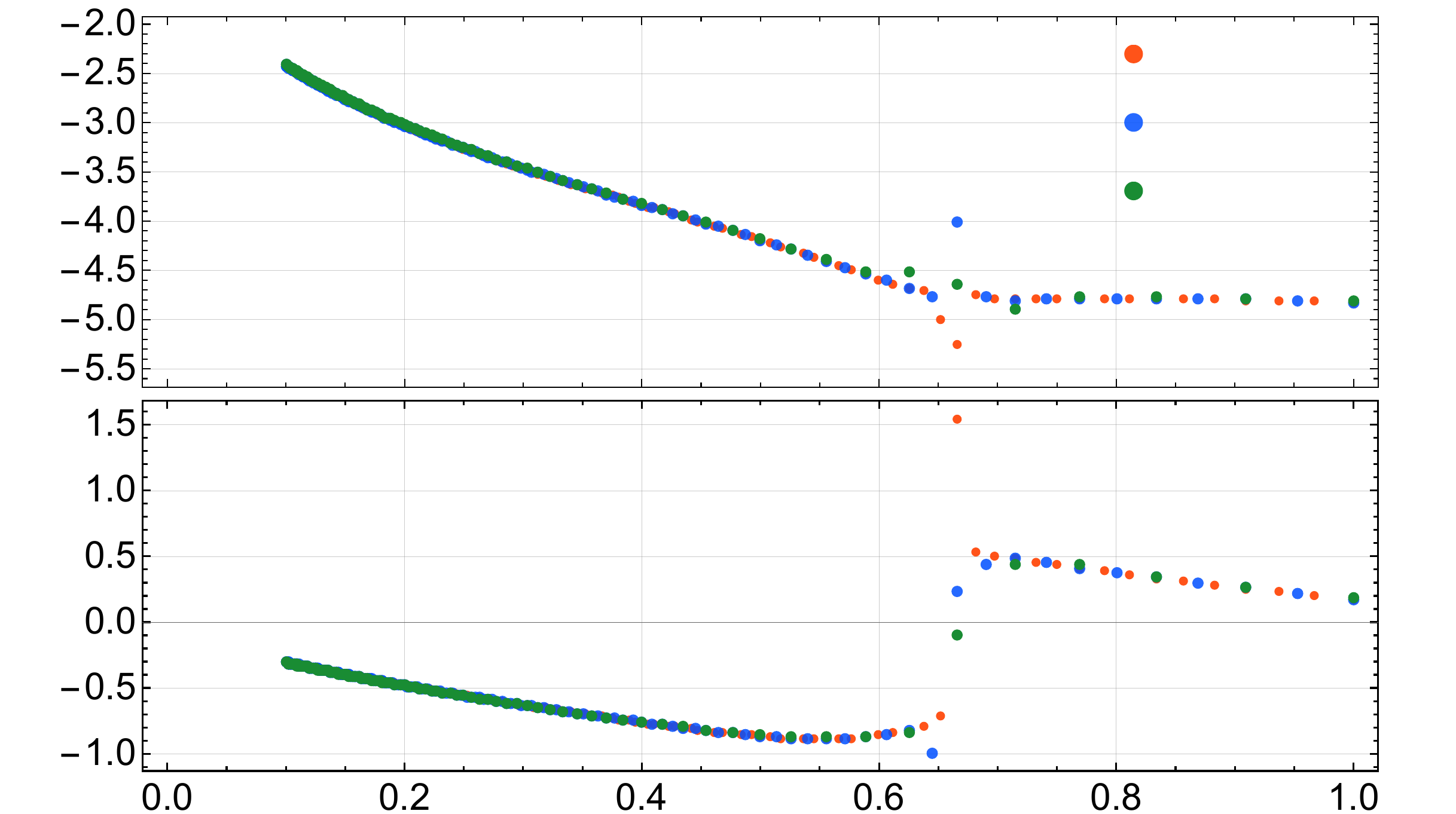} }

\put(150,0){$\gamma$}
\put(430,0){$\gamma$}

\put(3,75){\rotatebox{90}{$ {\rm Re} \langle A\rangle/( \Lambda \ell_P^2)$}}
\put(3,15){\rotatebox{90}{$ {\rm Im} \langle A\rangle/( \Lambda \ell_P^2) $}}
\put(258,98){\rotatebox{90}{$ {\rm Re} \langle \omega \rangle $}}
\put(258,40){\rotatebox{90}{$ {\rm Im} \langle \omega\rangle $}}

\put(430, 120){ \scriptsize   $\Lambda = 3 $ }
\put(430, 109.5){ \scriptsize   $\Lambda = 2 $ }
\put(430, 99){ \scriptsize   $\Lambda = 1 $ }

\end{picture}
\caption{ The left panel shows plots for the expectation value for the bulk spin label $j_A=A/(\gamma \ell_P^2)$ (normalized by $\Lambda$) and the right panel shows plots for the expectation value of the sum of the dihedral angles $\omega$ at the bulk triangle for the path integral that  sums over time-like and space-like irregular configurations. The boundary data induce a classical angle $\omega_c=0$.
\label{FigNullA}
} 
\end{figure}

In figure \ref{FigNullA} we show the expectation values for the  area  $A/( \Lambda \ell_P^2)$ of the bulk triangle  and the expectation value for the  sum of the dihedral angles $\omega= \sum_\sigma \theta^\sigma_t$  (whose classical values are real for both negative and \JPp{positive} $A^s$, and is zero for $A^s=0$)  for $\Lambda = 1,2,3$.   These expectation values basically reflect the shift of the peak in the absolute value of the amplitude $|G(j,\gamma)\times \exp(2\pi A)|$ with larger $\gamma$. We see that the area expectation values are approximated by the values $A_p(\gamma)$. E.g. for $\Lambda=1$ we have $A_ p(1)\approx 18 \ell_p^2$, $A_p(\gamma) \approx 8\ell_p^2$ and $A_p(0.01)\approx 4\ell_p^2$.  We can for instance require that the deviation from the classical value $A=0$ should not be more than, say half of the largest boundary area, which is given by $|C|=11\ell_p^2$. This would exclude $\gamma=1$ and even $\gamma=0.1$.

Thus, allowing for such causally irregular configurations, with an enhancing effect, and demanding a semi-classical regime for such configurations sets strong bounds on $\gamma$.

\begin{figure}[ht!]
\begin{picture}(500,140)
\put(1,7){ \includegraphics[scale=0.272]{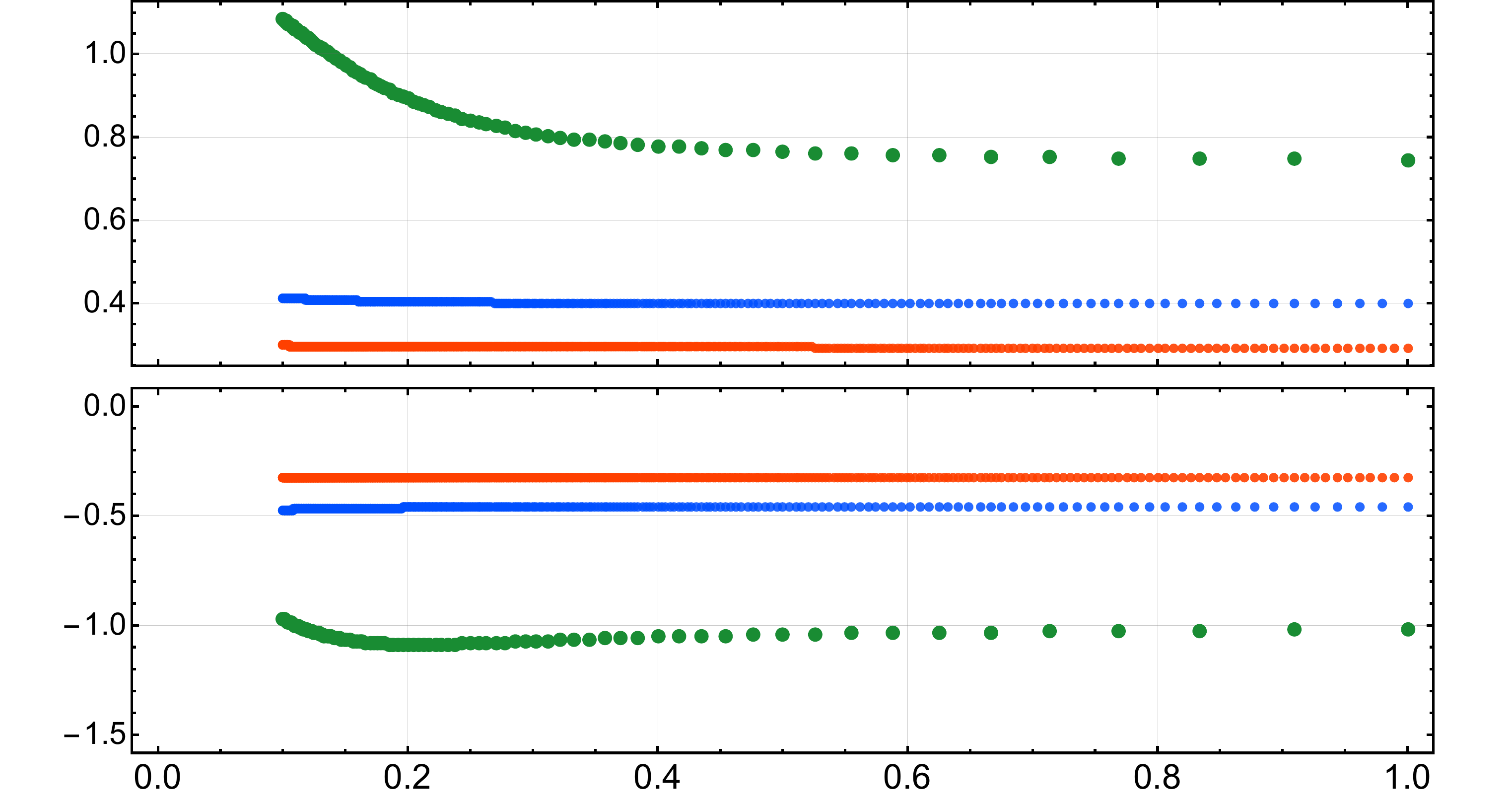} }
\put(255,7){ \includegraphics[scale=0.265]{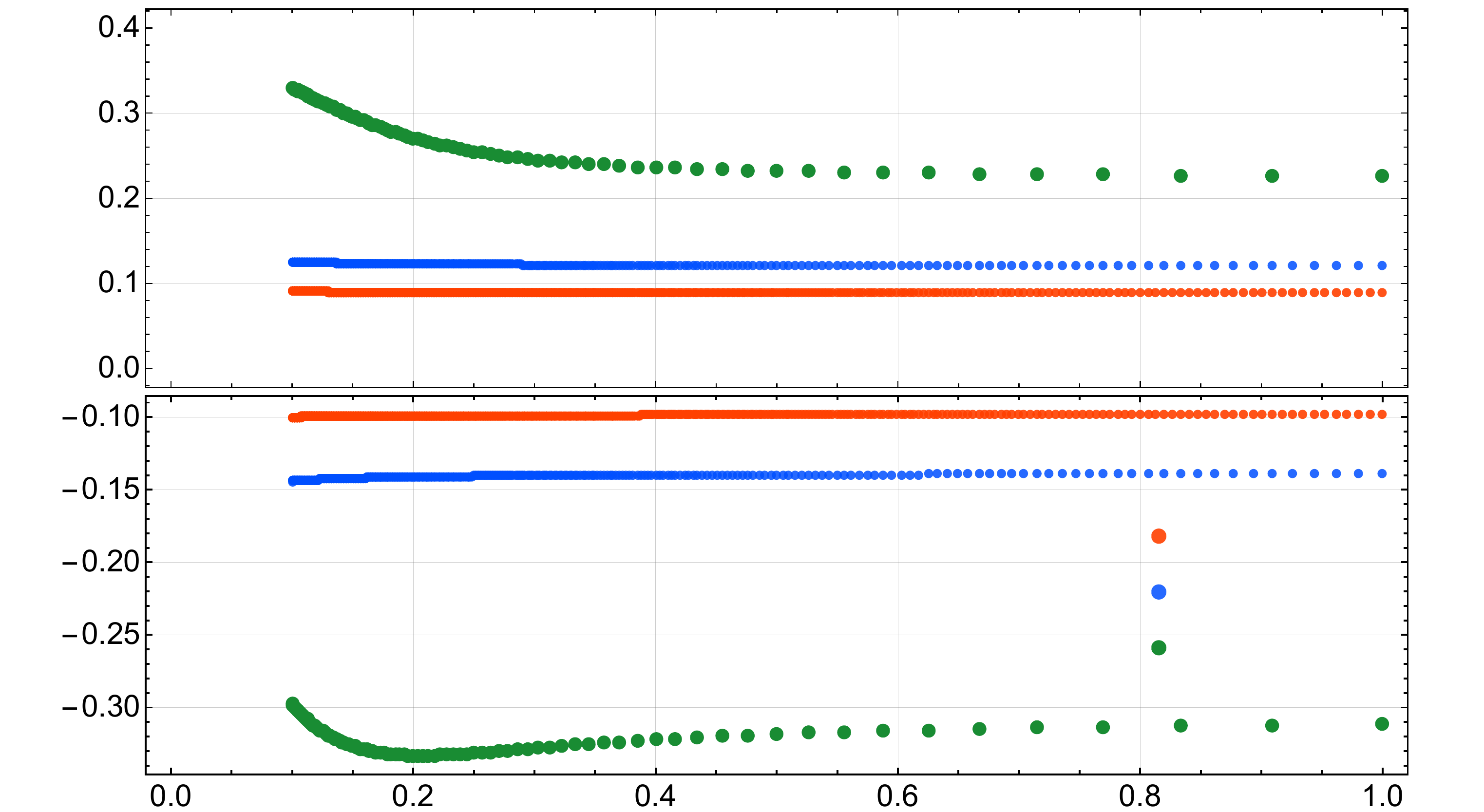} }

\put(150,0){$\gamma$}
\put(430,0){$\gamma$}

\put(3,75){\rotatebox{90}{$ {\rm Re} \langle A\rangle/( \Lambda \ell_P^2) $}}
\put(3,15){\rotatebox{90}{$ {\rm Im} \langle A\rangle/( \Lambda \ell_P^2)$}}
\put(258,98){\rotatebox{90}{$ {\rm Re} \langle \omega \rangle $}}
\put(258,40){\rotatebox{90}{$ {\rm Im} \langle \omega\rangle $}}

\put(441, 48){ \scriptsize   $\Lambda = 10 $ }
\put(441, 39){ \scriptsize   $\Lambda = 5 $ }
\put(441, 30){ \scriptsize   $\Lambda = 1 $ }

\end{picture}
\caption{ These figures show expectation values for the bulk spin label $j_A=A/(\gamma \ell_P^2)$, normalized by $\Lambda$ (left panel) and the expectation value of the sum of the dihedral angles $\omega$ (right panel) for the path integral excluding configurations with irregular light cone structure. 
\label{FigRegA}
} 
\end{figure}
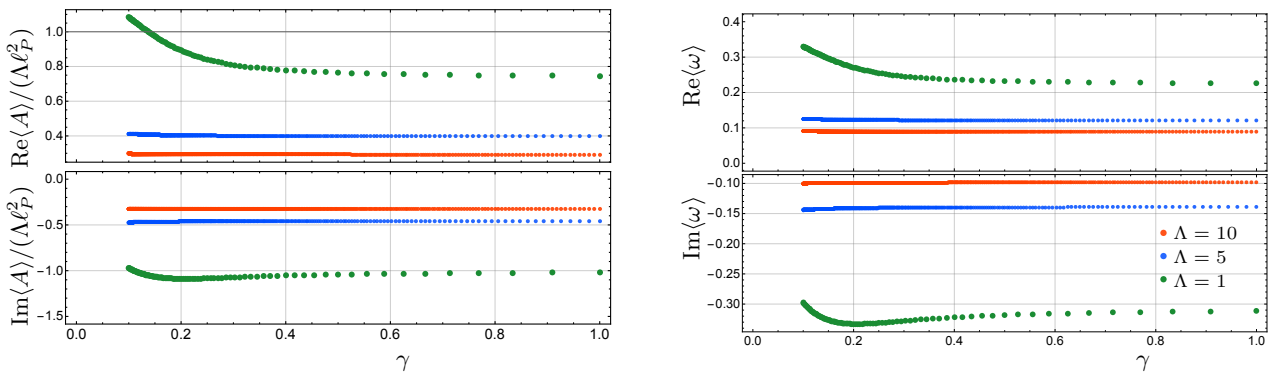

Excluding such configurations from the path integral, we \JPp{are left} with the sum over time like areas $A^s=-n^2$ with $n\in \mathbb{N}$.  Figure \ref{FigRegA} shows the expectation values for the area $A$ and the angle $\omega$ in this case. We see that the approximation to the classical value does surprisingly improve   for larger scales $\Lambda$, and is very stable over  $\gamma\in (0.1,1)$. For the area expectation values this has to be understood in reference to the boundary area values, i.e. we normalize the deviation from the classical value by $\Lambda$.

We have seen that allowing a sum over configurations with irregular light cone structure, and more precisely configurations  were the number of light cones at an inner triangle is less than two, can lead easily to a domination of the path integral by these configurations. This is of course only the case if we weight such configurations with the $\exp(N \pi A_t)$ factor, where $N=1$ if there is only one light cone at the bulk triangle $t$ and $N=2$ if there are zero light cones.  This weight is suggested by the Regge action, which becomes complex for causally irregular configurations \cite{Sorkin2019}. It is not clear yet \JPp{whether} such enhancing or suppressing weights do appear for e.g. the EPRL-FK spin foams. If they do appear, and one does not exclude such (enhanced) irregular configurations from the path integral, the results in this section suggest that one would have to choose  a very small $\gamma$ value \JPp{in} order to obtain a suitable classical limit.  The situation analyzed here does \JPp{not appear in this form} in Length Regge calculus -- here one would have \JPp{sharp} constraints and for the triangulation chosen here, no summation at all. But it would already show that we can easily construct  triangulations with  irregular light cone structure of an enhancing type, which might also dominate the path integral for Length Regge calculus.

On the other hand, excluding these causally irregular configurations, the expectation values do rather well approximate the classical value (if we consider the size of the deviations  relative to the boundary areas) for the  $\gamma$-range we tested, and the deviations (normalized by $\Lambda$) do even get better with larger $\Lambda$.

\section{Discussion}\label{sec:discuss}

In this work we defined  effective spin foam models for Lorentzian geometries. As for the corresponding Euclidean model \cite{EffSF1,EffSF2} we incorporate the key dynamical principles of spin foams: $(a)$  Areas are the fundamental variables and come with a discrete spectrum.  The parametrization of triangulations with areas does lead to a more general configuration space of geometries \JPp{than} the one obtained using length variables \cite{DittrichRyan1and2}, the spin foam path integral is therefore  a sum over area geometries. And $(b)$: To obtain a gravitational dynamics\JPp, constraints need to be imposed on these fundamental variables. Sourced by the discreteness of the area spectra, the constraint algebra is anomalous. The constraints are therefore imposed weakly but as strongly as allowed by \JPp{their non-commutativity}. 

The philosophy of effective spin foams has shown its flexibility for the Lorentzian case constructed here: we could construct the model for all possible signatures of simplices, in particular allow for space-like and time-like triangles.  To this end we have introduced a new more convenient form of the constraints.  

The Lorentzian model has also shown an additional role for the Barbero-Immirzi parameter. As it parametrizes the area gap for spatial triangles, but not for time-like triangles, it can be seen as an anisotropy parameter, similar to the one used in Causal Dynamical Triangulations \cite{CDT}. The Barbero-Immirzi parameter \JPp{also scales} the anomaly in the constraint algebra, and can be interpreted to determine how much fluctuations one allows in the path integral away from geometries with a consistent length \JPp{assignment}. 

Independently from the particularities of the spin foam approach, there are many open questions regarding the precise definition of the  Lorentzian path integral for quantum gravity. This does in particular concern the question of which kind of (generalized) geometries one allows in the path integral: Does one include configurations with irregular causal structure? Should one sum over orientations? Can we peak on configurations where many simplices are null?  What is the consequence of  restricting to triangulations \JPp{where} all tetrahedra are space-like, as in the (standard) EPRL-FK model? 

To answer these questions, it is important to have a computationally efficient model. Many of these questions can be asked in a number of approaches, and \JPp{also appear} to be independent from the precise choice of amplitudes for the model. 

Using effective spin foams we have performed a number of first tests on the viability of the model. We have firstly shown that for simple triangulations with one inner  space-like or time-like triangle, there is a regime of parameters for which the expectation values can reproduce the classical values. This includes configurations with (relatively large) curvature. We could also reproduce a semi-classical behaviour for a configuration peaked on a null area -- and an extreme curvature angle of $\epsilon=2\pi$. In this case, the semi-classical behaviour even improved (if the deviations are divided by the scale of the boundary areas), contrary to expectations resulting from the weak imposition of the constraints, with larger scales.

The simple triangulations we considered already allowed for configurations with irregular light cone structure and a first test whether such configurations should be allowed in the path integral. Such configurations have a complex action, suggesting that these are either enhanced or suppressed. We have seen that an enhancing effect would set very strong bounds on the Barbero-Immirzi parameter.  Excluding such configurations with irregular structure we could nevertheless peak on the area value (which in this case was null), which separated the regular from the irregular regime.  The enhancing effect grows exponentially with the size of the areas, and in our example this area could become arbitrary large, that is was not constrained by the triangle inequalities. This suggest that such causally  irregular configurations, leading to an enhancing effect, should be also excluded in length Regge calculus.   It is not clear yet how, e.g. the EPRL-FK  spin foams treat such causally irregular configurations. A better understanding of this issue would facilitate very much the extraction of physical properties encoded in the models.

In previous works \cite{EffSF1,EffSF2} and in this work we have also made use of symmetry reductions. In the cases considered here this symmetry reduction affected only the boundary data and not the dynamical variables. But it shows that such symmetry reductions can be easily implemented. That makes the Lorentzian effective spin foams ideally suited for applications to quantum cosmology \cite{CosmoReview} and black holes \cite{BHReview}.  It would be also interesting to introduce a further variant of the model that allows for homogeneously curved simplices \cite{CurvedS}, which is one way to incorporate a cosmological constant.

The numerical efficiency of the effective spin foam model, and the transparent way it encodes the key dynamical principles of spin foams, will also help to understand whether spin foams can have a satisfactory continuum limit \cite{DittrichReview14,CLimit}. Due to the minimal ingredients of the effective model, this would show whether having areas as independent variables with  discrete spectrum is a viable option for quantum gravity.

\appendix

\section{Generalized triangle \JPp{inequalities} and signed volume squares}\label{app:triangle}

The generalized triangle inequalities for a simplex $\sigma$ can be formulated as conditions on  the signed squared volumes $\text{Vol}_\sigma^s$ for $\sigma$  and the signed squared volumina $\text{Vol}_\rho^s$   for all its sub-simplices $\rho$ \cite{Sorkin74,Visser1,Visser2}.

The signed squared volume of a $d$-dimensional simplex  $\rho$ with vertices $(012\cdots d)$ can be computed via the \JPp{Cayley}-Menger determinant
\ba
\text{Vol}_\rho^s&\,=\,& \frac{ (-1)^{d+1} }{ 2^d (d!)^2}
\left(
\begin{matrix}
0 & 1 & 1 & 1 & \cdots & 1 \\
1 & 0 & l_{01}^s & l_{02}^s & \cdots & l_{0d}^s \\
1 & l_{01}^s & 0 & l_{12}^s & \cdots & l_{1d}^s \\
\vdots & \vdots & \vdots & \vdots & \ddots & \vdots \\
1 & l_{0d}^s & l_{1d}^s & l_{2d}^s & \cdots & 0
\end{matrix}
\right) \quad ,
\ea
where $l_{ij}^s$ is the signed squared length of the edge $e_{ij}$ between vertices $i$ and $j$. 
The signed squared volume gives  the square of the volume for a space-like simplex $\rho$, and minus the square of the volume for a time-like simplex $\rho$. The signed squared volume \JPp{therefore} also \JPp{determines} the space-like or time-like nature of a simplex. If the volume is zero, the $d$-simplex can be space-like or time-like, but degenerate\footnote{That is, the $d$-simplex can be embedded into an $(d-1)$-dimensional subspace of Euclidean or Minkowski space.}, or can be embedded into a null hyperplane of $(d+1)$-dimensional Minkowski-space. 

An Euclidean or space-like simplex $\sigma$ with signed squared edge lengths $l_e^s$ can be realized \JPp{if}  $\text{Vol}_\sigma^s\geq0$ and $\text{Vol}_\rho^s\geq0$ for all the  sub-simplices $\rho$ of $\sigma$. `Realizing' a Euclidean $d$--dimensional simplex means to find an embedding of $(d+1)$ vertices in  $d$--dimensional Euclidean space, so that the distance squares between the vertices coincide with the corresponding squared edge lengths $l_e^s$.

`Realizing' a Lorentzian or time-like $d$--dimensional simplex $\sigma$ means to find an embedding of $(d+1)$ vertices in  $d$--dimensional Minkowski space, so that the distance squares between the vertices coincide with the corresponding signed squared edge lengths $l_e^s$. The sub-simplices of a  Lorentzian simplex $\sigma$ can be time-like, space-like or null.  But if a sub-simplex $\rho'$ is  time-like, then all simplices $\rho$ containing this sub-simplex $\rho$ have to be also time-like.  For a time-like simplex $\sigma$ we  have thus the condition that $\text{Vol}^s_\sigma \leq 0$ and the following requirement: if there is a sub-simplex $\rho'$ with $V_{\rho'}^s<0$ then all sub-simplices $\rho$ with $\rho'\subset \rho$ need to satisfy $V_{\rho'}^s \leq 0$. There is a certain class of simplices $\rho$ where this condition is automatically satisfied \cite{Visser2}: If $\rho$ is an $n$-dimensional simplex with one and only one space-like $(d-1)$-dimensional sub-simplex $\rho''$ (that is $V_{\rho''}^s > 0$) and all edges $e$, which are not in $\rho''$ are time-like, then one has that $V_{\rho'}^s\leq 0$.

\section{Dihedral angles in a Lorentzian simplex}\label{app:angles}

\begin{figure}
\begin{tikzpicture}[scale=0.55]
\draw[gray!70] (-5,0)--(5,0) (0,-5)--(0,5);
\draw[dashed, red] (-4.5,-4.5)--(4.5,4.5) (-4.5,4.5)--(4.5,-4.5);
\draw[thick,<-] (-4,-5) parabola bend (0,-2.5) (4,-5);
\draw[thick,<-]  (4,5) parabola bend (0,2.5)(-4,5);
\draw[thick,rotate=90 ,->] (-4,-5) parabola bend (0,-2.5) (4,-5) ;
\draw[thick,rotate=90 ,->] (4,5) parabola bend (0,2.5)(-4,5);

\draw [->,red] (0,0)--(2.5,0.5) ;
\draw [->] (0,0)--(3.,1.9) ;
\draw [->] (0,0) -- (1.9,3.) ;
\draw [->] (0,0) -- (-2.8,1.3) ;
\draw [->] (0,0) -- (-0.8,2.6) ;
\draw [->] (0,0) -- (-0.8,-2.6) ;

\node[rotate = 45] at (3.6,5.2) {\scriptsize $+\infty - i \pi/2$};
\node[rotate = -45] at (-3.6,5.2) {\scriptsize $-\infty - i \pi/2$};
\node[below, rotate = -45] at (-4.6,3.8) {\scriptsize $-\infty - i \pi$};

\node[rotate = 45] at (-4.8,-3.2) {\scriptsize $+\infty - i \pi$};
\node[rotate = 45] at (-3.6,-5.2) {\scriptsize $+\infty - i 3\pi/2$};

\node[rotate = -45] at (3.6,-5.2) {\scriptsize $-\infty - i 3\pi/2$};
\node[below, rotate = -45] at (5.8,-3.8) {\scriptsize $-\infty - i 2\pi$};

\node at (4,0.5) {$C_1$};
\node at (0.5,4) {$C_2$};
\node at (-4,0.5) {$C_3$};
\node at (0.5,-4) {$C_4$};

\node at (1.7,0.6) {$\hat s$};
\node at (2.3,1.8) {$\hat s_1$};
\node at (0.8,2.) {$\hat t_1$};
\node at (-1.7,0.5) {$\hat s_2$};
\node at (-0.8,1.5) {$\hat t_2$};
\node at (-0.8,-1.5) {$\hat t_3$};

\end{tikzpicture}
\caption{Lorentzian angles defined by wedges in between vectors in 2D Minkowski space.}
\label{angles}
\end{figure}
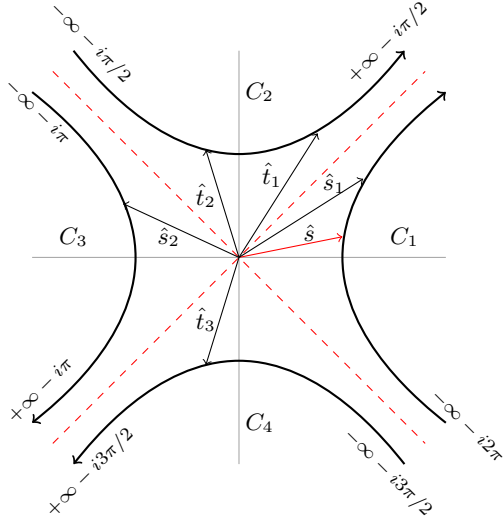

The notion of dihedral angle in a Lorentzian simplex is more subtle than its Euclidean counterpart. One reason is that whereas an angle in two-dimensional Euclidean geometry  parametrizes  distances on the circle, an angle in two-dimensional Lorentzian geometry has to somehow parametrize the distance between points on four disconnected hyperbolae, see Figure \ref{angles}. We will follow here Sorkin's convention \cite{Sorkin2019}, where the problem of having disconnected sheets is addressed by allowing the angles to have imaginary parts.

In  \cite{Sorkin2019} angles are defined for wedges, that are spanned between two vectors in $\mathbb M^2$. That is, given two vectors $(v_1,v_2)$ one has to specify whether the wedge is given by going anti-clockwise from $v_1$ to $v_2$ or from $v_2$ to $v_1$.  These angles have the property that they are finite and additive.  A  wedge, which in Euclidean geometry would have a $2\pi$ angle, has an angle of $-\imath 2\pi$ in $\mathbb M^2$. The notion of wedge and additivity allows also to make sense of angles with larger winding numbers, that is angles with values $-\imath 2\pi N$ with $N \in \mathbb{N}$. Such angles can appear in Regge calculus, as a deficit angle is formed from summing a number of dihedral angles. As discussed in  \cite{Sorkin2019}  and in section \ref{Sec:ImPart}   such angles occur for configurations which describe topology change (in time), e.g. a trouser-like space-time. 

Alexandrov \cite{Alexandrov2001}  has defined another notion of angle, namely oriented angles between two non-null vectors in Minkowski space.  Here angles  which differ by  multiples of $2\pi \imath$ are identified with each other, that is  information on winding  is not kept.

We refer for motivation and background to \cite{Sorkin2019} and will here just provide the definition of the Lorentzian (wedge) angles.

The null-cone splits $\mathbb M^2$ into space-like and time-like  regions (see Figure \ref{angles}). For simplicity, we only treat cases for which the vectors are either space-like or time\JPp-like and normalized. We also restrict to convex wedges, as only those can occur in a simplex.   
For a  space-like normalized vector $\hat s$ and  a space-like normalized 
\JPp{vector} $\hat s_i$ or  a time-like normalized vector $\hat t_i$, depicted in Figure \ref{angles},  the definitions  of angles for convex wedges are as follows:
\ba\label{ad1}
\theta_1  &=& \cosh^{-1} ({\hat s \cdot \hat s_1} )\q \q \q\q\q \text{if}  \q\q {\hat s \cdot \hat s}=+1 \;\&\;  {\hat s_1 \cdot \hat s_1}=+1  \;\&\; {\hat s \cdot \hat s_1} \geq +1 \,\, ,  \nn\\
\theta_2  &=& -\cosh^{-1} (-{\hat s \cdot \hat s_2} ) - {\pi \imath }  \,\q \text{if} \q\q {\hat s \cdot \hat s}=+1 \;\&\;  {\hat s_2 \cdot \hat s_2}=+1  \;\&\; {\hat s \cdot \hat s_2} \leq -1 \,\, , \nn\\
\theta_3  &=& \sinh^{-1} ({\hat s \cdot \hat t_1} ) - \tfrac{ \pi}{2} \imath   \q\q  \q \text{if} \q\q {\hat s \cdot \hat s}=+1 \;\&\;  {\hat t_1 \cdot \hat t_1}=-1 \q . 
\ea
Here we define $\cosh^{-1}(x) \in \mathbb{R}_+$ with $x \geq 1$.  The additivity property can be used to compute angles between time\JPp-like vectors. The angles between time-like vectors  of the type shown in Figure \ref{angles} are given by
\ba\label{ad2}
\theta_4  &=& -\cosh^{-1} (-{\hat t_1 \cdot \hat t_2} ) \q \q \q\q\q \text{if}  \q\q {\hat t_1 \cdot \hat t_1}=-1 \;\&\;  {\hat t_2 \cdot \hat t_2}=-1  \;\&\; {\hat t_1 \cdot \hat t_2} \leq -1 \,\, ,   \nn  \\
\theta_5  &=& \cosh^{-1} ({\hat t_2 \cdot \hat t_3} ) - \pi \imath \, \,\;\q \q\q\q \text{if}  \q\q {\hat t_2 \cdot \hat t_2}=-1 \;\&\;  {\hat t_3 \cdot \hat t_3}=-1  \;\&\; {\hat t_2 \cdot \hat t_3} \geq +1 \,\,  . 
\ea
Using these definitions, the angles in a triangle in $\mathbb M^2$ sum to $-\pi \imath$, and angles around a point in flat space sum to $-2\pi \imath$. 

{Figure \ref{angles} does also exemplify the notion of the number of light cone crossings: The wedges associated to the angles $\theta_1$ to $\theta_5$ include $N_1=0,N_2=2,N_3=1,N_4=0,N_5=2$ crossings of the light cone boundary, respectively. The imaginary parts of these angles are therefore given by $\text{Im}(\theta_i)= -\imath N_i (\pi/2).$

}

The  definitions (\ref{ad1},\ref{ad2}) are sufficient to define the dihedral angles in a Lorentzian 4-simplex. The internal dihedral angle at a triangle $t$  is defined to be the angle of the wedge at $t$, which one obtains after projecting to the hyperplane orthogonal to $t$. This  projection will map the 4-simplex to a triangle $t'$, and the triangle $t$ to a vertex $v'$ of this triangle. \JPp{Said} triangle is space-like if $t$ is time-like and time-like if  $t$ is space-like. Thus, if $t$ is space-like one  applies the definitions (\ref{ad1}) and (\ref{ad2}) for Lorentzian angles, and if $t$ is time-like the definition 
\ba\label{ad3}
\theta_E=\cos^{-1}(\hat s'\cdot \hat s'') \q \q \q\q\q \text{if}  \q\q {\hat s' \cdot \hat s'}=+1 \;\&\;  {\hat s'' \cdot \hat s''}=+1  \;\&\; {|\hat s' \cdot \hat s''|} \leq 1 
\ea
 for an Euclidean angle (with $\theta_E \in (0, \pi)$). 

 \vspace{3mm}

We will now provide more explicit formulas for the computation of the dihedral angles in terms of edge lengths. To this end we denote for a 4-simplex $\sigma$ and a triangle $t\subset \sigma$ the signed length square of the triangle $t'$, that results from the projection of $\sigma$,  by $(L^s_{\tau_1},L^s_{\tau_2},L^s_{\bar{t}})$  where  $L^s_{\bar{t}}$ is the length square of the edge in $t'$ opposite to the vertex $v'$ that results from projecting $t$.  $(\tau_1,\tau_2)$ are the two tetrahedra in $\sigma$ that share $t$. These lengths can then be computed as
\ba
L^s_{\tau_1}=9 \frac{  \text{Vol}_{\tau_1}^s }{\text{Vol}_{t}^s}  \, ,\q
L_{\tau_2}^s=9 \frac{  \text{Vol}_{\tau_2}^s }{ \text{Vol}_{t}^s }  \, ,\q 
L^s_{\bar{t}}= \left( 9 \text{Vol}_{\tau_1}^s +9 \text{Vol}_{\tau_2}^s -288  \frac{\partial \text{Vol}_{\sigma}^s}{\partial l_{\bar{t}}^s}\right) ( \text{Vol}_{t}^s )^{-1} \q .
\ea
where $\bar{t}$ denotes the edge in $\sigma$, that is opposite to $t$.

The inner product between the  normalized edge vectors $\hat e_{\tau_1}$ and $\hat e_{\tau_2}$ (which start from $v'$) of the triangle $t'$ is given by
\ba
\hat e_{\tau_1} \cdot \hat e_{\tau_2}& =& \frac{1}{2}\frac{L_{\tau_1}^s+ L_{\tau_2}^s-L_{\bar{t}}^s}{\sqrt{ |L_{\tau_1}^s L_{\tau_2}^s|}}
\, \,=\,\, \text{sign}(\text{Vol}^s_{t} )  \frac{4^2}{\sqrt{|  \text{Vol}_{\tau_1}^s  \text{Vol}_{\tau_2}^s         |}}
 \frac{\partial \text{Vol}_{\sigma}^s}{\partial l_{\bar{t}}^s}  
   \, , \ea
and the signatures of $\hat e_{\tau_1}$ and $\hat e_{\tau_2}$ are 
\ba
\hat e_{\tau_1} \cdot \hat e_{\tau_1}  & =&\text{sign}( \text{Vol}_{\tau_1}^s  \text{Vol}_{t}^s)\, ,\q\q
     \hat e_{\tau_2}  \cdot \hat e_{\tau_2} \,\, =\text{sign}( \text{Vol}_{\tau_1}^s  \text{Vol}_{t}^s)\,.
\ea
One can now apply the formulas (\ref{ad1},\ref{ad2}) or formula (\ref{ad3}) to compute the dihedral angle. 
This \JPp{also reproduces} the dihedral angles for Euclidean 4-simplices \cite{DittrichFreidelSpeziale}. 

 \vspace{3mm}

Similarly, we have for the 3D dihedral angle at an edge $e$ in a tetrahedron $\tau$
\ba
\hat e_{t_1} \cdot \hat e_{t_2}\,\, =\,\,   \text{sign}(l^s_{e} )  \frac{3^2}{\sqrt{|  \text{Vol}_{t_1}^s  \text{Vol}_{t_2}^s         |}}
 \frac{\partial \text{Vol}_{\tau}^s}{\partial l_{\bar{e}}^s}  \, ,\q 
 \hat e_{t_1} \cdot \hat e_{t_1}=\text{sign}( \text{Vol}_{t_1}^s l^s_e )  \, ,\q 
 \hat e_{t_2} \cdot \hat e_{t_2}=\text{sign}( \text{Vol}_{t_2}^s l^s_e ) 
\ea
where $(t_1,t_2)$ are the two triangles sharing $e$ and $\bar{e}$ is the edge opposite $e$. 

The projection of the tetrahedron to the plane orthogonal to the edge $e$ does not affect the 
normals $n_{t_1}$ and $n_{t_2}$ to the triangles $(123)$ and $(124)$.  These appear now as normals to the edges $e_{t_1}$ and $e_{t_2}$ of the triangle $t'$. The angles in this triangle $t'$ can be computed either from the edge vectors or the normals, as we have
\ba
\hat n_{t_1} \cdot \hat n_{t_2} = -\text{sign}(\text{Vol}_\tau^s \, l_e^s)\, \hat e_{t_1} \cdot \hat e_{t_2} 
\ea
Here we used that the signature of $t'$ is space-like if $e$ is time-like and vice versa, and that $  \hat n_{t_1} \cdot \hat n_{t_2}=-\hat e_{t_1} \cdot \hat e_{t_2}$ if $t'$ is space-like and $  \hat n_{t_1} \cdot \hat n_{t_2}=\hat e_{t_1} \cdot \hat e_{t_2}$ if $t'$ is time-like.

For the quantities $p_{t_1t_2}^\tau=\text{sign}( \text{Vol}_{\tau}^s) \,n_{t_1}\cdot n_{t_2}$ discussed in section \ref{Sec:Tetrahedron}, we therefore have
\ba
p_{t_1t_2}^\tau&=&-3^2 \frac{\partial \text{Vol}_{\tau}^s}{\partial l^s_{\bar{e}}} 
\ea
where $\bar{e}$ is the edge in $\tau$, which is not a sub-simplex of $t_1$ or $t_2$.

\section{Equations of motion from the constrained Area Regge action}\label{app:eom}

Here we will show that the constrained Area Regge action
\ba\label{CAR}
S_{CAR} \,=\, \sum_t a_t \epsilon_t (a) \,+\, \sum_\tau \sum_{i=1,2} \lambda^i_\tau \left( P_i^{\tau,\sigma}(a)-P_i^{\tau,\sigma'}(a)\right)
\ea
does lead to the Length Regge equation of motion (\ref{LREOM}). Here we assume a triangulation without boundary. Furthermore, we have that $P_i^{\tau,\sigma}(a)=p^\tau_i(L^\sigma(a))$, where the $p^\tau_i(l)$ are defined in section \ref{Sec:Tetrahedron}, and $L_e^\sigma(a)$ are 10 functions of the areas of the simplex $\sigma$ \JPp{that} determine the lengths of the edges of this simplex. 

Taking the variation of this action with respect to an area $a_t$, we obtain the equation of motion
\ba\label{C2}
\epsilon_t +  \sum_\tau \sum_{i=1,2} \lambda^i_\tau \left( \frac{\partial P_i^{\tau,\sigma}(a)}{\partial a_t} -\frac{\partial P_i^{\tau,\sigma'}(a)}{\partial a_t}\right) \, \stackrel{!}{=} \,0
\ea
Here we used the Schl\"afli identity \cite{Sorkin74} for the 4-simplices, which ensures that the variation of the deficit angles $\epsilon_{t}$ vanishes. 

Variation of the action $S_{CAR}$ with respect to the Lagrange \JPp{multipliers} $\lambda^i_\tau$ enforces the constraints
\ba
P_i^{\tau,\sigma}(a)-P_i^{\tau,\sigma'}(a)  \, \stackrel{!}{=} \,0 \q ,
\ea
that ensure that the lengths are well defined in terms of the edges, that is $L_e^\sigma(a)=L_e^{\sigma'}(a)$.  Given such a configuration of areas we multiply the equations of motion (\ref{C2}) with $\partial A_t(l)/\partial l_e$ and sum over all triangles (containing $e$).  We then use for the constraint terms
\ba\label{C4}
\sum_t \frac{\partial P_i^{\tau,\sigma}(a)}{\partial a_t} \frac{\partial A_t(l)}{\partial l_e} \,=\, 
\sum_t  \sum_{e'}\frac{\partial p_i^{\tau}(l)}{\partial l_{e'}} \frac{ \partial L_{e'}^\sigma(a) }{\partial a_t} \frac{\partial A_t(l)}{\partial l_e} \,=\, 
\frac{\partial p^\tau_i( l)}{\partial l_e} \q ,
\ea
where the right hand side of (\ref{C4}) is independent of the simplex $\sigma$. Thus, the constraint terms in (\ref{C2}) vanish \JPp{when} contracted with $\partial A_t(l)/\partial l_e$. We obtain the equations of \JPp{motion}
\ba
\sum_t  \frac{\partial A_t(l)}{\partial l_e} \epsilon_t   \, \stackrel{!}{=} \,0 \q ,
\ea
which agree with the Length Regge equation of motion. 

We obtained these equations of \JPp{motion} by contracting the derivative of the action (\ref{CAR}) with variations in the areas, which are tangential to the constraint hypersurface. This constraint hypersurface describes area configurations which result from a consistent lengths assignment to the edges. Thus  the tangent space to a given point in this constraint hypersurface is  spanned by $\{(V^e)_t=\partial A_t(l)/\partial l_e\}_e$. 

Contracting the equations (\ref{C2}) with vectors transversal to this tangent space, e.g. the gradients of the constraints in area configuration space, we obtain equations that fix the Lagrange \JPp{multipliers} $\lambda_\tau^i$.

\begin{acknowledgments}
\emph{Acknowledgments}.  BD thanks  Steffen Gielen, Hal Haggard, Wojciech Kaminski, Aldo Riello and Susanne Schander for discussions. 
Research at Perimeter Institute is supported in part by the Government of Canada through the Department of Innovation, Science and Economic Development Canada and by the Province of Ontario through the Ministry of Colleges and Universities.
\end{acknowledgments}

\providecommand{\href}[2]{#2}
\begingroup
\endgroup

\end{document}